\documentclass[superscriptaddress,noshowpacs,noshowkeys, twocolumn, floatfix,aps, prb,reprint]{revtex4-1}
\usepackage[english]{babel}
\usepackage{fontenc}
\usepackage{graphicx}
\usepackage{amsmath}
\usepackage{epstopdf}
\usepackage{amssymb}
\usepackage{amsbsy}
\usepackage{amscd}
\usepackage{color}
\usepackage{bbm}
\usepackage{braket}
\usepackage{bm}
\usepackage{siunitx}
\usepackage[normalem]{ulem} 
\usepackage{verbatim}
\usepackage{url}
\usepackage[verbose,hypertexnames=false,bookmarksopenlevel=1,filecolor=blue,
linkcolor=blue,citecolor=blue,urlcolor=blue,pdfstartview=FitH,bookmarksopen,bookmarksnumbered,
colorlinks,plainpages=false,linktocpage]{hyperref}
\DeclareSymbolFont{mathbold}{OML}{cmm}{b}{it}
\DeclareMathSymbol{\bsigma}{\mathord}{mathbold}{27}
\begin{document}
\renewcommand{\figurename}{Fig.}
\title{Spin relaxation in wurtzite nanowires}
\author{Michael Kammermeier}
\email{michael1.kammermeier@ur.de}
\affiliation{Institute for Theoretical
 Physics, University of Regensburg, 93040 Regensburg, Germany}
  \author{Paul Wenk}
\affiliation{Institute for Theoretical
 Physics, University of Regensburg, 93040 Regensburg, Germany}
 \author{Florian Dirnberger}
\affiliation{Institute for Experimental and Applied Physics, University of Regensburg, 93040 Regensburg, Germany}
 \author{Dominique Bougeard}
\affiliation{Institute for Experimental and Applied Physics, University of Regensburg, 93040 Regensburg, Germany}
\author{John Schliemann}
\affiliation{Institute for Theoretical
 Physics, University of Regensburg, 93040 Regensburg, Germany}
\date{\today }
\begin{abstract}
We theoretically investigate the D'yakonov-Perel' spin relaxation properties in diffusive wurtzite semiconductor nanowires and their impact on the quantum correction to the conductivity. 
Although the lifetime of the long-lived spin states is limited by the dominant $k$-linear spin-orbit contributions in the bulk, these terms show almost no effect in the finite-size nanowires.
Here, the spin lifetime is essentially determined by the small $k$-cubic spin-orbit terms and nearly independent of the wire radius.
At the same time,  these states possess in general a complex helical structure in real space that is modulated by the spin precession length induced by the $k$-linear terms.
For this reason, the experimentally detected spin relaxation largely depends on the ratio between the nanowire radius and the spin precession length as well as the type of measurement. 
In particular, it is shown that while a variation of the radius hardly affects the magnetoconductance correction, which is governed by the long-lived spin states, the change in the spin lifetime observed in optical experiments can be dramatic. 
We compare our results with recent experimental studies on wurtzite InAs nanowires.
\end{abstract}
\maketitle
%
%
\allowdisplaybreaks

\section{Introduction}

Even though nanowires have been intensively investigated over the past decades, these promising objects continue to attract a profound interest within the nanoscience community. \cite{Yang2010}
Aside from being the essential cornerstone for several fundamental  discoveries,\cite{Mourik2012,Das2012,Hofstetter2009,vanDam2006} they will constitute a key element in the realization of future electronic and spintronic devices.\cite{Heedt2013book,Greytak2005,Xiang2006,Nadj2010,Nadj2012,
 Krogstrup2013,Xing2014,FariaJunior2015,Xing2015}
To support this technical progress, a sound knowledge and reliable control of the system's transport parameters, the spin-orbit coupling (SOC), and the spin relaxation are essential.

In combination with disorder, the SOC usually randomizes the spin precession and therewith induces a spin relaxation process, called D'yakonov-Perel' (DP) mechanism, that significantly limits the spin lifetime.\cite{perel} 
Since this often unwanted effect strongly depends on the device geometry, the strength and structure of the SOC, as well as the initial polarization texture of the spin density, it can be efficiently manipulated. 
For instance, in 2D electron and hole systems special configurations of the SOC parameters even allow for a realization of persistent spin textures.\cite{Kammermeier2016PRL,Wenk2016,Schliemann2003,Bernevig2006,Trushin2007,Dollinger2014,Schliemann2016,Kohda2017}
Additionally, the presence of a narrow boundary in systems of finite-size can yield a further slowdown of the DP spin relaxation, which is known as motional narrowing.\cite{Malshukov2000,Kiselev2000,Schwab2006,Kettemann2007a,
Wenk2010,Wenk2011,Kammermeier2017,Schapers2006,Holleitner2007}

The SOC can be extrinsically induced by breaking the inversion symmetry, e.g., by applying an electric field or heterointerfaces. 
It is also intrinsically present in crystals without a center of inversion, which generically concerns nanowires with a zinc-blende or wurtzite lattice.
Nanowires built from III-V semiconductors, such as GaAs or InAs, are quite peculiar in the sense that their crystal structure can be found in the wurtzite phase even though the underlying material has a zinc-blende lattice in the bulk.
Over the last years, numerous groups have successfully developed sophisticated understanding and techniques which facilitate an excellent control of the crystal phase.\cite{Glas2007,Patriarche2008,Zhang2010a,Krogstrup2011,Rieger2013,
Schroth2015,FontcubertaiMorral2016,Jacobsson2016,Zhang2018}
Since the intrinsic SOC effects in these exceptional wurtzite systems are relatively unexplored, several recent studies have addressed this issue  theoretically\cite{De2010,Chei2011,Intronati2013,Gmitra2016,Campos2018,
FariaJunior2016} and experimentally.\cite{Furthmeier2016,Scheruebl2016,Jespersen2018}

\begin{figure}[b]
\includegraphics[width=1.0\columnwidth]{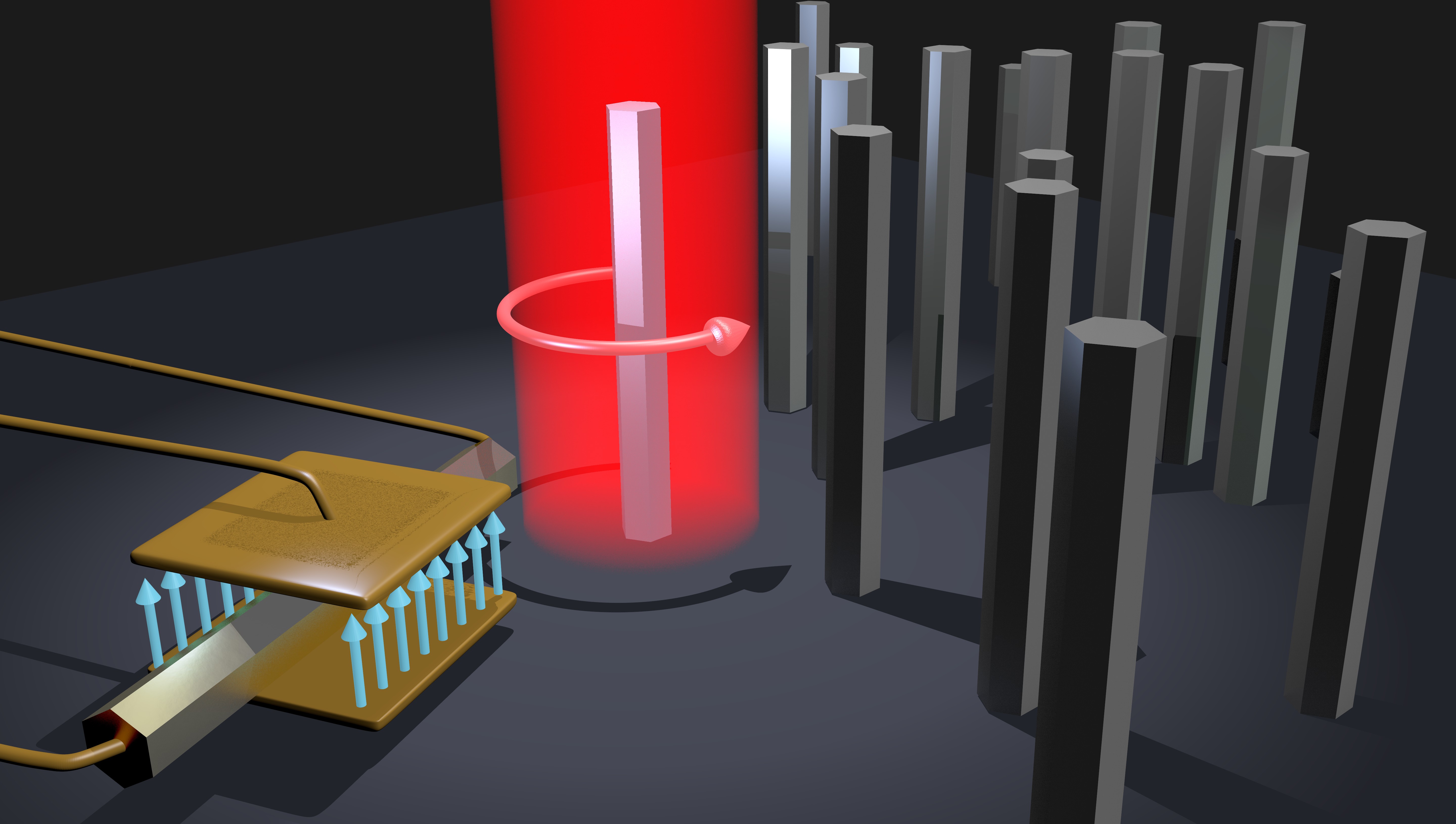}
\caption{Illustration of the two experimental techniques that can lead to very dissimilar results for the spin lifetime (cf. Secs.~\ref{sec:relax_nanowire} and \ref{sec:relax_nanowire_remarks}).  
While probing the magnetoconductance (left) under influence of a gate-induced electric field determines the lifetime of the long-lived helical spin states, micro-photoluminescence measurements\cite{Furthmeier2016} (center) follow the relaxation process of a homogeneous spin density which is excited by circularly polarized light.}
\label{fig:exp_methods}
\end{figure}

Among the diverse experimental methods, low-field magnetoconductance and optical orientation measurements provide convenient access to the desired information on nanowires (cf. Fig.~\ref{fig:exp_methods}).
On the one hand, magnetoconductance measurements enable to  gather transport parameters and identify lifetimes of the long-lived spin states without requiring any previous spin polarization.
On the other hand, they do not reveal details about the real space structure of the corresponding spin states, which can be rather complex and difficult to realize in other experiments.
Besides that, applying this method requires experimental data fitting with the appropriate theoretical model, which is sensitive to the mesoscopic features of the system.
Due to the lack of an adequate description, authors were compelled to use an existing theory which does not fully match with the mesoscopic details of the nanowire.\cite{Hansen2005,Dhara2009,XiaoJie2010,Weperen2015,
Roulleau2010,Liang2012,Scheruebl2016,Takase2017,Jespersen2018}
In line with our previous studies,\cite{Kammermeier2016,Kammermeier2017} which were focused on zinc-blende nanowires, we fill this gap by providing a compatible model for the wurtzite counterpart.

In comparison, optical experiments are feasible to monitor the relaxation process of a spin density whose real space distribution has a well-defined structure.
A downside of this technique is the limitation of possibilities concerning the initial polarization.
For instance, time-resolved micro-photoluminescence typically probes a homogeneous spin polarization pattern parallel to the laser beam.\cite{Furthmeier2016}
Other approaches also enable a wave-like real space modulation of the spin texture.\cite{Carter2006,Koralek2009,Wang2013b}
As demonstrated below, these different experimental methods can result in a  huge discrepancy in the extracted spin lifetime for wurtzite nanowires.

The main objective of this article is to understand the D'yakonov-Perel' spin relaxation properties in wurtzite nanowires. 
In analogy to our preceding article, Ref.~\onlinecite{Kammermeier2017}, we choose a Cooperon-based approach which enables a simultaneous determination of the magnetoconductance correction.
We use these results for a critical comparison of the often applied experimental techniques of magnetoconductance measurements and optical spin orientation (cf. Fig.~\ref{fig:exp_methods}), where we observe large discrepancies in the extracted spin relaxation rates.
The nanowires are oriented along the [0001]-axis and considered diffusive in three spatial dimensions, where the diffusive motion is subject to a radial spin-conserving and insulating boundary condition.
Aside from the intrinsic SOC effects resulting from the wurtzite lattice, an extrinsically side-gate-induced Rashba term is taken into account. 
The latter is relevant for a gate-dependent tuning of the SOC strength and spin relaxation rate in the magnetoconductance studies.
Explicit expressions for the leading-order magnetoconductance correction are derived in zero-mode approximation for the Cooperon.

We gained the following key insights.
In the bulk, the intrinsic spin relaxation is found to be limited by the dominant $k$-linear SOC terms, which agrees with experimental observations.\cite{Buss2011,Jahangir2012,Stefanowicz2014}
In contrast, due to the radial boundary condition for nanowires the intrinsic spin relaxation of the long-lived spin states is determined by the typically small $k$-cubic SOC terms and is nearly independent of the radius.
Since these relaxation rates enter the leading-order magnetoconductance correction, the experimentally observed intrinsic spin relaxation rates will be insignificant and a scaling with the radius hardly observable.
At the same time, the corresponding long-lived eigenstates can assume a complex helical structure in real space, which is largely influenced by the ratio of the wire radius the and spin precession length.
This has remarkable consequences for optical experiments, such as time-resolved micro-photoluminescence measurements, where spin densities are homogeneously polarized along the wire axis.\cite{Furthmeier2016} 
Here, the deviation of the spin density distribution from the long-lived eigenstates is highly sensitive to the radius.
As a consequence, we observe a dramatic radius-dependence of the spin relaxation rate.
This also constitutes a striking difference to zinc-blende nanowires, where the according eigenstate has been found to be independent of the wire radius.\cite{Kammermeier2017}
These insights are of crucial importance for the accurate interpretation of experimental results regarding spin relaxation properties in nanowires. 
To underline the significance of our results, we discuss the case of a wurtzite InAs nanowire and compare with to two recent publications,\cite{Scheruebl2016, Jespersen2018} which studied the spin relaxation in these systems by means of magnetoconductance measurements.

This paper is organized as follows.
The model Hamiltonian for bulk electrons in the wurtzite lattice and a generic expression for the weak (anti)localization correction are introduced in Secs.~\ref{sec:model} and \ref{sec:WAL}, respectively.
In Sec.~\ref{sec:BulkCooperon}, we compute the Cooperon for the bulk system, which is subject to a radial boundary condition for nanowires, to be discussed in Sec.~\ref{sec:finite-size}.
Next, we analyze the intrinsic DP spin relaxation properties for the bulk system as well as for nanowires in Sec.~\ref{sec:intr_spin_relax}.
The additional influence of a side-gate induced Rashba SOC on the lowest Cooperon eigenvalues (corresponding to the long-lived spin relaxation rates) is investigated in Sec.~\ref{sec:gate}.
In the last step, the results are used to derive an analytical expression for the magnetoconductance correction in Sec.~\ref{sec:mc}.

\section{Theoretical foundation}\label{sec:foundation}

\subsection{Electrons in the wurtzite lattice}\label{sec:model}

The bulk electrons in the $\Gamma_{7c}$ conduction band of a wurtzite type semiconductor with SOC are described by the Hamiltonian 
\begin{align}
\mathcal{H}&=\,\frac{\hbar^2k^2}{2 m}+\mathcal{H}_\text{so}^\text{ext}+\mathcal{H}_\text{so}^\text{int}.
\label{nanorodbulk}
\end{align}
The terms
\begin{align}
\mathcal{H}_\text{so}^\text{ext}&=\alpha_\text{R}^\text{ext}(k_x\sigma_z-k_z\sigma_x),\label{rashba}\\
\mathcal{H}_\text{so}^\text{int}&=\left[\gamma_\text{R}^\text{int}+\gamma_\text{D}\left(bk_z^2-k_\perp^2\right)\right](k_y\sigma_x-k_x\sigma_y), \label{dresselhausWZ}
\end{align}
with $k_\perp^2=k_x^2+k_y^2$ and  $\alpha_\text{R}^\text{ext}=\gamma_\text{R}^\text{ext}\mathcal{E}$ denote the extrinsic (ext) and intrinsic (int) Rashba (R) and the Dresselhaus
(D) SOC contributions with the material specific
parameters $\gamma_\text{R}^\text{int}, \gamma_\text{R}^\text{ext}, \gamma_\text{D}$, and $b$, the
electric field strength $\mathcal{E}$, the Pauli matrices $\sigma_i$, and the
effective electron mass $m$, which is here considered isotropic. \cite{Zutic2004a,Wu2010,Gmitra2016} 
In this notation, the $\mathbf{\hat{z}}$-axis corresponds to the [0001] crystal axis (c-axis).
Hereby, we assume that the electrons in the wire experience a nearly homogeneous electric field perpendicular to the wire axis. 
Without loss of generality, it is aligned with the system's $\mathbf{\hat{y}}$-axis, i.e., $\boldsymbol{\mathcal{E}}=\mathcal{E}\mathbf{\hat{y}}$, and results in the external Rashba contribution $\mathcal{H}_\text{so}^\text{ext}$. 

\subsection{Weak (Anti)localization correction}\label{sec:WAL}

The first-order correction to the Drude conductivity $\Delta\sigma$ in a disordered conductor is found within diagrammatic perturbation theory by
taking into account the quantum interference between self-crossing paths.
The random disorder potential $V_\text{imp}(\mathbf{r})$ is assumed to fulfill the following requirements: (i) We consider a standard white-noise
model, which implies that the disorder potential vanishes on average and is uncorrelated, i.e., $\left\langle V_\text{imp}(\mathbf{r})\right\rangle=0$ and $\left\langle V_\text{imp}(\mathbf{r})V_\text{imp}(\mathbf{r'})\right\rangle\propto\delta(\mathbf{r}-\mathbf{r'})$, respectively.
(ii) The localization due to disorder is weak, meaning that the \textit{Ioffe-Regel criterion} holds true, i.e., $\hbar/(\epsilon_F \tau_e)\ll 1$, where $\epsilon_F$ is the Fermi energy and $\tau_e$ is the mean elastic isotropic scattering time.  
Moreover, the electron motion is considered diffusive in all three spatial directions.  
Taking the average over all impurity configurations and summing up all maximally crossed ladder diagrams yields the quantum correction to the longitudinal static conductivity~\cite{nagaoka} to first order in $\hbar/(\epsilon_F \tau_e)$.
It is given by the real part of the Kubo-Greenwood formula
\begin{align}
\Delta\sigma=&\,\frac{2e^2 }{h}\frac{\hbar D_e}{ \mathcal{V}}\Re \text{e}\left(\sum_{\mathbf{Q},s,m_s}\chi_s\braket{s,m_s|\hat{\mathcal{C}}(\mathbf{Q})|s,m_s}\right).
\label{conductivity}
\end{align}
In this formula, $\mathcal{V}$ denotes the volume of the nanowire, $D_e$ the
3D diffusion constant, i.e., $D_e=v_F^2 \tau_e/3$, with
the Fermi velocity $v_F$, $\hat{\mathcal{C}}$ the Cooperon propagator, and
$\mathbf{Q}=\mathbf{k}+\mathbf{k'}$ the Cooperon wave vector, which is the sum of the wave vector of an electron with spin $\boldsymbol{\sigma}$ and the wave vector of an electron with spin $\boldsymbol{\sigma'}$.
The states $\ket{s,m_s}$ represent the singlet-triplet basis of the system with two electrons, that is, $s\in\{0,1\}$ is the total spin quantum number and $m_s\in\{0,\pm1\}$ the corresponding magnetic quantum number.
As shown in Refs.~\onlinecite{Wenk2010,Kammermeier2017}, there exists a unitary transformation between the spin diffusion equation and the Cooperon and, therefore, an according basis transformation between the components of the spin density $\mathbf{s}$ and the triplet components $\ket{1,m_s}$ of the Cooperon.
The respective transformation operator $U_{cd}$ is given in the App.~\ref{app:relation}.
Furthermore, the factor $\chi_s$, which is defined as $\chi_0=1$ and $\chi_1=-1$.
The sign indicates that, depending on the relative weight of the singlet and triplet matrix elements of the Cooperon, the conductivity correction can be either positive or negative, which refers to weak antilocalization or weak localization, respectively. 
Hereafter, we compute the Cooperon and the magnetoconductance correction following former approaches.\cite{Kettemann2007a, Wenk2010, Wenk2011,wenkbook,Kammermeier2016,Kammermeier2017}
%
%
\subsection{Cooperon in the bulk}\label{sec:BulkCooperon}
Treating SOC as a small perturbation to the kinetic part of the Hamiltonian $\mathcal{H}$, and noting that the main contribution to the Cooperon results from terms near $Q=0$, the Cooperon propagator $\hat{\mathcal{C}}$ can be written as 
\begin{equation}
\hat{\mathcal{C}}(\mathbf{Q})=\,\frac{\tau_e}{\hbar}\left(1-\int\frac{{\rm d}\Omega}{4\pi}\frac{1}{1-i\tau_e\hat{\Sigma}(\mathbf{Q})/\hbar}
\right)^{-1},
\label{cooperon1}
\end{equation}
where $\hat{\Sigma}(\mathbf{Q})=\,\mathcal{H}(\mathbf{Q}-\mathbf{k}_F,\boldsymbol{\sigma})-\mathcal{H}(\mathbf{k}_F,\boldsymbol{\sigma'})$.
Considering a sharp Fermi edge, the Fermi contour can be approximated in 3D by a sphere and the integral runs contiunously over the solid angle $\Omega$ of the Fermi wave vector $\mathbf{k}_F$ with constant length. 
Using the precondition  $\epsilon_F \tau_e/\hbar\gg 1$,
we may further simplify  $\hat{\Sigma}(\mathbf{Q})\approx -\mathbf{v}_F(\hbar\mathbf{Q}+2m\mathbf{\hat{a}}\mathbf{S}) $ with the total electron spin vector $\boldsymbol{S}$ in the singlet-triplet basis
as defined in App.~\ref{app:spin}. The matrix $\mathbf{\hat{a}}$ contains the SOC contributions, i.e.,
\begin{align}
\mathbf{\hat{a}}&=\,
\begin{pmatrix}
0&-a_\text{int}&a_\text{ext}\\
a_\text{int}&0&0\\
-a_\text{ext}&0&0
\end{pmatrix},
\end{align}
with 
\begin{align}
a_\text{ext}={}&\alpha_\text{R}^\text{ext}/\hbar,\quad
a_\text{int}={}\left[\gamma_\text{R}^\text{int}+\gamma_\text{D}\left(bk_z^2-k_\perp^2\right)\right]/\hbar\notag. 
\end{align}
For reasons of expediency and in accordance with previous publications, Refs.~\onlinecite{Wenk2010,Wenk2011,wenkbook,Kammermeier2016, Kammermeier2017}, we define the Cooperon Hamiltonian as $\hat{H}_c=(\hbar D_e \hat{\mathcal{C}})^{-1}$. 
An additional Taylor expansion of the integrand in Eq.~(\ref{cooperon1}) to second order in $(\hbar\mathbf{Q}+2m\mathbf{\hat{a}}\mathbf{S})$, yields 
\begin{align}
\hat{H}_c={}&\Big[\mathbf{Q}+2 e \mathbf{A}_s/\hbar\Big]^2+\Delta_s.
\label{eq:Hcbulk}
\end{align}
This approximation is valid in the diffusive regime when the SOC energy is small in comparison to the scattering energy $\hbar/\tau_e$, which is also the necessary precondition for the D'yakonov-Perel' spin relaxation. The impact of large SOC on the conductivity was studied in 2D systems in Refs.~\onlinecite{Golub2005a,Glazov2009,Araki2014}.
The effect of SOC becomes manifest in two different ways which origins from the distinct spherical harmonic decomposition of the SOC contributions in the wave vector $\mathbf{k}$.
(i) The SOC terms due to the first-degree spherical harmonics in the wave vector $\mathbf{k}$ lead to an effective vector potential $\mathbf{A}_s=\mathbf{A}_s^\text{ext}+\mathbf{A}_s^\text{int}$, where
\begin{align}
\mathbf{A}_s^\text{ext}={}&\alpha_\text{R}^\text{ext} m/(e\hbar)(S_z,0,-S_x)^\top,\label{eq:EffVecPot1}\\
\mathbf{A}_s^\text{int}={}&\left[\gamma_\text{R}^\text{int}+\delta_\text{D}^{(1)}\right]m/(e\hbar)(-S_y,S_x,0)^\top,\label{eq:EffVecPot2}
\end{align}
with $\delta_\text{D}^{(1)}= (b-4)\gamma_\text{D} k_F^2/5$,  and therefore couples to the Cooperon momentum.
(ii) In addition, we find an intrinsic spin-relaxation term $\Delta_s=\delta_\text{D}^{(3)}(S_x^2+S_y^2)$, where
\begin{align}
\delta_\text{D}^{(3)}={}&\frac{32}{175}\left((1+b)\gamma_\text{D} m k_F^2/\hbar^2\right)^2,
\label{eq:SRterm}
\end{align}
which is a result of the third-degree spherical harmonics in the Dresselhaus field and is independent of the Cooperon momentum.
The decomposition of the intrinsic SOC Hamiltonian $\mathcal{H}_\text{so}^\text{int}$, Eq.~(\ref{dresselhausWZ}), in terms of spherical harmonics is demonstrated in the App.~\ref{app:SOC}.
Notably, in the analogous zinc-blende system the intrinsic SOC contains only third-degree spherical harmonic terms and does not give rise to an effective vector potential but solely leads to a contribution $\Delta_s$ that is diagonal in the triplet basis.\cite{Kammermeier2017}

The minima of the triplet eigenmodes $E_{T,j}$ are direct measures of the spin relaxation rate $(1/\tau_s)_j$ of a certain polarized spin density $\mathbf{s}$ via the relation ${(1/\tau_s)_j=D_e E_{T,j}}$.
For this reason, the minima of the spectrum are of particular interest as they allow to identify long-lived spin density states.
In contrast to the term $\Delta_s$ in case (ii), the effective vector potential $\mathbf{A}_s$ is capable of shifting the global minimum of an eigenvalue to finite wave vectors $\mathbf{Q}$ and thereby giving rise to helical spin states with longer spin lifetimes than the homogeneous counterpart ($Q_z=0$).
Moreover, the effective vector potential $\mathbf{A}_s$ plays a crucial role in case of a boundary condition for the Cooperon as will be discussed in the following.

\subsection{Finite-size effects}\label{sec:finite-size}
Owing to the finite-size geometry of the nanowire, the Cooperon has to be complemented by a boundary condition.
The impact of the boundary becomes relevant if the dephasing length is larger than the nanowire diameter.
As the length of the nanowire typically largely exceeds its radial extension, we assume periodic boundary conditions along the wire axis for simplicity.
Considering spin-conserving and specular scattering at the insulating lateral surface, the boundary condition for a cylindrical nanowire reads as\cite{AltshulerAronov1981,Aleiner2001,Meyer2002,Kettemann2007a}
\begin{align}
\boldsymbol{\hat{\rho}}\cdot\left(\boldsymbol{\nabla}+2i e  \mathbf{A}_s/\hbar\right)\hat{\mathcal{C}}\vert_{\rho=R}={}&0,
\label{eq:boundary}
\end{align}
where $R$ denotes the radius of the wire and we introduced the standard cylindrical coordinates $(\rho,\phi,z)$ with the corresponding basis vectors $\{\boldsymbol{\hat{\rho}},\boldsymbol{\hat{\phi}},\boldsymbol{\hat{z}} \}$.
It is practical to simplify the above equation to a Neumann boundary condition, i.e., $\boldsymbol{\hat{\rho}}\cdot(\boldsymbol{\nabla}\hat{\mathcal{C}}')\vert_{\rho=R}=0$. 
This can be achieved by performing a non-Abelian gauge transformation of the Cooperon (and simultaneously the Cooperon Hamiltonian), that is, $\hat{\mathcal{C}}\rightarrow\hat{\mathcal{C}}'=U_A\hat{\mathcal{C}}U_A^\dag$, with the unitary transformation operator $U_A=\exp[i 2 e\,(\boldsymbol{\hat{\rho}} \cdot\mathbf{A}_s)\rho/\hbar ]$.
As a downside of the mutual interplay of intrinsic and extrinsic SOC effects, the transformed Cooperon Hamiltonian $\hat{H}_c'$ has an ample and complex structure.
Dealing with the resulting symbolic expressions is a delicate task and we  shall discuss only specific situations analytically.

A suitable and generic real space basis for the transformed Cooperon (Hamiltonian) which satisfies the Neumann boundary condition is
\begin{align}
\braket{\mathbf{r}|n,l,Q_z}={}&J_l^{(n)}(\rho)e^{il\phi}e^{iQ_z z}/N_{nl},
\label{eq:basis}
\end{align}
with the angular momentum quantum number $l\in \mathbb{Z}$, the continuous plane wave number $Q_z$ along the wire axis, and an appropriate normalization constant $N_{nl}$.
The radial dependence is given by the Bessel function of the first kind $J_l^{(n)}:=J_l(\rho\,\zeta_{n,\vert l\vert}/R)$, where $\zeta_{n,\vert l\vert}$ signifies the $n$-th radial extremum ($n\in \mathbb{N}_+$) of the Bessel function of $J_l(\rho)$. 
Additionally, we  define $J_l^{(0)}=\delta_{l,0}$ which corresponds to a constant solution in the cross-sectional plane  and constitutes the lowest mode of $\hat{H}_c'$, usually denoted as zero-mode $\ket{0}$, i.e., $\ket{0}\equiv \ket{n=0,l=0,Q_z}$.

The zero-mode is of central interest since it allows to determine the spin states with the longest spin lifetime in narrow wires.
These states are also characteristic to the conductance correction in transport as they yield the predominant contribution.
In particular, if the wire is thin enough that the lowest Cooperon mode is well separated from the others, the transformed Cooperon Hamiltonian $\hat{H}_c'$ can be evaluated only for the lowest mode, i.e., $\braket{0|\hat{H}_c'|0}$.
This approach, which is often termed zero-mode approximation,\cite{Aleiner2001,Meyer2002,Kettemann2007a} is used in the following to obtain analytical expressions for the spin relaxation rates and compute the magnetoconductance correction.

However, it is essential to notice that, due to the gauge transformation, the lowest mode is position-dependent in the (untransformed) system.
More precisely, the real space representation of the lowest mode of the  Cooperon Hamiltonian $\hat{H}_c$ is in fact $U_A^\dag \braket{\mathbf{r}|0}$.
Consequently, the corresponding long-lived spin states have in general a rather complex helical structure in real space and are, therefore, often experimentally not accessible.
Only in narrow wires, if the spin precession length is much larger than the boundary separation, the eigenstates are nearly homogeneous in real space.
Exemplary in this context are optical orientation measurements or spin lasers, where the spin densities are homogeneously excited along the wire axis.\cite{Furthmeier2016,FariaJunior2015,FariaJunior2017}
For this reason, we will pay special attention to this scenario in Sec.~\ref{sec:hom}.

\section{Intrinsic Spin Relaxation}\label{sec:intr_spin_relax}

The dynamics of a local spin density $\mathbf{s}=\mathbf{s}(\mathbf{r},t)$ follows the spin-diffusion equation\cite{Wenk2010}
\begin{align}
0={}&\partial_t\, \mathbf{s}+D_e \hat{H}_\text{SD}\,  \mathbf{s}.
\label{eq:spindiff_eq}
\end{align}
An initial spin density $\mathbf{s}_0$ evolves in time as ${\mathbf{s}_t=\exp(-D_e \hat{H}_\text{SD}t)\,\mathbf{s}_0}$.
The spin-diffusion Hamiltonian $\hat{H}_\text{SD}$ is related to the Cooperon Hamiltonian $\hat{H}_c$ via the unitary transformation $\hat{H}_\text{SD}=U_{cd}^\dag \hat{H}_c U_{cd}$, where $U_{cd}$ is defined in App.~\ref{app:relation}.
Consequently, by analyzing the Cooperon Hamiltonian we can study the temporal and spatial evolution of a spin density.
In the following subsections, we omit the effects of a lateral gate electrode, i.e., $\alpha_\text{R}^\text{ext}=0$.

\subsection{Spin relaxation in the bulk}

In the bulk, the Cooperon Hamiltonian $\hat{H}_c$, Eq.~(\ref{eq:Hcbulk}),  can be diagonalized in the basis of plane waves ${\braket{\mathbf{r}|\mathbf{Q}}\propto\exp(i \mathbf{Q}\cdot\mathbf{r})}$ with the continuous wave vectors $Q_i$.
Then, the eigenvalues read as
\begin{align}
E_S={}&\mathbf{Q}^2,\label{bulk_eigenvalue_singlet}\\
E_{T,\pm}={}&\mathbf{Q}^2+\frac{3}{2}\left(Q_\text{so}^2+\delta_\text{D}^{(3)}\right)\notag\\
&\pm\frac{1}{2}\sqrt{16 Q_\perp^2 Q_\text{so}^2+\left(Q_\text{so}^2+\delta_\text{D}^{(3)}\right)^2},\label{bulk_eigenvalue_pm}\\
E_{T,0}={}&\mathbf{Q}^2+Q_\text{so}^2+\delta_\text{D}^{(3)},\label{bulk_eigenvalue_0}
\end{align}
where $Q_\perp^2=Q_x^2+Q_y^2$  and  $Q_\text{so}=2m(\gamma_\text{R}^\text{int}+\delta_\text{D}^{(1)})/\hbar^2=2\pi/L_\text{so}$, where $L_\text{so}$ denotes the spin precession length due to the intrinsic SOC.
Consequently, the spin relaxation rates for homogeneously polarized spin densities, i.e., $\mathbf{Q}=\mathbf{0}$, are
\begin{align}
\left(\tau_s^{-1}\right)_\perp^\text{hom}{}&=\left(\tau_s^{-1}\right)_{z}^\text{hom}/2=D_e (Q_\text{so}^2+\delta_\text{D}^{(3)}),
\label{eq:hom_bulk_rate}
\end{align} 
where the \textit{z-}polarized densities decay twice as fast as the states in the \textit{x-y}-plane ($\perp$).
Yet, for $\delta_\text{D}^{(3)}<3Q_\text{so}^2$ (which is usually fulfilled) the spin densities with the longest spin lifetime are homogeneous along the c-axis but have helical structure in the \textit{x-y}-plane.
Their spin decays according to $(1/\tau_s)^\text{helix}=D_e {E_{T,-}(Q_\perp=Q_0,Q_z=0)}$, that is,
\begin{align}
\left(\frac{1}{\tau_s}\right)^\text{helix}={}&D_e\left[\frac{7}{16}Q_\text{so}^2+\frac{11}{8}\delta_\text{D}^{(3)}-\frac{1}{16}\left(\frac{\delta_\text{D}^{(3)}}{Q_\text{so}}\right)^2\right],
\label{eq:helix_bulk_rate}
\end{align}
at the finite wave vectors perpendicular to the c-axis 
\begin{align}
Q_0={}&\frac{1}{4}\sqrt{15 Q_\text{so}^2-2\delta_\text{D}^{(3)}-\left(\frac{\delta_\text{D}^{(3)}}{Q_\text{so}}\right)^2}.
\end{align}
Disregarding the typically small cubic SOC term $\propto \delta_\text{D}^{(3)}$, the relaxation rate is about half as large as for the homogeneous long-lived state.  
We can identify the corresponding helical spin density as
\begin{align}
\mathbf{s}(\mathbf{r},t)\propto{}&\left[ \frac{\mathbf{q}}{\Vert\mathbf{q}\Vert}\Sigma\cos(\mathbf{q}\cdot\mathbf{r})+ \boldsymbol{\hat{z}} \sin(\mathbf{q}\cdot\mathbf{r})\right]\exp\left(-t/ \tau_s^\text{helix}\right)
\end{align}
with $\Sigma\approx (15Q_\text{so}^2+4\delta_\text{D}^{(3)})/(3\sqrt {15}Q_\text{so}^2)$ to lowest non-vanishing order in $\delta_\text{D}^{(3)}$.
The wave vector $\mathbf{q}$ lies in the \textit{x-y}-plane and has the length $\Vert\mathbf{q}\Vert=Q_0$.
For $\delta_\text{D}^{(3)}\rightarrow 0$ the solutions coincide with the result for the 2D Rashba system as discussed in Refs.~\onlinecite{Schwab2006,Wenk2010,Pershin2010}.

\subsection{Spin dynamics in the nanowire}\label{sec:relax_nanowire}

As described in Sec.~\ref{sec:finite-size}, in order to simplify the boundary condition, required by the finite-size geometry of the wire, we apply a gauge transformation to the Cooperon Hamiltonian.
The transformed Cooperon Hamiltonian $\hat{H}_c'$ is found as
\begin{align}
\hat{H}_c'={}&\boldsymbol{Q}^2+\frac{\delta_\text{D}^{(3)}}{4}\bigg\{3\mathbf{S}^2-S_z^2-\frac{1}{2}(S_+^2 e^{-2i\phi}+S_-^2 e^{2i\phi})\notag\\
&+\Big[ \mathbf{S}^2-3S_z^2+\frac{1}{2}(S_+^2 e^{-2i\phi}+S_-^2 e^{2i\phi})\Big]\cos(2 Q_\text{so}\rho)\notag\\
&-2\Big[ \{S_x,S_z\}\cos(\phi)+ \{S_y,S_z\}\sin(\phi)\Big]\sin(2 Q_\text{so}\rho)\bigg\}
\label{eq:Hc_transf}
\end{align}
with $S_\pm=S_x\pm i S_y$.
We stress that the gauge transformation removes the effective vector potential $\mathbf{A}_s^\text{int}$ completely and only quadratic wave vectors $\mathbf{Q}^2$ remain.
As in the bulk, the global minimum with respect to the wave vector $Q_z$ of the spectrum is found at $Q_z=0$.

\subsubsection{Long-lived spin states and diffusive-ballistic crossover}

An analytical result for the lowest eigenvalues can be obtained by evaluating the transformed Cooperon Hamiltonian in \textit{zero-mode approximation}.
The boundary-induced shift of the first excited mode is of the order of $\braket{1,0,0|Q_\perp^2|1,0,0}\propto R^{-2}$.
On the other hand, the spin-orbit broadening within each mode is of the order of $\delta_\text{D}^{(3)}$.
Consequently, we can estimate the zero-mode to be well separated if $\delta_\text{D}^{(3)}R^2\ll 1$ holds.
Under these circumstances, the eigenvalues of $\braket{0|\hat{H}_c'|0}$ read as
\begin{align}
E_S^{(0)}={}&Q_z^2,\label{eigenvalue1}\\
E^{(0)}_{T,\pm}={}&Q_z^2+\delta_\text{D}^{(3)}\left(\frac{5}{4}+\frac{a_\text{so}}{2}\right),\label{eigenvalue2}\\
E^{(0)}_{T,0}={}&Q_z^2+\delta_\text{D}^{(3)}\left(\frac{3}{2}-a_\text{so}\right),\label{eigenvalue3}
\end{align}
where we introduced 
\begin{align}
a_\text{so}={}&\left[1-\cos(2R_\text{so})-2 R_\text{so}\sin( 2 R_\text{so})\right]/ (2R_\text{so})^2,
\end{align}
and $R_\text{so}=Q_\text{so}R$. 
Asymptotically, we obtain $a_\text{so}\rightarrow -1/2$ for $R_\text{so}\rightarrow 0$ and $a_\text{so}\rightarrow 0$ for $R_\text{so}\rightarrow \infty$.
\begin{figure}[t]
\includegraphics[width=1.0\columnwidth]{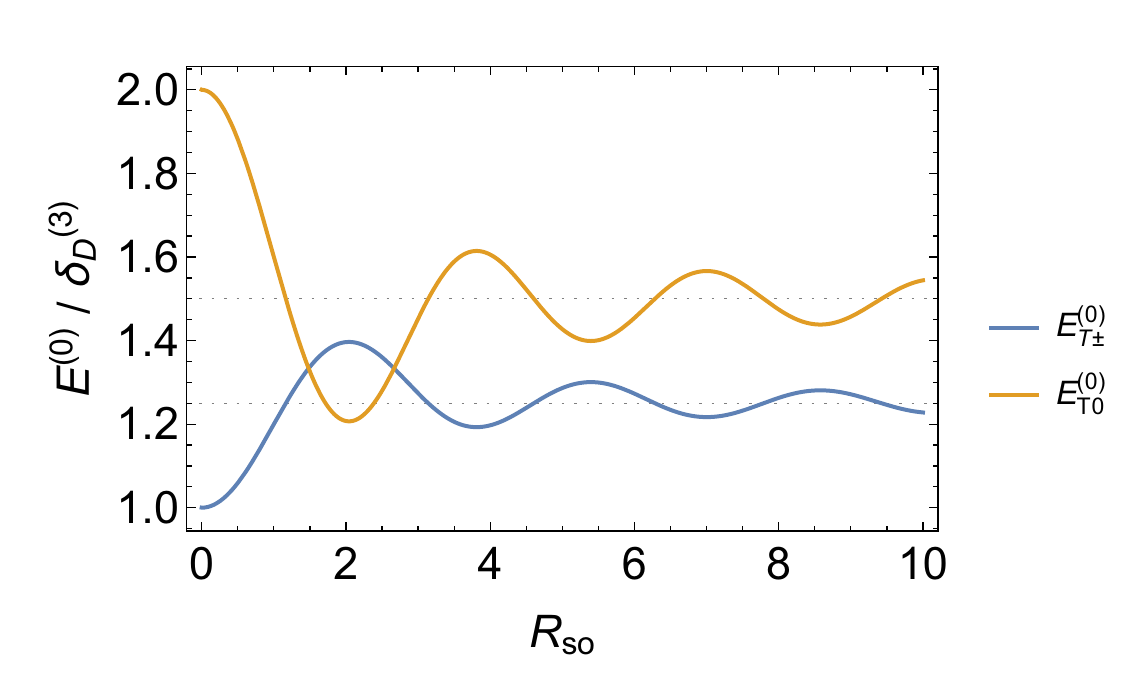}
\caption{Lowest eigenvalues of the Cooperon Hamiltonian for $Q_z=0$ in zero-mode-approximation in dependence of the dimensionless radius $R_\text{so}$ and in absence of external electric fields, i.e., $\alpha_\text{R}^\text{ext}=0$.}
\label{fig:zero_mode_spec}
\end{figure}

We focus again on the long-lived spin states, which are found for a homogeneous spin polarization along the c-axis, i.e., $Q_z=0$.
The eigenvalues are displayed in Fig.~\ref{fig:zero_mode_spec}  in dependence of $R_\text{so}$.
Besides the slight increase (decrease) of the eigenvalue $E^{(0)}_{T,\pm}$  ($E^{(0)}_{T,0}$) for small $R_\text{so}$, the eigenvalues show $R_\text{so}$-periodic oscillations with decreasing amplitude.
We emphasize that the amplitudes depend solely on the term $\delta_\text{D}^{(3)}$, which is usually small compared to $Q_\text{so}^2$.
Hence, the resulting spin relaxation rates show very little dependence on the thickness of the nanowire.
Since these rates enter the leading-order conductance correction, the latter will be hardly affected by any changes in the nanowire radius either.
Owing to the gauge transformation, the according eigenvectors of $\hat{H}_c$ are position-dependent in the cross-sectional plane.
More precisely, the (unnormalized) eigenvectors $\mathbf{a}_j$, which are associated with the triplet eigenvalues $E^{(0)}_{T,j}$ in Eqs.~ (\ref{eigenvalue2}) and (\ref{eigenvalue3})  for $Q_z=0$, take the form
\begin{align}
\mathbf{a}_+&={}(\cos(\phi),\sin(\phi),-\tan(Q_\text{so}\rho))^\top,\\
\mathbf{a}_-&={}(-\sin(\phi),\cos(\phi),0)^\top,\\
\mathbf{a}_0&={}(\cos(\phi)\tan(Q_\text{so}\rho),\sin(\phi)\tan(Q_\text{so}\rho),1)^\top,
\end{align}
in the basis of spin-density components $\{s_x,s_y,s_z\}$.
The eigenvectors $\mathbf{a}_\pm$ are not uniquely defined as the corresponding eigenvalues are degenerate.

In the \textit{1D-diffusive} limit, i.e., $R_\text{so}\ll 1$, we can write
$\mathbf{a}_+=\boldsymbol{\hat{\rho}}$, $\mathbf{a}_-=\boldsymbol{\hat{\phi}}$, and $\mathbf{a}_0=\mathbf{\hat{z}}$ since $Q_\text{so}\rho \leq R_\text{so}$.  
We stress that for $R_\text{so}\rightarrow 0$ the corresponding eigenvalues 
are identical to the ones resulting from bulk spin relaxation term $\Delta_s$ in Eq.~(\ref{eq:Hcbulk}) giving rise to the spin relaxation rates in Eq.~(\ref{eq:hom_bulk_rate}) for $Q_\text{so}=0$, i.e.,
\begin{align}
\left(\tau_s^{-1}\right)_\perp^\text{1D}{}&=\left(\tau_s^{-1}\right)_{z}^\text{1D}/2=D_e\delta_\text{D}^{(3)}.
\label{eq:1D_rate}
\end{align}
The equivalent result is obtained by considering only the DP spin relaxation tensor\cite{Zutic2004a} for the bulk system and taking only into account the Dresselhaus contribution due to the higher spherical harmonics $(\mathcal{H}_\text{so}^\text{int})_{(3)}$ (cf. App.~\ref{app:SOC}).
Hence, the spin relaxation resulting from the first-degree spherical harmonic contribution $(\mathcal{H}_\text{so}^\text{int})_{(1)}$ is absent for $R_\text{so}\rightarrow 0$.
As for small densities the $k$-linear contribution, which is comprised in $(\mathcal{H}_\text{so}^\text{int})_{(1)}$, is expected to be dominant, the spin lifetime is significantly enhanced in wires with small radii.
Aside from that, it is to mention that for third-degree spherical harmonic SOC terms the mean free scattering time $\tau_e$ is lowered to $\tau_e/u$, where $1\leq u\leq 9$ depending on the type of scattering process, e.g., $u=1$ for isotropic and $u=9$ for small-angle scattering. \cite{Knap1996,Zutic2004a} 
This can further reduce the spin relaxation rate of the long-lived spin states in the nanowire.
The impact on the bulk spin relaxation rate, e.g., Eqs.~(\ref{eq:hom_bulk_rate}) and (\ref{eq:helix_bulk_rate}), is less important due to the dominance of the spin relaxation rate resulting from $k$-linear SOC terms.

\begin{figure}[t]
\includegraphics[width=.9\columnwidth]{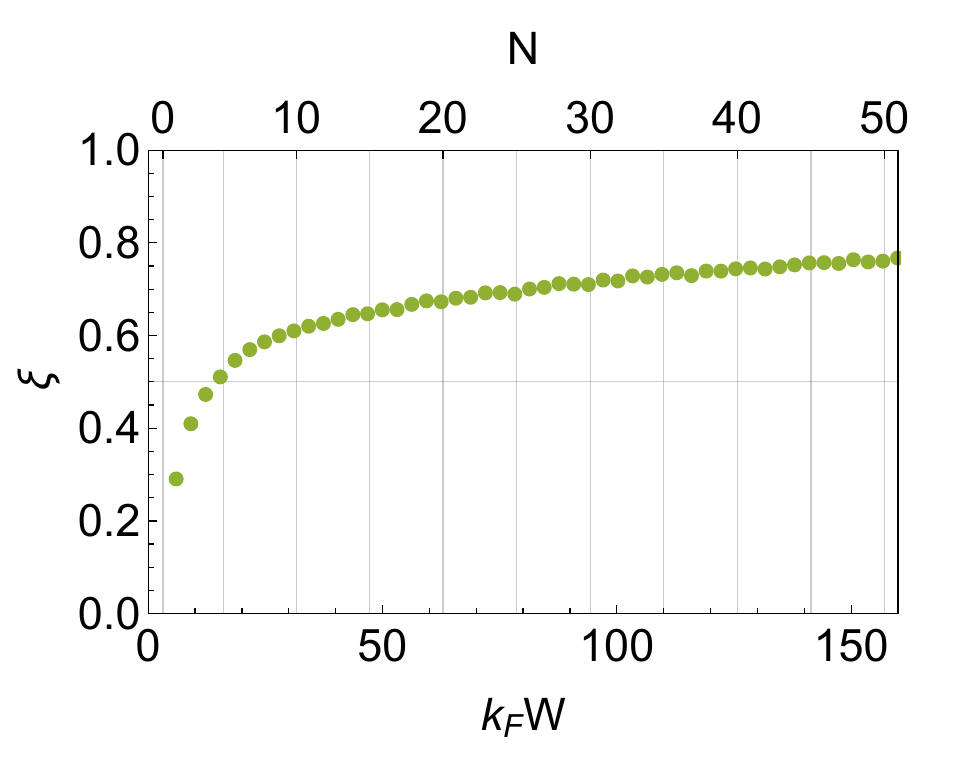}
\caption{Factor of reduction $\xi$ of the spin relaxation contribution due to diffusive-ballistic crossover, i.e., $\delta_{D}^{(3)}\rightarrow\xi \delta_\text{D}^{(3)}$, in dependence of $k_F W$ or the maximal quantum number $N$.}
\label{fig:matsubara}
\end{figure}

At last, we discuss the \textit{diffusive-ballistic transition} regime, in which the nanowire radius is not only much smaller than the spin precession length but also of the order of the mean free path $l_e$, i.e., $ R_\text{so}\ll 1 \wedge R/l_e \sim 1$. 
As shown in Ref.~\onlinecite{Wenk2011}, the number of the conducting channels decreases with the reduction of the wire width.
This leads to a suppression of the cubic SOC terms $(\mathcal{H}_\text{so}^\text{int})_{(3)}$, which are responsible for the spin relaxation rate for $R_\text{so}\ll 1$.
We can account for the diffusive-ballistic crossover by replacing the integral over the Fermi surface in Eq.~(\ref{cooperon1}) by a sum over all modes as shown in detail in App.~\ref{app:Crossover}. 
For simplicity, we treat the size-quantization according to a square wire along $\mathbf{\hat{z}}$ with side lengths $W$ and hard-wall boundaries.
Consequently, two quantum numbers occur, which are labeled by $n$ and $p$ with $n,p \in [1,N]$ where $N$ denotes the maximum quantum number.
In Fig.~\ref{fig:matsubara}, we show how the parameter $\delta_\text{D}^{(3)}$ decreases to $\xi \delta_\text{D}^{(3)}$ due to the reduction of contributing modes via the wire side length $W$ or maximum quantum number $N$.
The decay can be well fitted with $\xi \propto \ln(k_F W)$.

We stress that in the diffusive-ballistic crossover regime the above modifications are plausible and explain further decrease of the spin relaxation rate.
However, in the pure transversal ballistic regime, the subband structure of the system is fully resolved, which has dramatic consequences on the DP spin relaxation mechanism.
Owing to $k_z$-mirror symmetry of the Hamiltonian of the wurtzite nanowires, the spin degeneracy is \textit{not} lifted along the crystal c-axis.
As a consequence, there is obviously no spin-rotation about a spin-orbit induced effective magnetic field (spin-orbit field) and hence no DP spin-relaxation.
This is a remarkable difference to, e.g., the transversal ballistic planar quantum wires with Rashba SOC.
In a strictly one-dimensional limit, there are two kinds of persistent spin states, that is, (a) a homogeneous spin density which is polarized along the (uni-directional) spin-orbit field and (b) the persistent spin helix perpendicular to it.\cite{wenkdiss}
In Ref.~\onlinecite{Hachiya2014} it is shown that in the multisubband Rashba wire the persistent spin helix disappears.
Responsible for this are inter-subband transitions which lead to a non-commutativity of the time-evolution operator $U(k_z)$ for reversed paths along the channel, i.e., $[U(k_z),U(-k_z)]\neq 0$.
In a multisubband wurtzite nanowire the commutativity is trivially given since $U(k_z)=U(-k_z)$.

\subsubsection{Decay of a homogeneous spin density}\label{sec:hom}

Optical spin injection in semiconductor nanowires typically generates collective spin excitations, that are polarized along the wire axis and homogeneously distributed throughout the entire volume.\cite{Furthmeier2016} 
In general, such spin densities do not constitute eigenstates of the spin-diffusion/Cooperon Hamiltonian and one has to solve the respective initial value problem.

Regarding this, we can set $Q_z=0$ and only focus on the dynamics in the cross-sectional plane ($\perp$).
Then the initial spin density $\mathbf{s}_0$ at the time $t=0$ is defined as
\begin{align}
\mathbf{s}_0(\mathbf{r})={}&\mathbf{\hat{z}} \,\Theta (R-\rho)/(\pi R^2),
\label{eq:init_cond}
\end{align}
where $\Theta$ denotes the Heaviside function and the total average spin $\boldsymbol{\mathcal{S}}(t)=\int {\rm d}^2r_\perp\,\mathbf{s}(\mathbf{r},t)$ is normalized at $t=0$ with respect to the cross-sectional plane, i.e., $\Vert\boldsymbol{\mathcal{S}}(0)\Vert=1$. 
The temporal and spatial evolution of the spin density according to Eq.~(\ref{eq:spindiff_eq}) yields
\begin{align}
\mathbf{s}(\mathbf{r},t)={}&U_{cd}^\dag U_{A}^\dag \exp(-D_e \hat{H}_c' t)\cdot \mathbf{s}_0', 
\end{align}
where $\mathbf{s}_0'=U_{A} U_{cd} \mathbf{s}_0$ or explicitely 
\begin{align}
\mathbf{s}_0'(\mathbf{r})={}&\frac{\Theta (R-\rho)}{\pi R^2}\Bigg[\frac{\sin(Q_\text{so}\rho)}{\sqrt{2}}\left(e^{i \phi}\ket{1,-1}-e^{-i \phi}\ket{1,1}\right)\notag\\
&\phantom{\frac{\Theta (R-\rho)}{\pi R^2}\Big[}+\cos(Q_\text{so}\rho)\ket{1,0}\Bigg],
\end{align}
represents the initial state in the singlet-triplet basis in the gauge-transformed system.
It is practical, to expand $\mathbf{s}_0'$ in the basis $\braket{\mathbf{r}|n,l,Q_z=0}$, Eq.~(\ref{eq:basis}), that fulfills the boundary condition of $\hat{H}_c'$.
Apparently, the deviation of the initial state from the zero-mode $\braket{\mathbf{r}|0}$, which is constant in real space, becomes stronger with increasing values of $R_\text{so}$.
As a consequence, the inclusion of higher modes and thereby larger spin relaxation rates in the expansion becomes more relevant.
In absence of the SOC terms in  $\hat{H}_c'$ the functions $\braket{\mathbf{r}|n,l,0}$ constitute the eigenbasis.
Hence, we can estimate the boundary-induced spin relaxation rates by $(1/\tau_s)_{n,\vert l \vert}:=D_e\braket{n,l,0|Q_\perp^2|n,l,0}=D_e(\zeta_{n,\vert l \vert}/R)^2$. 
This has a significant impact on the total spin relaxation rate even for small values of $R_\text{so}$.

\begin{figure}[t]
\includegraphics[width=.95\columnwidth]{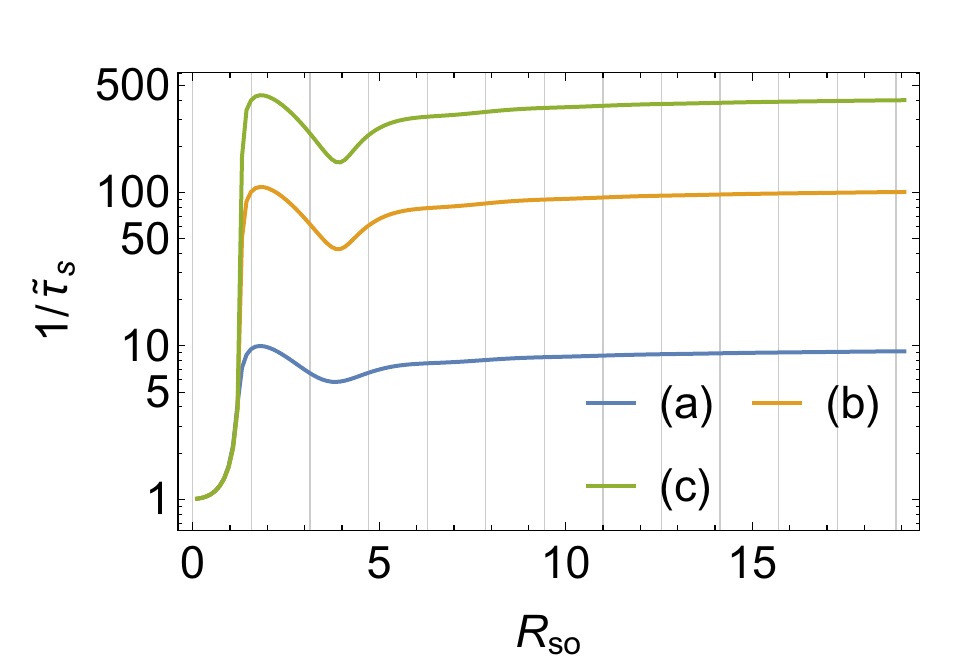}
\caption{Total spin relaxation rate   in terms of the 1D-diffusive rate, i.e.,  $1/\tilde{\tau}_s:=(\tau_s)^{1D}_z/(\tau_s)_z$, for a homogeneously $z$-polarized spin density in dependence of the radius $R_\text{so}$ for (a) $Q_\text{so}^2 =18 \,\delta_\text{D}^{(3)}$, (b) $Q_\text{so}^2 =220\,\delta_\text{D}^{(3)}$, and (c) $Q_\text{so}^2 =880\,\delta_\text{D}^{(3)}$.}
\label{fig:sz_rates}
\end{figure}

\begin{figure}[t]
\includegraphics[width=.95\columnwidth]{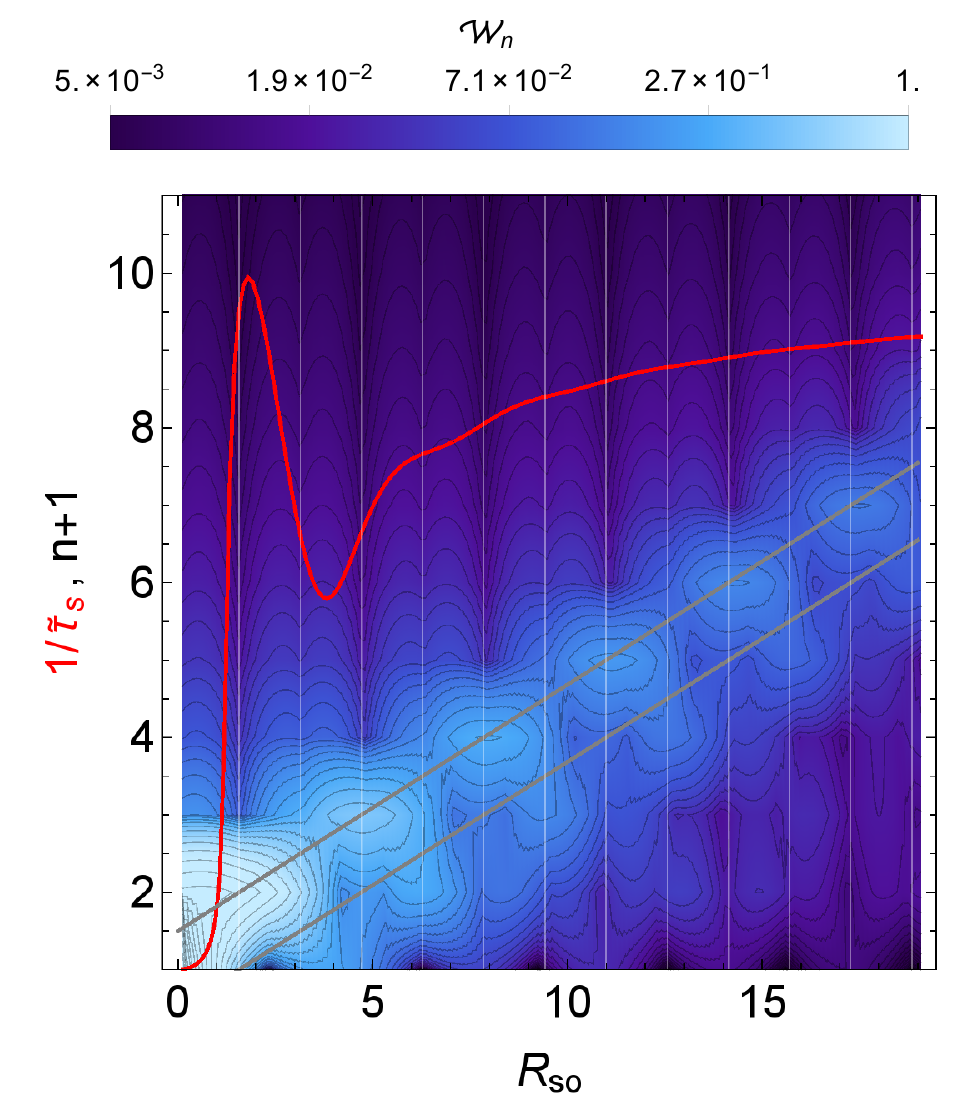}
\caption{The red solid line shows again the total spin relaxation rate $1/\tilde{\tau}_s$ as displayed in Fig.~\ref{fig:sz_rates}(a) in dependence of the radius $R_\text{so}$.
The density plot in the background visualizes the relative weight $\mathcal{W}_n$ of the $n$-th radial Cooperon modes $J_l^{(n)}$, that gives the dominant contribution in the expansion of the initial state $\mathbf{s}_0'$.
For better perceptibility, we summed over all contributing angular momentum quantum numbers in the expansion coefficients $c_{nl}$, i.e., $\mathcal{W}_n\propto\sum_{l\in\{0,\pm 1\}}\vert c_{nl}\vert $, where $c_{nl}=\int {\rm d}^2r_\perp\, \braket{n,l,0|\mathbf{r}}\mathbf{s}_0'$. 
The gray solid lines illustrate that the dominance of $J_l^{(n)}$ increases with the radius $R_\text{so}$ in discrete steps of approximately $R_\text{so}=n \pi/2$ for even and odd $n$, respectively.}
\label{fig:sz_density_plot}
\end{figure}

\begin{figure}[t]
\includegraphics[width=.95\columnwidth]{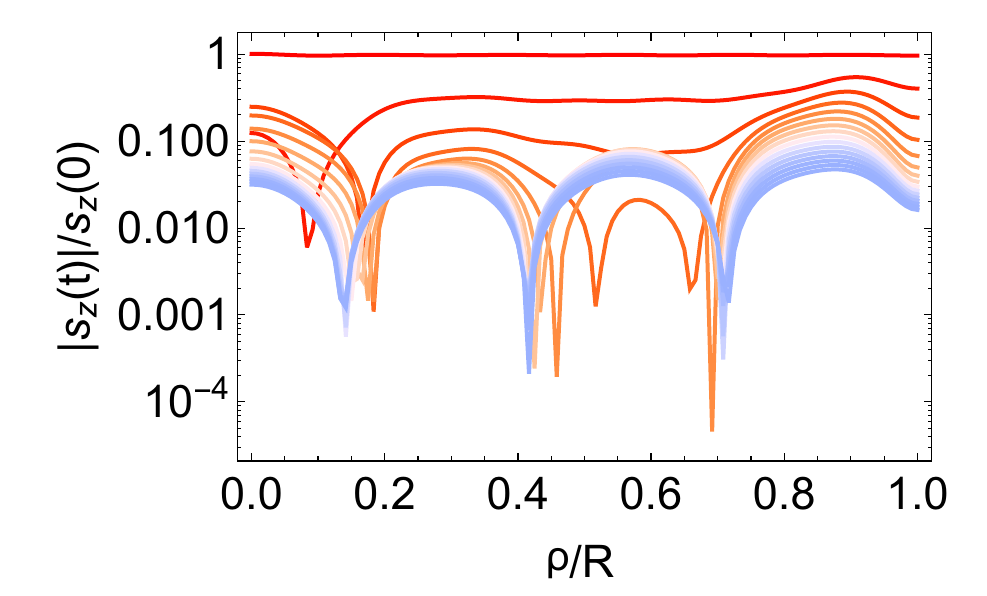}
\caption{Temporal and spatial evolution of a homogeneously $z$-polarized spin density for $Q_\text{so}^2=18\delta_\text{D}^{(3)}$ and $R_\text{so}=10$ in time-steps of $\Delta t=\tau_s$ from red to blue from $t=0$ to $t=15 \tau_s$, respectively. }
\label{fig:sz_time_ev}
\end{figure}

In Fig.~\ref{fig:sz_rates}, we display the numerically computed total spin relaxation rate $(1/\tau_s)_z$ in terms of the 1D-diffusive rate $(1/\tau_s)^{1D}_z$, Eq.~(\ref{eq:1D_rate}), in dependence of $R_\text{so}$ and for different ratios of $Q_\text{so}^2/\delta_\text{D}^{(3)}$.
The rate $(1/\tau_s)_z$ is defined by the time, after which the \textit{z}-component of the total spin is decayed to the factor $\mathcal{S}_z(t)/\mathcal{S}_z(0)=e^{-1}$ of its initial value.
Notice that, here a single-exponential fit is not necessarily reliable for the extraction of the spin relaxation rate since a single-exponential decay is only given for an eigenstate.
Most striking is the massive increase of the spin relaxation rate for small values of $R_\text{so}$.
The peak in the relaxation rate occurs almost precisely at $R_\text{so}=\pi/2$.
We can understand this behavior by noting that for $R_\text{so}=\pi/2$ the $\ket{1,\pm 1}$-components of $\mathbf{s}_0'$  can be well represented by the basis functions $\braket{\mathbf{r}|n=1,l=\mp 1,0}$.
The respective boundary-induced relaxation rate is given by $(1/\tau_s)_{1,\vert 1\vert}=D_e(2 \zeta_{1,1}/  \pi)^2 Q_\text{so}^2$, which is remarkably of the order of magnitude of the \textit{bulk} spin relaxation rate.
Similar but less pronounced resonances occur at larger integer values of $R_\text{so}/ (\pi/2)$.
As  the radius $R_\text{so}$ further increases, the influence of higher modes gains more and more weight and the mixing of the modes becomes larger, which is depicted in Fig.~\ref{fig:sz_density_plot}.
Nevertheless, the total increment is weakened by the simultaneously decreasing significance of the boundary-induced relaxation rates, which scale with $\propto R_\text{so}^{-1}$.
At last, we illustrate in Fig.~\ref{fig:sz_time_ev} the dynamical evolution of a spin density for the radius $R_\text{so}=10$, where the corresponding (gauge-transformed) initial state $\mathbf{s}_0'$ strongly deviates from a spatially homogeneous distribution.
Similar characteristic behavior was observed in planar quantum wires.\cite{Schwab2006}
The relaxation process of the local spin density $\mathbf{s}(\mathbf{r},t)$ is strongly inhomogeneous and locally  accelerated due to the fast-decaying modes.
As the optical measurement typically provides information about the average spin $\boldsymbol{\mathcal{S}}(t)$, the long-lived spin states are masked by the fast-decaying modes.
Note, that also in 2D systems an accelerated decay can be found if the initial state is spatially not homogeneous.\cite{Froltsov2001}

In conclusion, we found a dramatic change of the total average spin relaxation rate for an initially homogeneously $z$-polarized spin density with the wire radius. 
Within the range of ${0< R/L_\text{so} \leq 1/4}$ (with $l_e< R$) the spin relaxation rate varies from the very small 1D-diffusive rate to a rate which is of the order of the bulk spin relaxation rate.
This peculiar feature should be directly detectable in optical spin  injection measurements.\cite{Furthmeier2016}
We stress that, this behavior cannot be observed in zinc-blende nanowires since the homogeneous initial state, Eq.~(\ref{eq:init_cond}), constitutes an eigenstate that is independent of the wire radius.\cite{Kammermeier2017}
This is a consequence of the missing effective vector potential in Eq.~(\ref{eq:boundary}) which in turn is due to the lack of first-degree spherical harmonic SOC terms, in particular, the $k$-linear contribution.

\subsection{Conclusive remarks and example}\label{sec:relax_nanowire_remarks}

The intrinsic spin relaxation in bulk wurtzite semiconductors is dominated by the $k$-linear SOC terms.
In nanowires, however, owing to the interplay of the particular form of the wurtzite SOC Hamiltonian and the finite-size geometry, there exist special long-lived spin states.
The lifetimes of these states are mainly determined by the $k$-cubic SOC terms and are, thus, much longer than what is found in the bulk. 
At the same time, the long-lived spin states have in general a complex helical texture in real space, which is very sensitive to the system parameters, especially, the ratio of the spin precession length $L_\text{so}$ to the  nanowire radius $R$.
Magnetoconductance measurements of the weak (anti)localization always detect the lifetimes of the long-lived spin states irrespective of their texture.
In contrast, optical spin orientation determines the lifetime of some specifically configured state, which in most cases strongly deviates from the long-lived spin states.
Therefore, the extracted lifetimes in both experiments can differ drastically.
In particular, opposed to the magnetoconductance measurement the optical measured lifetime is highly sensitive to the nanowire radius.
They alter from the very long lifetime in narrow wires, which coincides with the lifetime of the long-lived spin states, to a very short lifetime, which is of the order of magnitude of the bulk lifetime.

\subsubsection*{Example: InAs nanowire in wurtzite phase}

In order to emphasize the significance of the results, we provide a concrete example of a wurtzite InAs nanowire grown along the [0001]-axis.
The spin relaxation in these systems has been experimentally investigated recently in Refs.~\onlinecite{Jespersen2018,Scheruebl2016} by means of magnetoconductance measurements.
Both studies use nanowires with diameters of about $\SI{80}{nm}$ and carrier densities which correspond to a 3D electron density $n\sim \SI{e17}{cm^{-3}}$.
The authors extract values for the spin relaxation length from fitting using different theoretical models.
Ref.~\onlinecite{Jespersen2018} applies the model of Kettemann\cite{Kettemann2007a} developed for diffusive planar wires with DP spin  relaxation.
On the other hand, Ref.~\onlinecite{Scheruebl2016} uses the 1D magnetoconductance model of Kurdak \textit{et al.},\cite{Kurdak1992} which is developed for ballistic planar wires.
As already pointed out in Ref.~\onlinecite{Scheruebl2016}, we emphasize that in both situations the utilized model does not include an accurate description of the wurtzite nanowire.
Ref.~\onlinecite{Jespersen2018} observes spin relaxation lengths of $\SI{75}{nm}$  and $\SI{100}{nm}$ for two different samples and a fixed gate voltage. 
In Ref.~\onlinecite{Scheruebl2016} various gating techniques are used which yield spin relaxation lengths of 150-$\SI{170}{nm}$ for low gate voltages.

\begin{figure}
\includegraphics[width=\columnwidth]{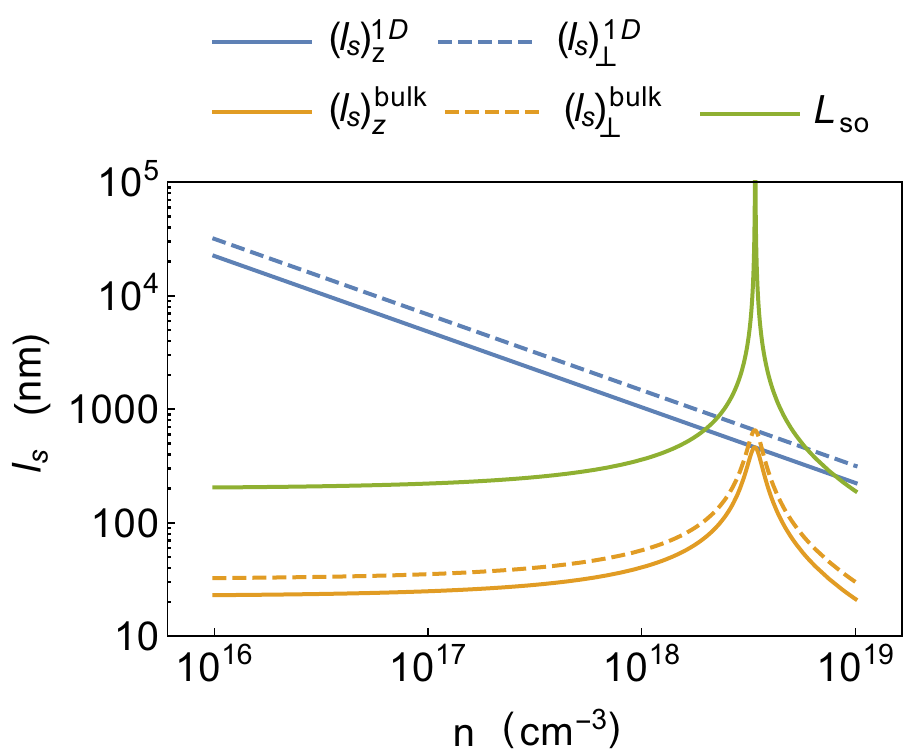}
\caption{Dependence of the spin precession length $L_\text{so}$ and spin relaxation lengths $l_s$ on the 3D electron density $n$ in wurtzite InAs.  The relaxation lengths of the bulk and long-lived spin states in the 1D-diffusive limit are labeled with $(l_s)^\text{bulk}_{z,\perp}$ and $(l_s)^\text{1D}_{z,\perp}$, respectively. }
\label{fig:InAs}
\end{figure}

For comparison with our findings, we consider an average effective mass $m$ of the $\Gamma_7$ conduction band of wurtzite InAs as $m=(2m_\perp+m_\parallel)/3$, where $m_\parallel=0.042\, m_0$, $m_\perp=0.037 \,m_0$, and $m_0$ denotes the bare electron mass.\cite{FariaJunior2016,Campos2018}
The respective SOC coefficients read as $\gamma_\text{R}^\text{int}=\SI{0.3}{eV\AA}$, $\gamma_\text{D}=\SI{132.5}{eV\AA^3}$, and $b=-1.24$.\cite{Gmitra2016}
The Fermi wave vector $k_F$ can be estimated from the 3D electron density $n$ as $k_F=(3\pi^2 n)^{1/3}$.
The DP spin relaxation length is related to the spin lifetime $\tau_s$ as $l_s=\sqrt{D_e\tau_s}$.
Let us concentrate on the relaxation of spin states that are homogeneously polarized in real space since the bulk eigenstates coincide with the nanowire eigenstates in the 1D-diffusive limit.
In Fig.~\ref{fig:InAs}, we compare the spin precession length and the spin relaxation lengths of the bulk and the long-lived spin states in the 1D diffusive limit with the spin relaxation rates Eqs.~(\ref{eq:hom_bulk_rate}) and (\ref{eq:1D_rate}), respectively.
In general, the density-modulation enters through the parameters $\delta_\text{D}^{(1)}$ and $\delta_\text{D}^{(3)}$, which result from the $k$-cubic SOC terms.
Remarkably, the spin precession length, i.e., $L_\text{so}=\pi\hbar^2/[m(\gamma_\text{R}^\text{int}+\delta_\text{D}^{(1)})]$, diverges for a large density of $n=\SI{3.4e18}{cm^{-3}}$ since the coefficients $\delta_\text{D}^{(1)}$ and $\gamma_\text{R}$ cancel each other.
In this case, the bulk spin relaxation lengths are solely determined by the $k$-cubic terms and, therefore, the relaxation lengths of bulk and long-lived spin states coincide.

Focusing on the regime of low to moderate electron densities, i.e., $n<\SI{e18}{cm^{-3}}$, the spin precession length alters only insignificantly, i.e., $L_\text{so}= 200-\SI{350}{nm}$. 
Moreover, the spin relaxation lengths of the long-lived spin states ($>\SI{1}{\mu m}$) are at least two orders of magnitude larger than the bulk spin relaxation lengths ($<\SI{60}{n m}$).
As we have seen above, for nanowires with diameter $d>L_\text{so}/2=100-\SI{175}{nm}$ the optical measurement will detect a spin relaxation length that is of the order of magnitude of the relaxation length in the bulk.
This is in strong contrast to the magnetoconductance measurement, which probes the spin relaxation of the long-lived spin states and hardly changes with the radius (cf. Fig.~\ref{fig:zero_mode_spec}).
Hence, there is a large discrepancy between experimental characterization methods.
These findings also indicate that in Refs.~\onlinecite{Scheruebl2016,Jespersen2018} the obtained spin relaxation lengths predominantly result from the externally induced Rashba SOC, assuming that the results do not largely deviate due to the 
employed magnetoconductance model.
This reasoning is also in agreement with the presumptions made in Ref.~\onlinecite{Jespersen2018}.
Last, it should be mentioned that for InAs in the low-density range additional SOC effects due to Fermi level surface pinning may become relevant.\cite{Bringer2011,Kammermeier2016,Degtyarev2017}
Their impact on the intrinsic spin relaxation in wurtzite nanowires shall be discussed elsewhere.

We conclude that it will be a delicate task to gain information about the intrinsic spin relaxation and the SOC coefficients from both experimental techniques.
In magnetoconductance measurements owing to the long-lived spin states the intrinsic relaxation features can be easily covered by the externally induced Rashba terms due to electrical gating.
On the other hand, in optical spin orientation the long-lived spin states are only excited in the 1D diffusive limit, where $R/ L_\text{so}\ll 1$.
Beyond this regime, the measured lifetime corresponds to a superposition of states and can strongly differ from the one of the long-lived spin states.

%

\section{Magnetoconductance correction}

\subsection{Nanowire with lateral gate electrode}\label{sec:gate}

To establish a connection to transport experiments and, thereby, enable a different experimental approach, we shall focus on the impact of the extrinsic SOC on the Cooperon modes in the following.
The external spin manipulation by electrical gating is a central component in magnetoconductance measurements as well as for the realization of all-electrical spintronic devices.

\begin{figure}[t]
\includegraphics[width=.95\columnwidth]{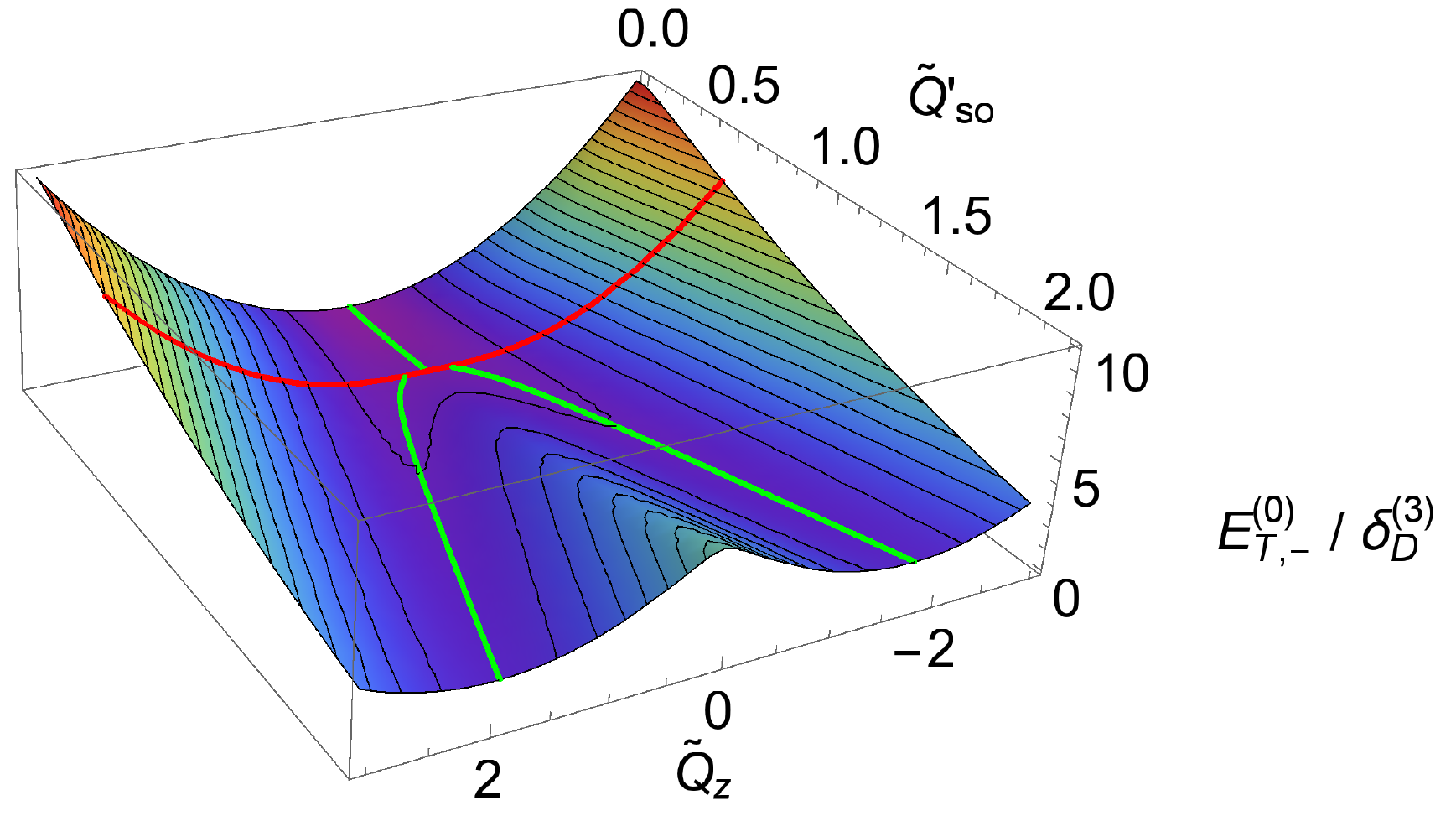}
\caption{Eigenvalue $E^{(0)}_{T,-}$  in terms of $\delta_\text{D}^{(3)}$ in dependence of $\tilde{Q}_\text{so}'=Q_\text{so}'/\sqrt{\delta_\text{D}^{(3)}}$ and $\tilde{Q}_z=Q_z/\sqrt{\delta_\text{D}^{(3)}}$. The green lines depict the minimum $E^{(0)}_{T,-}(Q_z=0)$, Eq.~(\ref{Em}), for $4Q_\text{so}'^2 < \delta_\text{D}^{(3)}$ and $E^{(0)}_{T,-}(\vert Q_{z,0} \vert)$, Eq.~(\ref{Em2}), elsewise. The red line marks the bifurcation point $4Q_\text{so}'^2 = \delta_\text{D}^{(3)}$.}
\label{fig:minima}
\end{figure}

\begin{figure}[t]
\includegraphics[width=\columnwidth]{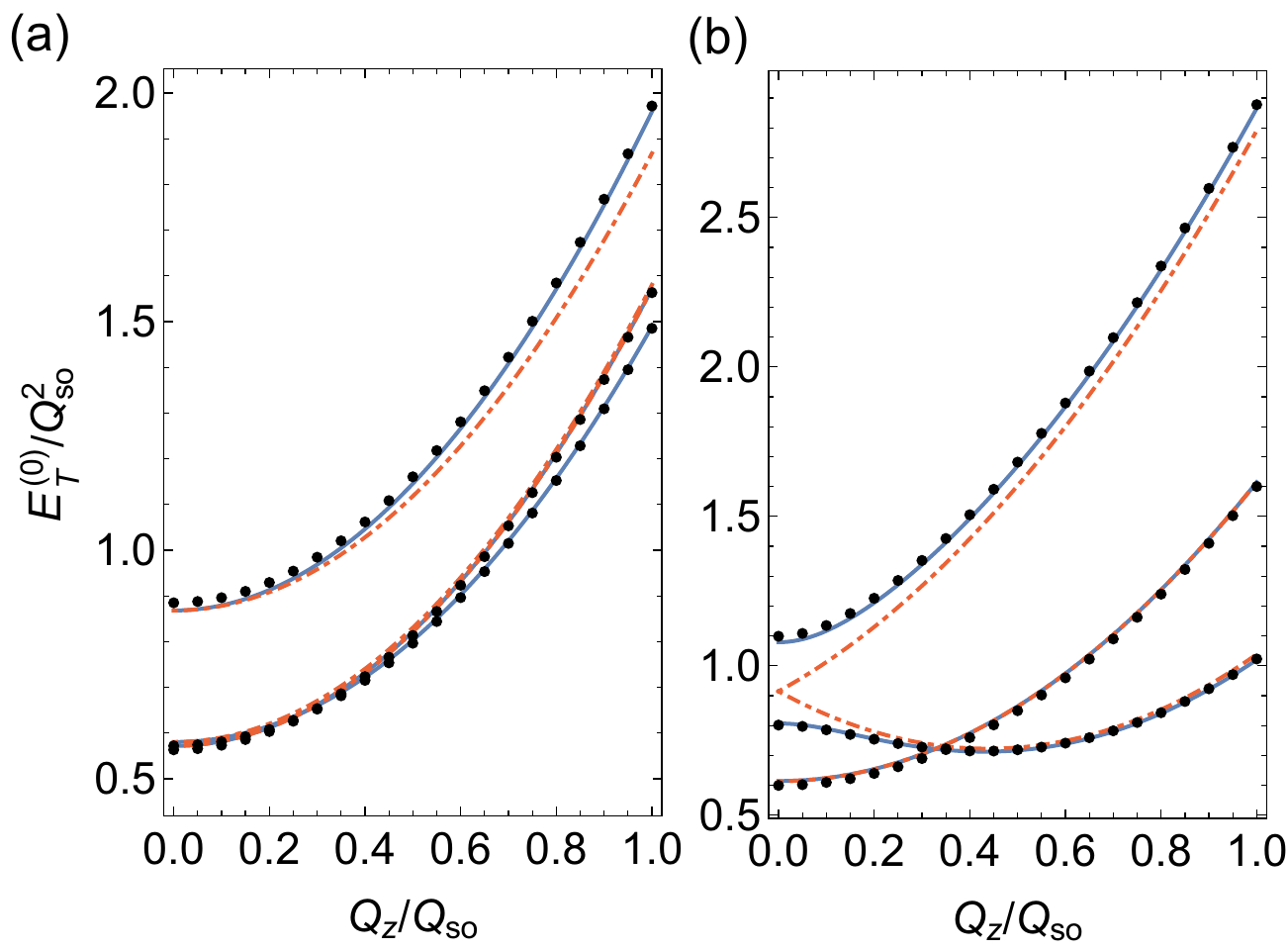}
\caption{Triplet eigenvalues $E^{(0)}_{T,j}$ in terms of $Q_\text{so}^2$ for $R_\text{so}=0.75$, $\delta_\text{D}^{(3)}/Q_\text{so}^2=0.5$ in the case of (a) dominant intrinsic SOC, i.e., $\eta=12.5$, or (b) dominant extrinsic SOC, i.e., $\eta=0.5$. The black dotted lines correspond to the exact eigenvalues of the full gauge-transformed Cooperon Hamiltonian $\hat{H}_c'$ in zero-mode approximation. The blue solid lines depict the approximative analytic solution for the eigenvalues, Eqs.~(\ref{EV2}) and (\ref{EV3}), and the red dot-dashed lines to the simplified solutions in the limiting cases, 
cf. Sec.~\ref{subsec:limit1} and \ref{subsec:limit2}.}
\label{fig:spec}
\end{figure}

Due the combination of intrinsic and extrinsic SOC contributions a straightforward gauge transformation of the Cooperon Hamiltonian is impractical.
In order to yet still obtain a useful analytical result, we may approximate the gauge-transformed Cooperon Hamitonian $\hat{H}_c'$ by expanding it in terms of $Q_\text{so}\rho$ ($Q_\text{so}'\rho$) up to second order, which is well justified for wires of width smaller than the spin precession length, i.e., $Q_\text{so}\rho(Q_\text{so}'\rho)\leq R_\text{so}(R_\text{so}')\ll 1$.
Here we defined $R_\text{so}'=Q_\text{so}'R$ with $Q_\text{so}'=2m\alpha_\text{R}^\text{ext}/\hbar^2$, which is related to the spin precession length $L_\text{so}^\text{ext}$ induced by the extrinsic SOC via $L_\text{so}^\text{ext}=2\pi/Q_\text{so}^\text{ext}$.
Using this simplification, the triplet eigenvalues in zero-mode approximation read as
\begin{align}
E^{(0)}_{T,0}={}&Q_z^2+\frac{1}{4}\left[Q_\text{so}'^2\varrho_\text{so}^2+\delta_\text{D}^{(3)}\left(4+R_\text{so}^2\right)\right],\label{EV2}\\
E^{(0)}_{T,\pm}={}&Q_z^2+\frac{1}{8}\Big[Q_\text{so}'^2(8-\varrho_\text{so}^2)\pm\sqrt{\kappa(Q_z)}\notag\\
&\phantom{Q_z^2+\frac{1}{8}[}+\delta_\text{D}^{(3)}(12-R_\text{so}^2)\Big]\label{EV3},
\end{align}
where $\varrho_\text{so}=\sqrt{R_\text{so}^2+R_\text{so}'^2}$ and
\begin{align}
\kappa(Q_z)={}&4Q_z^2 Q_\text{so}'^2(\varrho_\text{so}^2-8)^2+Q_\text{so}'^4\varrho_\text{so}^4\notag\\
&+2\delta_\text{D}^{(3)}Q_\text{so}'^2\left[4R_\text{so}'^2-R_\text{so}^2(4+R_\text{so}'^2)+3R_\text{so}^4\right]\notag\\
&+(\delta_\text{D}^{(3)})^2
\left[\left(4-3 R_\text{so}^2\right)^2+R_\text{so}^2R_\text{so}'^2\right].
\end{align}
By expanding up to second order in $R_\text{so}(R_\text{so}')$, one can easily verify that the correct results are obtained for the pure intrinsic and pure extrinsic SOC cases (cf. Sec.~\ref{sec:intr_spin_relax} and Ref.~\onlinecite{Kammermeier2017}, respectively).
In order to derive a closed-form expression for the magnetoconductivity, we consider below the two limiting cases, where either the extrinsic or intrinsic SOC dominates and the eigenvalues $E^{(0)}_{T,\pm}$ can be approximated by parabolas.
More precisely, for $\eta:=\delta_\text{D}^{(3)}/(4Q_\text{so}'^2)>1$ the eigenvalue $E^{(0)}_{T,-}$ exhibits one or otherwise two  minima (cf. Fig.~\ref{fig:minima}).
The derived expressions are compared in Fig.~\ref{fig:spec} to the numerical calculation of the spectrum with the full gauge-transformation and to the approximated spectrum in Eqs.~(\ref{EV2}) and (\ref{EV3}).

\subsubsection{Low extrinsic SOC and homogeneous spin density}\label{subsec:limit1}

For small external fields, i.e., $Q_\text{so}'/Q_\text{so}\ll 1$, the term in $\kappa$, which couples to the wave vector $Q_z$ can be neglected and the global minimum of the spectrum is found at $Q_z=0$.
In this case, the triplet eigenvalues $E^{(0)}_{T,\pm}$ simplify to gaped unit parabolas, i.e.,  
\begin{align}
E^{(0)}_{T,-}={}&Q_z^2+Q_\text{so}'^2\left(1-R_\text{so}'^2/4\right)+\delta_\text{D}^{(3)}\left(1+R_\text{so}^2/4\right)\label{Em},\\
E^{(0)}_{T,+}={}&Q_z^2+Q_\text{so}'^2\left(1-R_\text{so}^2/4\right)+\delta_\text{D}^{(3)}\left(2-R_\text{so}^2/2\right)
\label{Ep},
\end{align}
to second order in $R_\text{so}(R_\text{so}')$.

In analogy and for better comparison to many other previous works,\cite{Iordanskii1994,Knap1996,Kettemann2007a,Kammermeier2016,Kammermeier2017} the spin relaxation time is defined here by the global minimum of the spectrum at $Q_z=0$, which describes the decay of a spin density, that is homogeneously excited along the wire axis.
Even though it is determined by the relative strength of the extrinsic and intrinsic SOC, in the limit $R_\text{so}\rightarrow 0$ and $R_\text{so}' \rightarrow 0$ the lowest eigenvalue is always given by $E^{(0)}_{T,0}(0)$.
Therefore, we define here
\begin{align}
\frac{1}{\tau_s}={}&D_e E^{(0)}_{T,0}(0).
\label{eq:tau_s}
\end{align}
The eigenvectors of $\hat{H}_c$, that correspond to the eigenvalues $E^{(0)}_{T,j}(0)$ are $\mathbf{b}_0={}\mathbf{ \hat{x}},\mathbf{b}_-={}\mathbf{ \hat{y}},$ and $\mathbf{b}_+={}\mathbf{ \hat{z}}$ in the basis of spin density components to lowest order in $Q_\text{so}\rho$ ($Q_\text{so}'\rho$).

\subsubsection{Strong extrinsic SOC}\label{subsec:limit2}

In our previous work,\cite{Kammermeier2017} we have seen that for zinc-blende wires a dominating external field was necessary to observe WAL characteristics.
The latter are urgent for an unambigious parameter fitting.
For $\eta<1$,  the minimum of $E^{(0)}_{T,-}$ moves to finite wave vectors 
\begin{align}
\vert Q_{z,0}\vert={}&\frac{Q_\text{so}'}{16\sqrt{1-\eta^2}}\Big[\eta R_\text{so}'^2\left(1+2\eta\right)-\eta R_\text{so}^2\left(1+10\eta\right)\notag\\&\phantom{\frac{Q_\text{so}'}{16\sqrt{1-\eta^2}}\Big[}+2\left(8\eta^2-8+\rho_\text{so}^2\right)\Big]
\end{align}
to second order in $R_\text{so}(R_\text{so}')$, which yields the gap
\begin{align}
E^{(0)}_{T,-}(\vert Q_{z,0}\vert)={}&
\frac{Q_\text{so}'^2}{8}\big[\varrho_\text{so}^2+\eta\left(48-3R_\text{so}^2-R_\text{so}'^2\right)\notag\\
&\phantom{\frac{Q_\text{so}'^2}{8}\big[}+2\eta^2\left(5R_\text{so}^2-R_\text{so}'^2-4\right)\Big]
\label{Em2}
\end{align}
to second order in $R_\text{so}(R_\text{so}')$. 
Using this, we can rewrite the eigenvalues $E^{(0)}_{T,\pm}$ as
\begin{align}
E^{(0)}_{T,\pm}={}&\left(\vert Q_{z,0}\vert\pm\vert Q_z\vert\right)^2+E^{(0)}_{T,-}(\vert Q_{z,0}\vert)\label{Epm}.
\end{align}
For large extrinsic SOC, the gap $E^{(0)}_{T,-}(\vert Q_{z,0}\vert)$ turns into the global minimum of the spectrum, which underlines again the superiority of helical spin states and was also seen in other systems.\cite{Kettemann2007a,Schwab2006,Wenk2010,Wenk2011,Kammermeier2016,Kammermeier2016PRL,Kammermeier2017}
Neglecting the term $\propto  \eta^2$, we can estimate the transition to occur at 
\begin{align}
\eta\approx{}&\frac{\varrho_\text{so}^2}{16-11R_\text{so}^2-R_\text{so}'^2}\leq1/2, 
\end{align}
for $R_\text{so}\wedge R_\text{so}'\leq 1$.
Note that for $\eta \ll 1$,  the gap is about half as large as the global minimum for $Q_z=0$, i.e., $E^{(0)}_{T,0}(0)$.

\subsection{Zero-mode magnetoconductance correction}\label{sec:mc}

To support the experimental probing by means of  transport measurements, we provide analytical formulas for the magnetoconductance correction $\Delta G(B)$ in wurtzite nanowires.
We can write the leading-order correction of $\Delta G(B)=(\pi R^2/L)\Delta \sigma(B)$ for $\delta_\text{D}^{(3)}R^2 \ll 1$ in zero-mode approximation as
\begin{align}
\Delta G^{(0)}(B)={}&\frac{2e^2}{h}\frac{1}{L \pi}\int_0^{1/l_e}{\rm d}Q_z\,\Bigg(\frac{1}{Q_z^2+l_\phi^{-2}+l_B^{-2}}\notag\\
&-\sum_{j\in\{0,\pm\}}\frac{1}{E_{T,j}^{(0)}(Q_z)+l_\phi^{-2}+l_B^{-2}}\Bigg),
\label{eq:MC1}
\end{align}
where $L$ denotes the nanowire length, $l_\phi$  the electron dephasing length, and $l_e$ the mean free path.
The magnetic dephasing length $l_B$ depends on the orientation of the external magnetic field.
For a magnetic field perpendicular~$\perp$ or parallel~$\parallel$ to the nanowire axis, the magnetic length reads $l_{B,\perp}=\hbar/(eB R)$ or  $l_{B,\parallel}=\sqrt{2}l_{B,\perp}$, respectively.\cite{Kammermeier2017}
The $E_{T,j}^{(0)}(Q_z)$ represent the triplet eigenvalues of the Cooperon Hamiltonian in zero-mode approximation.  
In diffusive approximation $l_e$ is the shortest of all length scales.
In order to make the effects of the radial boundary relevant the dephasing lengths $l_\phi$ and $l_B$ should exceed the diameter $d$ of the nanowire.
As $l_B$  is computed within zero-mode approximation, we shall additionally demand that the magnetic field should be chosen small enough that the free magnetic length $\widetilde{l}_B=\sqrt{\hbar/(2e\vert B\vert)}$ is larger than the nanowire diameter $d$.\cite{Altshuler1981,Kammermeier2017} 

In the limiting cases of purely intrinsic as well as either dominant intrinsic or extrinsic SOC and neglecting the upper limit of the integral, we obtain the closed-form expression
\begin{align}
\Delta G^{(0)}(B)={}&\frac{2e^2}{h}\frac{1}{2 L}\left(\frac{1}{\sqrt{l_\phi^{-2}+l_B^{-2}}}\right.\notag\\
&\left.-\sum_{i}\frac{1}{\sqrt{l_\phi^{-2}+l_B^{-2}+l_{s,i}^{-2}}} \right),
\label{eq:MC2}
\end{align}
where  $l_{s,i}:=\left(E_{T,i,min}^{(0)}\right)^{-1/2}$ is the spin relaxation length of the $i$-th long-lived spin state according to the three lowest minima of the triplet spectrum.
(i) For purely intrinsic SOC and $\delta_\text{D}^{(3)}R^2 \ll 1$, the minima can be replaced by Eqs.~(\ref{eigenvalue2}) and (\ref{eigenvalue3}) for $Q_z=0$.
Regarding small radii $R_\text{so}(R_\text{so}')$ and
(ii) dominating intrinsic SOC, the $E_{T,i,min}^{(0)}$ are given by the gaps at $Q_z=0$, i.e., Eqs.~(\ref{EV2}), (\ref{Em}), and (\ref{Ep}), or
(iii) for dominating extrinsic SOC, we find one minima at $E_{T,0}^{(0)}(0)$, Eq.~(\ref{EV2}), and the other two both at $E^{(0)}_{T,-}(\vert Q_{z,0}\vert)$, Eq.~(\ref{Em2}).

On the other hand, considering small radii $R_\text{so}(R_\text{so}')$ but arbitrary ratios of extrinsic and intrinsic SOC, the integral in Eq.~(\ref{eq:MC1})  has to be solved numerically by using Eqs.~(\ref{EV2}) and  (\ref{EV3}).
Each of these cases allows a direct comparison with low-field magnetoconductance measurements and the extraction of transport parameters of the individual systems.
As an important aspect, we emphasize that the leading-order magnetoconductance correction is governed by the minimum in the spin relaxation rate. 
The corresponding long-lived spin states can, however, be difficult to realize in other experimental approaches.

Going beyond zero-mode approximation  requires the numerical diagonalization of the full multiband Cooperon (Hamiltonian).
As a result, writing down a closed-form expression as in Eq.~(\ref{eq:MC2}) is not possible anymore.
Yet, if the wire diameter is small enough 
and the separation between the modes is much larger than the broadening due to SOC, we might neglect the SOC-induced intermode mixing.
In this case, we can simply write 
\begin{align}
\Delta G=\sum_q \Delta G^{(q)},
\end{align}
where for each $\Delta G^{(q)}$ the Cooperon (Hamiltonian) is, analogously to the calculation of $\Delta G^{(0)}$, projected on the $q$-th Cooperon mode, i.e., $\braket{q|\hat{H}_c'|q}$ where $\ket{q}=\ket{n,l,Q_z}$ 
and $n\in\mathbbm{N}_0$, $l\in\mathbbm{Z}$, and $Q_z\in\mathbbm{R}$ 
as defined in Sec.~\ref{sec:finite-size}.
The impact of small magnetic fields can be treated by including the corresponding magnetic vector potential $\mathbf{A}$ via minimal coupling in the Cooperon Hamiltonian, i.e., $\mathbf{Q}\rightarrow \mathbf{Q}+2e\mathbf{A}/\hbar$ in Eq.~(\ref{eq:Hcbulk}).

\section{Summary and Conclusion}

We have studied the effects of a cylindrical boundary on the spin relaxation properties in wurtzite semiconductor nanowires.
The nanowires were assumed to be grown along the [0001] crystal axis and of approximately cylindrical shape.
The electron motion was considered diffusive transversally as well as longitudinally with respect to the nanowire axis.
In addition to the intrinsic SOC, the influence of an additional side-gate induced extrinsic Rashba SOC was taken into account.
Within zero-mode approximation for the Cooperon we derived explicit expressions for the leading-order magnetoconductance correction.

At this point, we summarize the previous observations and discuss the differences and similarities to zinc-blende semiconductor nanowires and planar quantum wires focusing primarily on the boundary effects on the intrinsic spin relaxation.\cite{Malshukov2000,Schwab2006,Kettemann2007a,Wenk2010,Wenk2011,Kammermeier2017} 
In general, the SOC terms can be sorted in terms of spherical harmonics.
Only the first-degree spherical harmonics give rise to an effective vector potential $\mathbf{A}_s$, which constitutes the key element in the boundary condition for the Cooperon, Eq.~(\ref{eq:boundary}).
In order to fulfill the boundary condition for the Cooperon, the component of the effective vector potential normal to the boundary, is removed by gauge transformation, e.g., $\boldsymbol{\hat{\rho}}\cdot\mathbf{A}_s$ in case of the cylindrical wire. 
This has two important consequences. 
(i) 
The spin relaxation rates, associated with the first-degree spherical harmonics of the removed vector potential, are suppressed. 
This gives rise to long-lived spin states with lifetimes much longer than in the bulk.
(ii) At the same time, these states assume a complex helical structure in real space, which depends on the spin precession length induced by the first-degree spherical harmonics SOC terms.

In zinc-blende nanowires, the Dresselhaus SOC consists solely of third-degree spherical harmonics.
Due to the absence of an effective vector potential, the boundary condition for the Cooperon is independent of the SOC and the lowest eigenstates (zero-mode) are constant in real space with respect to the cross-sectional plane.
The according intrinsic spin relaxation is therefore independent of the wire radius and identical with the bulk system.
The situation is fundamentally different in both wurtzite nanowires and planar zinc-blende quantum wires.
Owing to the presence of an effective vector potential, the boundary effect strongly reduces the minimal spin relaxation rates.
In wurtzite wires, the intrinsic vector potential lies completely in the cross-sectional plane.
Therefore, it is entirely removed by the gauge-transformation and the spin relaxation rate of the long-lived spin states is purely limited by the third-degree spherical harmonic SOC terms.
This rate is also hardly affected by any changes in the radius.
In quantum wires, the impact of the boundary is less significant since a share of the vector potential remains.
The respective minimal spin relaxation rate still depends on first-degree spherical harmonic terms.
However, it can be further suppressed in the 1D-diffusive limit leading to the well-known $1/\tau_s\propto W^2$ scaling with the wire width $W$.\cite{Kettemann2007a}

As stated above, the corresponding long-lived spin states exhibit, in general, a complex helical structure across the cross-section.
An experimental preparation of such states can be challenging.
In Sec.~\ref{sec:intr_spin_relax}, it was demonstrated that in wurtzite nanowires the optically-measured spin relaxation rate for a homogeneously $z$-polarized spin density shows a significant dependence on the wire radius whereas the spin relaxation rates of the long-lived eigenstates hardly varies.
More precisely, below the critical radius $R=L_\text{so}/4$ the spin relaxation rate massively decreases from the large bulk-like rate, mainly defined by the $k$-linear SOC terms, to a tiny rate, that is given by the $k$-cubic SOC terms and corresponds to the long-lived spin states. 
The reason is that, depending on the radius and the spin precession length, the real space structure of the  initial state can strongly deviate from the long-lived eigenstate.
Therefore, a comparison between the experimentally-extracted spin relaxation rates may be delusive.
Similar results can be expected for planar quantum wires.
Remarkably, however, this does not apply to zinc-blende nanowires since  the homogeneous initial state corresponds to a long-lived eigenstate and is independent of the wire radius.

On the other hand, the minima in the relaxation rate play a crucial role as they enter the leading-order quantum correction to the conductivity.
In wurtzite systems with purely intrinsic SOC, the minimum is determined by the parameter $\delta^{(3)}_\text{D}$, which results from the cubic Dresselhaus terms and is, thus, typically very small. 
As a consequence, the characteristic weak antilocalization minimum, which is often required for unambiguous parameter fitting,\cite{Kammermeier2017} is expected to appear at very low magnetic fields.
An exemplary comparison in Sec.~\ref{sec:relax_nanowire_remarks} of our predictions with recent experiments\cite{Scheruebl2016,Jespersen2018}   indicates that the intrinsic SOC effects can be easily obscured by the extrinsic effects  due to the utilization of an external gate.
To avoid this situation, we suggest transport experiments in which the electron density is modulated.
Since the spin relaxation rate is via $\delta^{(3)}_\text{D}$ highly sensitive to variations in the electron density, the magnetoconductance correction can be manipulated efficiently, e.g., by doping.
In case of a constant elastic scattering time $\tau_e$, a dependence of $\left(\tau_s\right)_{z}^\text{1D}\propto n_{3D}^{-2}$ should be observed, similar to a bulk zinc-blende system,\cite{Kikkawa1998,Dzhioev2002} but in contrast to a bulk wurtzite system.\cite{Buss2011}
For $\tau_s/\tau_\phi < 1.14$, where $\tau_s$ is defined in Eq.~(\ref{eq:tau_s}), a crossover from positive to negative magnetoconductance should be found.\cite{Kammermeier2016,Kammermeier2017}

To conclude, magnetoconductance measurements of the weak (anti)localization correction are convenient to extract transport parameters of the system.
They  constitute also a practical tool to identify the lowest possible spin relaxation rates and determine parameter configurations, which minimize them.
However, these experiments do not provide any information on the structure of the corresponding eigenstates.
Therefore, drawing general conclusions for the spin relaxation rate can be sometimes misleading.
The spin relaxation rate depends always  on the device geometry as well as the structure and orientation of the prepared state, where the latter can be controlled in optical experiments.
Therefore, optical and transport experiments are complementary tools, which together enable a reliable overall picture.

\section{Acknowledgments}

The authors thank Paulo E. Faria Junior, Martin Gmitra, and J. Carlos Egues for fruitful discussions.
This work was supported by Deutsche Forschungsgemeinschaft via Grants No. SFB 689, No. SFB 1277, and No.~336985961.

\appendix

\section{Intrinsic spin-orbit coupling}\label{app:SOC}
A spherical harmonic decomposition of the  intrinsic SOC Hamiltonian $\mathcal{H}_\text{so}^\text{int}=\sum_l(\mathcal{H}_\text{so}^\text{int})_{(l)}$, Eq.~(\ref{dresselhausWZ}), with respect to the angular momentum $l$, results in the two contributions, i.e., $l\in\{1,3\}$,
\begin{align}
(\mathcal{H}_\text{so}^\text{int})_{(1)}=&{}\left[\gamma_\text{R}^\text{int}+\frac{\gamma_\text{D}(b-4) k^2}{5}\right](k_y\sigma_x-k_x\sigma_y),\label{1SH}\\
(\mathcal{H}_\text{so}^\text{int})_{(3)}=&{}\frac{\gamma_\text{D}(b+1)}{5}\left(4k_z^2-k_\perp^2\right)(k_y\sigma_x-k_x\sigma_y).\label{3SH}
\end{align}
where $k_\perp^2=k_x^2+k_y^2$ and $k^2=k_x^2+k_y^2+k_z^2$.
In the ungated nanowire, the contribution $(\mathcal{H}_\text{so}^\text{int})_{(1)}$ is completely removed by the gauge transformation due to the boundary condition Eq.~(\ref{eq:boundary}).
Thus, the second term $(\mathcal{H}_\text{so}^\text{int})_{(3)}$ is  responsible for the DP spin relaxation in narrow nanowires.
It gives rise to the bulk spin relaxation term in Eq.~(\ref{eq:SRterm}).

\section{Spin matrices}\label{app:spin}

In a system with two electrons, the spin-1 matrices in the singlet-triplet basis $\ket{s,m_s}$, with total spin quantum number $s \in\{0,1\}$ and according magnetic quantum number $m_s \in\{0,\pm1\}$, read as

\begin{align}
S_x={}&
\,\frac{1}{\sqrt{2}}
\begin{pmatrix}
0&0&0&0\\
0&0&1&0\\
0&1&0&1\\
0&0&1&0
\end{pmatrix},\notag\\ 
S_y={}&
\,\frac{i}{\sqrt{2}}
\begin{pmatrix}
0&0&0&0\\
0&0&-1&0\\
0&1&0&-1\\
0&0&1&0
\end{pmatrix},\notag\\ 
S_z={}&
\,
\begin{pmatrix}
0&0&0&0\\
0&1&0&0\\
0&0&0&0\\
0&0&0&-1
\end{pmatrix},
\label{spinst}
\end{align}
in the order $\{\ket{0,0},\ket{1,1},\ket{1,0},\ket{1,-1}\}$. 
The singlet and triplet sectors are decoupled in this representation.

\section{Relation  between triplet basis and spin density components}\label{app:relation}

As shown in Ref.~\onlinecite{Wenk2010}, there exists a unitary transformation between the spin-diffusion equation and the Cooperon.
Therefore, we obtain an according transformation between the spin density $\mathbf{s}=(s_x,s_y,s_z)^\top$ and the triplet vector $\mathbf{\tilde{s}}=(\ket{1,1},\ket{1,0},\ket{1,-1})^\top$  of the Cooperon, which reads as
\begin{align}
\mathbf{\tilde{s}}&={}U_{cd}\,\mathbf{s},
\end{align}
with the unitary operator
\begin{align}
U_{cd}&={}\begin{pmatrix}
-1&i&0\\
0&0&\sqrt{2}\\
1&i&0
\end{pmatrix}/\sqrt{2}.
\end{align}
In Sec.~\ref{sec:intr_spin_relax}, we make use of this relation to
identify long-lived spin states and compute the decay of a certain  well-defined initial spin polarization.

\section{Diffusive-ballistic crossover}\label{app:Crossover}
As soon as the wire width becomes comparable to the mean free path, i.e. $W\sim l_e$, the condition of the transverse diffusivity is no more well fulfilled.
In the diffusive-ballistic crossover regime, the number of states for scattering becomes finite. 
Depending on the confinement, the number of available states will decrease with reduction of the wire width. 
Hence, we can include the crossover to the quasi-ballistic case by replacing the continuous integration over the Fermi surface in Eq.~(\ref{cooperon1}) by a sum over all discrete modes.\cite{Wenk2011} 
More precisely, when computing the Cooperon we are dealing with integrals $I$ of the form
\begin{align}
I{}&=\frac{1}{4\pi k_F^2}\int {\rm d}^3 k\, \delta(k_F-\vert\mathbf{k}\vert)f(\mathbf{k}),
\end{align}
where the Fermi contour is approximated to be spherical.
Due to symmetry, odd terms in $k_i$ vanish after integration.
Consequently, we can write $I$ as an integral over the unit sphere $\mathbf{u}=(u_x,u_y,u_z)=(k_x,k_y,k_z)/k_F$ in Cartesian coordinates, that is,
\begin{align}
I{}&=\frac{2}{\pi}\int_0^1 {\rm d}u_x \int_0^{\sqrt{1-u_x^2}}{\rm d}u_y\, \frac{f(u_x,u_y,\sqrt{1-u_x^2-u_y^2})}{\sqrt{1-u_x^2-u_y^2}}.\label{eq:surface_int}
\end{align}
For simplicity, we treat the size-quantization according to a square wire along $\mathbf{\hat{z}}$ with side lengths $W$ and hard-wall boundaries along the $\mathbf{\hat{x}}$ and $\mathbf{\hat{y}}$ axes.
The maximum number of modes $N$ along  $\mathbf{\hat{x}}$ (or $\mathbf{\hat{y}}$) is approximately $N=\lfloor \sqrt{s^2-1}\rfloor$ where $s=k_F W/\pi$ and $\lfloor \chi\rfloor$ denotes the integer part of $\chi$.
Thus, by replacing $u_x=n/s$ and $u_y=p/s$ with $n,p \in [1,N]$ we can express the (continuous) integral in Eq.~(\ref{eq:surface_int}) by a (discrete) sum over all channels, that is,
\begin{align}
I{}&=\frac{2}{\pi s}\sum_{n=1}^N\sum_{p=1}^{\sqrt{1+N^2-n^2}}\frac{f(\frac{n}{s},\frac{p}{s},\sqrt{1-\left(\frac{n}{s}\right)^2-\left(\frac{p}{s}\right)^2})}{\sqrt{s^2-n^2-p^2}}.
\end{align}
In Fig.~\ref{fig:matsubara}, we demonstrate the impact of the discretization on the parameter $\delta_\text{D}^{(3)}$, which is responsible for the finite spin relaxation rate even for $R_\text{so}\rightarrow0$.

\bibliographystyle{apsrev4-1}
\bibliography{WK}

\def\url#1{}
\begin{thebibliography}{87}%
\makeatletter
\providecommand \@ifxundefined [1]{%
 \@ifx{#1\undefined}
}%
\providecommand \@ifnum [1]{%
 \ifnum #1\expandafter \@firstoftwo
 \else \expandafter \@secondoftwo
 \fi
}%
\providecommand \@ifx [1]{%
 \ifx #1\expandafter \@firstoftwo
 \else \expandafter \@secondoftwo
 \fi
}%
\providecommand \natexlab [1]{#1}%
\providecommand \enquote  [1]{``#1''}%
\providecommand \bibnamefont  [1]{#1}%
\providecommand \bibfnamefont [1]{#1}%
\providecommand \citenamefont [1]{#1}%
\providecommand \href@noop [0]{\@secondoftwo}%
\providecommand \href [0]{\begingroup \@sanitize@url \@href}%
\providecommand \@href[1]{\@@startlink{#1}\@@href}%
\providecommand \@@href[1]{\endgroup#1\@@endlink}%
\providecommand \@sanitize@url [0]{\catcode `\\12\catcode `\$12\catcode
  `\&12\catcode `\#12\catcode `\^12\catcode `\_12\catcode `\%12\relax}%
\providecommand \@@startlink[1]{}%
\providecommand \@@endlink[0]{}%
\providecommand \url  [0]{\begingroup\@sanitize@url \@url }%
\providecommand \@url [1]{\endgroup\@href {#1}{\urlprefix }}%
\providecommand \urlprefix  [0]{URL }%
\providecommand \Eprint [0]{\href }%
\providecommand \doibase [0]{http://dx.doi.org/}%
\providecommand \selectlanguage [0]{\@gobble}%
\providecommand \bibinfo  [0]{\@secondoftwo}%
\providecommand \bibfield  [0]{\@secondoftwo}%
\providecommand \translation [1]{[#1]}%
\providecommand \BibitemOpen [0]{}%
\providecommand \bibitemStop [0]{}%
\providecommand \bibitemNoStop [0]{.\EOS\space}%
\providecommand \EOS [0]{\spacefactor3000\relax}%
\providecommand \BibitemShut  [1]{\csname bibitem#1\endcsname}%
\let\auto@bib@innerbib\@empty
\bibitem [{\citenamefont {Yang}\ \emph {et~al.}(2010)\citenamefont {Yang},
  \citenamefont {Yan},\ and\ \citenamefont {Fardy}}]{Yang2010}%
  \BibitemOpen
  \bibfield  {author} {\bibinfo {author} {\bibfnamefont {P.}~\bibnamefont
  {Yang}}, \bibinfo {author} {\bibfnamefont {R.}~\bibnamefont {Yan}}, \ and\
  \bibinfo {author} {\bibfnamefont {M.}~\bibnamefont {Fardy}},\ }\href
  {\doibase 10.1021/nl100665r} {\bibfield  {journal} {\bibinfo  {journal} {Nano
  Lett.}\ }\textbf {\bibinfo {volume} {10}},\ \bibinfo {pages} {1529} (\bibinfo
  {year} {2010})}\BibitemShut {NoStop}%
\bibitem [{\citenamefont {Mourik}\ \emph {et~al.}(2012)\citenamefont {Mourik},
  \citenamefont {Zuo}, \citenamefont {Frolov}, \citenamefont {Plissard},
  \citenamefont {Bakkers},\ and\ \citenamefont {Kouwenhoven}}]{Mourik2012}%
  \BibitemOpen
  \bibfield  {author} {\bibinfo {author} {\bibfnamefont {V.}~\bibnamefont
  {Mourik}}, \bibinfo {author} {\bibfnamefont {K.}~\bibnamefont {Zuo}},
  \bibinfo {author} {\bibfnamefont {S.~M.}\ \bibnamefont {Frolov}}, \bibinfo
  {author} {\bibfnamefont {S.~R.}\ \bibnamefont {Plissard}}, \bibinfo {author}
  {\bibfnamefont {E.~P. A.~M.}\ \bibnamefont {Bakkers}}, \ and\ \bibinfo
  {author} {\bibfnamefont {L.~P.}\ \bibnamefont {Kouwenhoven}},\ }\href
  {\doibase 10.1126/science.1222360} {\bibfield  {journal} {\bibinfo  {journal}
  {Science}\ }\textbf {\bibinfo {volume} {336}},\ \bibinfo {pages} {1003}
  (\bibinfo {year} {2012})}\BibitemShut {NoStop}%
\bibitem [{\citenamefont {Das}\ \emph {et~al.}(2012)\citenamefont {Das},
  \citenamefont {Ronen}, \citenamefont {Most}, \citenamefont {Oreg},
  \citenamefont {Heiblum},\ and\ \citenamefont {Shtrikman}}]{Das2012}%
  \BibitemOpen
  \bibfield  {author} {\bibinfo {author} {\bibfnamefont {A.}~\bibnamefont
  {Das}}, \bibinfo {author} {\bibfnamefont {Y.}~\bibnamefont {Ronen}}, \bibinfo
  {author} {\bibfnamefont {Y.}~\bibnamefont {Most}}, \bibinfo {author}
  {\bibfnamefont {Y.}~\bibnamefont {Oreg}}, \bibinfo {author} {\bibfnamefont
  {M.}~\bibnamefont {Heiblum}}, \ and\ \bibinfo {author} {\bibfnamefont
  {H.}~\bibnamefont {Shtrikman}},\ }\href {http://dx.doi.org/10.1038/nphys2479}
  {\bibfield  {journal} {\bibinfo  {journal} {Nat. Phys.}\ }\textbf {\bibinfo
  {volume} {8}},\ \bibinfo {pages} {887} (\bibinfo {year} {2012})}\BibitemShut
  {NoStop}%
\bibitem [{\citenamefont {Hofstetter}\ \emph {et~al.}(2009)\citenamefont
  {Hofstetter}, \citenamefont {Csonka}, \citenamefont {Nyg{\aa}rd},\ and\
  \citenamefont {Sch{\"o}nenberger}}]{Hofstetter2009}%
  \BibitemOpen
  \bibfield  {author} {\bibinfo {author} {\bibfnamefont {L.}~\bibnamefont
  {Hofstetter}}, \bibinfo {author} {\bibfnamefont {S.}~\bibnamefont {Csonka}},
  \bibinfo {author} {\bibfnamefont {J.}~\bibnamefont {Nyg{\aa}rd}}, \ and\
  \bibinfo {author} {\bibfnamefont {C.}~\bibnamefont {Sch{\"o}nenberger}},\
  }\href {http://dx.doi.org/10.1038/nature08432} {\bibfield  {journal}
  {\bibinfo  {journal} {Nature}\ }\textbf {\bibinfo {volume} {461}},\ \bibinfo
  {pages} {960} (\bibinfo {year} {2009})}\BibitemShut {NoStop}%
\bibitem [{\citenamefont {van Dam}\ \emph {et~al.}(2006)\citenamefont {van
  Dam}, \citenamefont {Nazarov}, \citenamefont {Bakkers}, \citenamefont
  {De~Franceschi},\ and\ \citenamefont {Kouwenhoven}}]{vanDam2006}%
  \BibitemOpen
  \bibfield  {author} {\bibinfo {author} {\bibfnamefont {J.~A.}\ \bibnamefont
  {van Dam}}, \bibinfo {author} {\bibfnamefont {Y.~V.}\ \bibnamefont
  {Nazarov}}, \bibinfo {author} {\bibfnamefont {E.~P. A.~M.}\ \bibnamefont
  {Bakkers}}, \bibinfo {author} {\bibfnamefont {S.}~\bibnamefont
  {De~Franceschi}}, \ and\ \bibinfo {author} {\bibfnamefont {L.~P.}\
  \bibnamefont {Kouwenhoven}},\ }\href {http://dx.doi.org/10.1038/nature05018}
  {\bibfield  {journal} {\bibinfo  {journal} {Nature}\ }\textbf {\bibinfo
  {volume} {442}},\ \bibinfo {pages} {667} (\bibinfo {year}
  {2006})}\BibitemShut {NoStop}%
\bibitem [{\citenamefont {Heedt}\ \emph {et~al.}(2013)\citenamefont {Heedt},
  \citenamefont {Wehrmann}, \citenamefont {Weis}, \citenamefont {Calarco},
  \citenamefont {Hardtdegen}, \citenamefont {Gr\"utzmacher}, \citenamefont
  {\mbox{Th}. Sch\"apers}, \citenamefont {Morgan},\ and\ \citenamefont
  {B\"urgler}}]{Heedt2013book}%
  \BibitemOpen
  \bibfield  {author} {\bibinfo {author} {\bibfnamefont {S.}~\bibnamefont
  {Heedt}}, \bibinfo {author} {\bibfnamefont {I.}~\bibnamefont {Wehrmann}},
  \bibinfo {author} {\bibfnamefont {K.}~\bibnamefont {Weis}}, \bibinfo {author}
  {\bibfnamefont {R.}~\bibnamefont {Calarco}}, \bibinfo {author} {\bibfnamefont
  {H.}~\bibnamefont {Hardtdegen}}, \bibinfo {author} {\bibfnamefont
  {D.}~\bibnamefont {Gr\"utzmacher}}, \bibinfo {author} {\bibnamefont
  {\mbox{Th}. Sch\"apers}}, \bibinfo {author} {\bibfnamefont {C.}~\bibnamefont
  {Morgan}}, \ and\ \bibinfo {author} {\bibfnamefont {D.~E.}\ \bibnamefont
  {B\"urgler}},\ }\enquote {\bibinfo {title} {Toward spin electronic devices
  based on semiconductor nanowires},}\ in\ \href {\doibase
  10.1002/9781118678107.ch25} {\emph {\bibinfo {booktitle} {Future Trends in
  Microelectronics}}}\ (\bibinfo  {publisher} {John Wiley {\&} Sons, Inc.},\
  \bibinfo {year} {2013})\ pp.\ \bibinfo {pages} {328--339}\BibitemShut
  {NoStop}%
\bibitem [{\citenamefont {Greytak}\ \emph {et~al.}(2005)\citenamefont
  {Greytak}, \citenamefont {Barrelet}, \citenamefont {Li},\ and\ \citenamefont
  {Lieber}}]{Greytak2005}%
  \BibitemOpen
  \bibfield  {author} {\bibinfo {author} {\bibfnamefont {A.~B.}\ \bibnamefont
  {Greytak}}, \bibinfo {author} {\bibfnamefont {C.~J.}\ \bibnamefont
  {Barrelet}}, \bibinfo {author} {\bibfnamefont {Y.}~\bibnamefont {Li}}, \ and\
  \bibinfo {author} {\bibfnamefont {C.~M.}\ \bibnamefont {Lieber}},\ }\href
  {\doibase 10.1063/1.2089157} {\bibfield  {journal} {\bibinfo  {journal}
  {Appl. Phys. Lett.}\ }\textbf {\bibinfo {volume} {87}},\ \bibinfo {eid}
  {151103} (\bibinfo {year} {2005})}\BibitemShut {NoStop}%
\bibitem [{\citenamefont {Xiang}\ \emph {et~al.}(2006)\citenamefont {Xiang},
  \citenamefont {Lu}, \citenamefont {Hu}, \citenamefont {Wu}, \citenamefont
  {Yan},\ and\ \citenamefont {Lieber}}]{Xiang2006}%
  \BibitemOpen
  \bibfield  {author} {\bibinfo {author} {\bibfnamefont {J.}~\bibnamefont
  {Xiang}}, \bibinfo {author} {\bibfnamefont {W.}~\bibnamefont {Lu}}, \bibinfo
  {author} {\bibfnamefont {Y.}~\bibnamefont {Hu}}, \bibinfo {author}
  {\bibfnamefont {Y.}~\bibnamefont {Wu}}, \bibinfo {author} {\bibfnamefont
  {H.}~\bibnamefont {Yan}}, \ and\ \bibinfo {author} {\bibfnamefont {C.~M.}\
  \bibnamefont {Lieber}},\ }\href {\doibase 10.1038/nature04796} {\bibfield
  {journal} {\bibinfo  {journal} {Nature}\ }\textbf {\bibinfo {volume} {441}},\
  \bibinfo {pages} {489} (\bibinfo {year} {2006})}\BibitemShut {NoStop}%
\bibitem [{\citenamefont {Nadj-Perge}\ \emph {et~al.}(2010)\citenamefont
  {Nadj-Perge}, \citenamefont {Frolov}, \citenamefont {Bakkers},\ and\
  \citenamefont {Kouwenhoven}}]{Nadj2010}%
  \BibitemOpen
  \bibfield  {author} {\bibinfo {author} {\bibfnamefont {S.}~\bibnamefont
  {Nadj-Perge}}, \bibinfo {author} {\bibfnamefont {S.~M.}\ \bibnamefont
  {Frolov}}, \bibinfo {author} {\bibfnamefont {E.~P. A.~M.}\ \bibnamefont
  {Bakkers}}, \ and\ \bibinfo {author} {\bibfnamefont {L.~P.}\ \bibnamefont
  {Kouwenhoven}},\ }\href {http://dx.doi.org/10.1038/nature09682} {\bibfield
  {journal} {\bibinfo  {journal} {Nature}\ }\textbf {\bibinfo {volume} {468}},\
  \bibinfo {pages} {1084} (\bibinfo {year} {2010})}\BibitemShut {NoStop}%
\bibitem [{\citenamefont {Nadj-Perge}\ \emph {et~al.}(2012)\citenamefont
  {Nadj-Perge}, \citenamefont {Pribiag}, \citenamefont {van~den Berg},
  \citenamefont {Zuo}, \citenamefont {Plissard}, \citenamefont {Bakkers},
  \citenamefont {Frolov},\ and\ \citenamefont {Kouwenhoven}}]{Nadj2012}%
  \BibitemOpen
  \bibfield  {author} {\bibinfo {author} {\bibfnamefont {S.}~\bibnamefont
  {Nadj-Perge}}, \bibinfo {author} {\bibfnamefont {V.~S.}\ \bibnamefont
  {Pribiag}}, \bibinfo {author} {\bibfnamefont {J.~W.~G.}\ \bibnamefont
  {van~den Berg}}, \bibinfo {author} {\bibfnamefont {K.}~\bibnamefont {Zuo}},
  \bibinfo {author} {\bibfnamefont {S.~R.}\ \bibnamefont {Plissard}}, \bibinfo
  {author} {\bibfnamefont {E.~P. A.~M.}\ \bibnamefont {Bakkers}}, \bibinfo
  {author} {\bibfnamefont {S.~M.}\ \bibnamefont {Frolov}}, \ and\ \bibinfo
  {author} {\bibfnamefont {L.~P.}\ \bibnamefont {Kouwenhoven}},\ }\href
  {\doibase 10.1103/PhysRevLett.108.166801} {\bibfield  {journal} {\bibinfo
  {journal} {Phys. Rev. Lett.}\ }\textbf {\bibinfo {volume} {108}},\ \bibinfo
  {pages} {166801} (\bibinfo {year} {2012})}\BibitemShut {NoStop}%
\bibitem [{\citenamefont {Krogstrup}\ \emph {et~al.}(2013)\citenamefont
  {Krogstrup}, \citenamefont {J{\o}rgensen}, \citenamefont {Heiss},
  \citenamefont {Demichel}, \citenamefont {Holm}, \citenamefont {Aagesen},
  \citenamefont {Nyg\aa{}rd},\ and\ \citenamefont {{Fontcuberta i
  \mbox{Morral}}}}]{Krogstrup2013}%
  \BibitemOpen
  \bibfield  {author} {\bibinfo {author} {\bibfnamefont {P.}~\bibnamefont
  {Krogstrup}}, \bibinfo {author} {\bibfnamefont {H.~I.}\ \bibnamefont
  {J{\o}rgensen}}, \bibinfo {author} {\bibfnamefont {M.}~\bibnamefont {Heiss}},
  \bibinfo {author} {\bibfnamefont {O.}~\bibnamefont {Demichel}}, \bibinfo
  {author} {\bibfnamefont {J.~V.}\ \bibnamefont {Holm}}, \bibinfo {author}
  {\bibfnamefont {M.}~\bibnamefont {Aagesen}}, \bibinfo {author} {\bibfnamefont
  {J.}~\bibnamefont {Nyg\aa{}rd}}, \ and\ \bibinfo {author} {\bibfnamefont
  {A.}~\bibnamefont {{Fontcuberta i \mbox{Morral}}}},\ }\href {\doibase
  10.1038/nphoton.2013.32} {\bibfield  {journal} {\bibinfo  {journal} {Nat.
  Photonics}\ }\textbf {\bibinfo {volume} {7}},\ \bibinfo {pages} {306}
  (\bibinfo {year} {2013})}\BibitemShut {NoStop}%
\bibitem [{\citenamefont {Dai}\ \emph {et~al.}(2014)\citenamefont {Dai},
  \citenamefont {Zhang}, \citenamefont {Wang}, \citenamefont {Adamo},
  \citenamefont {Liu}, \citenamefont {Huang}, \citenamefont {Couteau},\ and\
  \citenamefont {Soci}}]{Xing2014}%
  \BibitemOpen
  \bibfield  {author} {\bibinfo {author} {\bibfnamefont {X.}~\bibnamefont
  {Dai}}, \bibinfo {author} {\bibfnamefont {S.}~\bibnamefont {Zhang}}, \bibinfo
  {author} {\bibfnamefont {Z.}~\bibnamefont {Wang}}, \bibinfo {author}
  {\bibfnamefont {G.}~\bibnamefont {Adamo}}, \bibinfo {author} {\bibfnamefont
  {H.}~\bibnamefont {Liu}}, \bibinfo {author} {\bibfnamefont {Y.}~\bibnamefont
  {Huang}}, \bibinfo {author} {\bibfnamefont {C.}~\bibnamefont {Couteau}}, \
  and\ \bibinfo {author} {\bibfnamefont {C.}~\bibnamefont {Soci}},\ }\href
  {\doibase 10.1021/nl5006004} {\bibfield  {journal} {\bibinfo  {journal} {Nano
  Lett.}\ }\textbf {\bibinfo {volume} {14}},\ \bibinfo {pages} {2688} (\bibinfo
  {year} {2014})}\BibitemShut {NoStop}%
\bibitem [{\citenamefont {Faria~Junior}\ \emph {et~al.}(2015)\citenamefont
  {Faria~Junior}, \citenamefont {Xu}, \citenamefont {Lee}, \citenamefont
  {Gerhardt}, \citenamefont {Sipahi},\ and\ \citenamefont {\ifmmode
  \check{Z}\else \v{Z}\fi{}uti\ifmmode~\acute{c}\else
  \'{c}\fi{}}}]{FariaJunior2015}%
  \BibitemOpen
  \bibfield  {author} {\bibinfo {author} {\bibfnamefont {P.~E.}\ \bibnamefont
  {Faria~Junior}}, \bibinfo {author} {\bibfnamefont {G.}~\bibnamefont {Xu}},
  \bibinfo {author} {\bibfnamefont {J.}~\bibnamefont {Lee}}, \bibinfo {author}
  {\bibfnamefont {N.~C.}\ \bibnamefont {Gerhardt}}, \bibinfo {author}
  {\bibfnamefont {G.~M.}\ \bibnamefont {Sipahi}}, \ and\ \bibinfo {author}
  {\bibfnamefont {I.}~\bibnamefont {\ifmmode \check{Z}\else
  \v{Z}\fi{}uti\ifmmode~\acute{c}\else \'{c}\fi{}}},\ }\href {\doibase
  10.1103/PhysRevB.92.075311} {\bibfield  {journal} {\bibinfo  {journal} {Phys.
  Rev. B}\ }\textbf {\bibinfo {volume} {92}},\ \bibinfo {pages} {075311}
  (\bibinfo {year} {2015})}\BibitemShut {NoStop}%
\bibitem [{\citenamefont {Dai}\ \emph {et~al.}(2015)\citenamefont {Dai},
  \citenamefont {Messanvi}, \citenamefont {Zhang}, \citenamefont {Durand},
  \citenamefont {Eymery}, \citenamefont {Bougerol}, \citenamefont {Julien},\
  and\ \citenamefont {Tchernycheva}}]{Xing2015}%
  \BibitemOpen
  \bibfield  {author} {\bibinfo {author} {\bibfnamefont {X.}~\bibnamefont
  {Dai}}, \bibinfo {author} {\bibfnamefont {A.}~\bibnamefont {Messanvi}},
  \bibinfo {author} {\bibfnamefont {H.}~\bibnamefont {Zhang}}, \bibinfo
  {author} {\bibfnamefont {C.}~\bibnamefont {Durand}}, \bibinfo {author}
  {\bibfnamefont {J.}~\bibnamefont {Eymery}}, \bibinfo {author} {\bibfnamefont
  {C.}~\bibnamefont {Bougerol}}, \bibinfo {author} {\bibfnamefont {F.~H.}\
  \bibnamefont {Julien}}, \ and\ \bibinfo {author} {\bibfnamefont
  {M.}~\bibnamefont {Tchernycheva}},\ }\href {\doibase
  10.1021/acs.nanolett.5b02900} {\bibfield  {journal} {\bibinfo  {journal}
  {Nano Lett.}\ }\textbf {\bibinfo {volume} {15}},\ \bibinfo {pages} {6958}
  (\bibinfo {year} {2015})}\BibitemShut {NoStop}%
\bibitem [{\citenamefont {D'yakonov}\ and\ \citenamefont
  {Perel'}(1972)}]{perel}%
  \BibitemOpen
  \bibfield  {author} {\bibinfo {author} {\bibfnamefont {M.~I.}\ \bibnamefont
  {D'yakonov}}\ and\ \bibinfo {author} {\bibfnamefont {V.~I.}\ \bibnamefont
  {Perel'}},\ }\href@noop {} {\bibfield  {journal} {\bibinfo  {journal} {Sov.
  Phys. Solid State}\ }\textbf {\bibinfo {volume} {13}},\ \bibinfo {pages}
  {3023} (\bibinfo {year} {1972})},\ \bibinfo {note} {[Fiz. Tverd. Tela {\bf
  13}, 3581 (1971)]}\BibitemShut {NoStop}%
\bibitem [{\citenamefont {Kammermeier}\ \emph
  {et~al.}(2016{\natexlab{a}})\citenamefont {Kammermeier}, \citenamefont
  {Wenk},\ and\ \citenamefont {Schliemann}}]{Kammermeier2016PRL}%
  \BibitemOpen
  \bibfield  {author} {\bibinfo {author} {\bibfnamefont {M.}~\bibnamefont
  {Kammermeier}}, \bibinfo {author} {\bibfnamefont {P.}~\bibnamefont {Wenk}}, \
  and\ \bibinfo {author} {\bibfnamefont {J.}~\bibnamefont {Schliemann}},\
  }\href {\doibase 10.1103/PhysRevLett.117.236801} {\bibfield  {journal}
  {\bibinfo  {journal} {Phys. Rev. Lett.}\ }\textbf {\bibinfo {volume} {117}},\
  \bibinfo {pages} {236801} (\bibinfo {year} {2016}{\natexlab{a}})}\BibitemShut
  {NoStop}%
\bibitem [{\citenamefont {Wenk}\ \emph {et~al.}(2016)\citenamefont {Wenk},
  \citenamefont {Kammermeier},\ and\ \citenamefont {Schliemann}}]{Wenk2016}%
  \BibitemOpen
  \bibfield  {author} {\bibinfo {author} {\bibfnamefont {P.}~\bibnamefont
  {Wenk}}, \bibinfo {author} {\bibfnamefont {M.}~\bibnamefont {Kammermeier}}, \
  and\ \bibinfo {author} {\bibfnamefont {J.}~\bibnamefont {Schliemann}},\
  }\href {\doibase 10.1103/PhysRevB.93.115312} {\bibfield  {journal} {\bibinfo
  {journal} {Phys. Rev. B}\ }\textbf {\bibinfo {volume} {93}},\ \bibinfo
  {pages} {115312} (\bibinfo {year} {2016})}\BibitemShut {NoStop}%
\bibitem [{\citenamefont {Schliemann}\ \emph {et~al.}(2003)\citenamefont
  {Schliemann}, \citenamefont {Egues},\ and\ \citenamefont
  {Loss}}]{Schliemann2003}%
  \BibitemOpen
  \bibfield  {author} {\bibinfo {author} {\bibfnamefont {J.}~\bibnamefont
  {Schliemann}}, \bibinfo {author} {\bibfnamefont {J.~C.}\ \bibnamefont
  {Egues}}, \ and\ \bibinfo {author} {\bibfnamefont {D.}~\bibnamefont {Loss}},\
  }\href {\doibase 10.1103/PhysRevLett.90.146801} {\bibfield  {journal}
  {\bibinfo  {journal} {Phys. Rev. Lett.}\ }\textbf {\bibinfo {volume} {90}},\
  \bibinfo {pages} {146801} (\bibinfo {year} {2003})}\BibitemShut {NoStop}%
\bibitem [{\citenamefont {Bernevig}\ \emph {et~al.}(2006)\citenamefont
  {Bernevig}, \citenamefont {Orenstein},\ and\ \citenamefont
  {Zhang}}]{Bernevig2006}%
  \BibitemOpen
  \bibfield  {author} {\bibinfo {author} {\bibfnamefont {B.~A.}\ \bibnamefont
  {Bernevig}}, \bibinfo {author} {\bibfnamefont {J.}~\bibnamefont {Orenstein}},
  \ and\ \bibinfo {author} {\bibfnamefont {S.-C.}\ \bibnamefont {Zhang}},\
  }\href {\doibase 10.1103/PhysRevLett.97.236601} {\bibfield  {journal}
  {\bibinfo  {journal} {Phys. Rev. Lett.}\ }\textbf {\bibinfo {volume} {97}},\
  \bibinfo {pages} {236601} (\bibinfo {year} {2006})}\BibitemShut {NoStop}%
\bibitem [{\citenamefont {Trushin}\ and\ \citenamefont
  {Schliemann}(2007)}]{Trushin2007}%
  \BibitemOpen
  \bibfield  {author} {\bibinfo {author} {\bibfnamefont {M.}~\bibnamefont
  {Trushin}}\ and\ \bibinfo {author} {\bibfnamefont {J.}~\bibnamefont
  {Schliemann}},\ }\href {http://stacks.iop.org/1367-2630/9/i=9/a=346}
  {\bibfield  {journal} {\bibinfo  {journal} {New J. Phys.}\ }\textbf {\bibinfo
  {volume} {9}},\ \bibinfo {pages} {346} (\bibinfo {year} {2007})}\BibitemShut
  {NoStop}%
\bibitem [{\citenamefont {Dollinger}\ \emph {et~al.}(2014)\citenamefont
  {Dollinger}, \citenamefont {Kammermeier}, \citenamefont {Scholz},
  \citenamefont {Wenk}, \citenamefont {Schliemann}, \citenamefont {Richter},\
  and\ \citenamefont {Winkler}}]{Dollinger2014}%
  \BibitemOpen
  \bibfield  {author} {\bibinfo {author} {\bibfnamefont {T.}~\bibnamefont
  {Dollinger}}, \bibinfo {author} {\bibfnamefont {M.}~\bibnamefont
  {Kammermeier}}, \bibinfo {author} {\bibfnamefont {A.}~\bibnamefont {Scholz}},
  \bibinfo {author} {\bibfnamefont {P.}~\bibnamefont {Wenk}}, \bibinfo {author}
  {\bibfnamefont {J.}~\bibnamefont {Schliemann}}, \bibinfo {author}
  {\bibfnamefont {K.}~\bibnamefont {Richter}}, \ and\ \bibinfo {author}
  {\bibfnamefont {R.}~\bibnamefont {Winkler}},\ }\href {\doibase
  10.1103/PhysRevB.90.115306} {\bibfield  {journal} {\bibinfo  {journal} {Phys.
  Rev. B}\ }\textbf {\bibinfo {volume} {90}},\ \bibinfo {pages} {115306}
  (\bibinfo {year} {2014})}\BibitemShut {NoStop}%
\bibitem [{\citenamefont {Schliemann}(2017)}]{Schliemann2016}%
  \BibitemOpen
  \bibfield  {author} {\bibinfo {author} {\bibfnamefont {J.}~\bibnamefont
  {Schliemann}},\ }\href {\doibase 10.1103/RevModPhys.89.011001} {\bibfield
  {journal} {\bibinfo  {journal} {Rev. Mod. Phys.}\ }\textbf {\bibinfo {volume}
  {89}},\ \bibinfo {pages} {011001} (\bibinfo {year} {2017})}\BibitemShut
  {NoStop}%
\bibitem [{\citenamefont {Kohda}\ and\ \citenamefont
  {Salis}(2017)}]{Kohda2017}%
  \BibitemOpen
  \bibfield  {author} {\bibinfo {author} {\bibfnamefont {M.}~\bibnamefont
  {Kohda}}\ and\ \bibinfo {author} {\bibfnamefont {G.}~\bibnamefont {Salis}},\
  }\href {http://stacks.iop.org/0268-1242/32/i=7/a=073002} {\bibfield
  {journal} {\bibinfo  {journal} {Semicond. Sci. Technol.}\ }\textbf {\bibinfo
  {volume} {32}},\ \bibinfo {pages} {073002} (\bibinfo {year}
  {2017})}\BibitemShut {NoStop}%
\bibitem [{\citenamefont {Mal'shukov}\ and\ \citenamefont
  {Chao}(2000)}]{Malshukov2000}%
  \BibitemOpen
  \bibfield  {author} {\bibinfo {author} {\bibfnamefont {A.~G.}\ \bibnamefont
  {Mal'shukov}}\ and\ \bibinfo {author} {\bibfnamefont {K.~A.}\ \bibnamefont
  {Chao}},\ }\href {\doibase 10.1103/PhysRevB.61.R2413} {\bibfield  {journal}
  {\bibinfo  {journal} {Phys. Rev. B}\ }\textbf {\bibinfo {volume} {61}},\
  \bibinfo {pages} {R2413} (\bibinfo {year} {2000})}\BibitemShut {NoStop}%
\bibitem [{\citenamefont {Kiselev}\ and\ \citenamefont
  {Kim}(2000)}]{Kiselev2000}%
  \BibitemOpen
  \bibfield  {author} {\bibinfo {author} {\bibfnamefont {A.~A.}\ \bibnamefont
  {Kiselev}}\ and\ \bibinfo {author} {\bibfnamefont {K.~W.}\ \bibnamefont
  {Kim}},\ }\href {\doibase 10.1103/PhysRevB.61.13115} {\bibfield  {journal}
  {\bibinfo  {journal} {Phys. Rev. B}\ }\textbf {\bibinfo {volume} {61}},\
  \bibinfo {pages} {13115} (\bibinfo {year} {2000})}\BibitemShut {NoStop}%
\bibitem [{\citenamefont {Schwab}\ \emph {et~al.}(2006)\citenamefont {Schwab},
  \citenamefont {Dzierzawa}, \citenamefont {Gorini},\ and\ \citenamefont
  {Raimondi}}]{Schwab2006}%
  \BibitemOpen
  \bibfield  {author} {\bibinfo {author} {\bibfnamefont {P.}~\bibnamefont
  {Schwab}}, \bibinfo {author} {\bibfnamefont {M.}~\bibnamefont {Dzierzawa}},
  \bibinfo {author} {\bibfnamefont {C.}~\bibnamefont {Gorini}}, \ and\ \bibinfo
  {author} {\bibfnamefont {R.}~\bibnamefont {Raimondi}},\ }\href {\doibase
  10.1103/PhysRevB.74.155316} {\bibfield  {journal} {\bibinfo  {journal} {Phys.
  Rev. B}\ }\textbf {\bibinfo {volume} {74}},\ \bibinfo {pages} {155316}
  (\bibinfo {year} {2006})}\BibitemShut {NoStop}%
\bibitem [{\citenamefont {Kettemann}(2007)}]{Kettemann2007a}%
  \BibitemOpen
  \bibfield  {author} {\bibinfo {author} {\bibfnamefont {S.}~\bibnamefont
  {Kettemann}},\ }\href {http://link.aps.org/abstract/PRL/v98/e176808}
  {\bibfield  {journal} {\bibinfo  {journal} {Phys. Rev. Lett.}\ }\textbf
  {\bibinfo {volume} {98}},\ \bibinfo {pages} {176808} (\bibinfo {year}
  {2007})}\BibitemShut {NoStop}%
\bibitem [{\citenamefont {Wenk}\ and\ \citenamefont
  {Kettemann}(2010{\natexlab{a}})}]{Wenk2010}%
  \BibitemOpen
  \bibfield  {author} {\bibinfo {author} {\bibfnamefont {P.}~\bibnamefont
  {Wenk}}\ and\ \bibinfo {author} {\bibfnamefont {S.}~\bibnamefont
  {Kettemann}},\ }\href {\doibase 10.1103/PhysRevB.81.125309} {\bibfield
  {journal} {\bibinfo  {journal} {Phys. Rev. B}\ }\textbf {\bibinfo {volume}
  {81}},\ \bibinfo {pages} {125309} (\bibinfo {year}
  {2010}{\natexlab{a}})}\BibitemShut {NoStop}%
\bibitem [{\citenamefont {Wenk}\ and\ \citenamefont
  {Kettemann}(2011)}]{Wenk2011}%
  \BibitemOpen
  \bibfield  {author} {\bibinfo {author} {\bibfnamefont {P.}~\bibnamefont
  {Wenk}}\ and\ \bibinfo {author} {\bibfnamefont {S.}~\bibnamefont
  {Kettemann}},\ }\href {\doibase 10.1103/PhysRevB.83.115301} {\bibfield
  {journal} {\bibinfo  {journal} {Phys. Rev. B}\ }\textbf {\bibinfo {volume}
  {83}},\ \bibinfo {pages} {115301} (\bibinfo {year} {2011})}\BibitemShut
  {NoStop}%
\bibitem [{\citenamefont {Kammermeier}\ \emph {et~al.}(2017)\citenamefont
  {Kammermeier}, \citenamefont {Wenk}, \citenamefont {Schliemann},
  \citenamefont {Heedt}, \citenamefont {\mbox{Th}. Gerster},\ and\
  \citenamefont {\mbox{Th}. Sch\"apers}}]{Kammermeier2017}%
  \BibitemOpen
  \bibfield  {author} {\bibinfo {author} {\bibfnamefont {M.}~\bibnamefont
  {Kammermeier}}, \bibinfo {author} {\bibfnamefont {P.}~\bibnamefont {Wenk}},
  \bibinfo {author} {\bibfnamefont {J.}~\bibnamefont {Schliemann}}, \bibinfo
  {author} {\bibfnamefont {S.}~\bibnamefont {Heedt}}, \bibinfo {author}
  {\bibnamefont {\mbox{Th}. Gerster}}, \ and\ \bibinfo {author} {\bibnamefont
  {\mbox{Th}. Sch\"apers}},\ }\href {\doibase 10.1103/PhysRevB.96.235302}
  {\bibfield  {journal} {\bibinfo  {journal} {Phys. Rev. B}\ }\textbf {\bibinfo
  {volume} {96}},\ \bibinfo {pages} {235302} (\bibinfo {year}
  {2017})}\BibitemShut {NoStop}%
\bibitem [{\citenamefont {\mbox{Th}. Sch\"apers}\ \emph
  {et~al.}(2006)\citenamefont {\mbox{Th}. Sch\"apers}, \citenamefont {Guzenko},
  \citenamefont {Pala}, \citenamefont {Z\"ulicke}, \citenamefont {Governale},
  \citenamefont {Knobbe},\ and\ \citenamefont {Hardtdegen}}]{Schapers2006}%
  \BibitemOpen
  \bibfield  {author} {\bibinfo {author} {\bibnamefont {\mbox{Th}.
  Sch\"apers}}, \bibinfo {author} {\bibfnamefont {V.~A.}\ \bibnamefont
  {Guzenko}}, \bibinfo {author} {\bibfnamefont {M.~G.}\ \bibnamefont {Pala}},
  \bibinfo {author} {\bibfnamefont {U.}~\bibnamefont {Z\"ulicke}}, \bibinfo
  {author} {\bibfnamefont {M.}~\bibnamefont {Governale}}, \bibinfo {author}
  {\bibfnamefont {J.}~\bibnamefont {Knobbe}}, \ and\ \bibinfo {author}
  {\bibfnamefont {H.}~\bibnamefont {Hardtdegen}},\ }\href {\doibase
  10.1103/PhysRevB.74.081301} {\bibfield  {journal} {\bibinfo  {journal} {Phys.
  Rev. B}\ }\textbf {\bibinfo {volume} {74}},\ \bibinfo {pages} {081301}
  (\bibinfo {year} {2006})}\BibitemShut {NoStop}%
\bibitem [{\citenamefont {Holleitner}\ \emph {et~al.}(2007)\citenamefont
  {Holleitner}, \citenamefont {Sih}, \citenamefont {Myers}, \citenamefont
  {Gossard},\ and\ \citenamefont {Awschalom}}]{Holleitner2007}%
  \BibitemOpen
  \bibfield  {author} {\bibinfo {author} {\bibfnamefont {A.~W.}\ \bibnamefont
  {Holleitner}}, \bibinfo {author} {\bibfnamefont {V.}~\bibnamefont {Sih}},
  \bibinfo {author} {\bibfnamefont {R.~C.}\ \bibnamefont {Myers}}, \bibinfo
  {author} {\bibfnamefont {A.~C.}\ \bibnamefont {Gossard}}, \ and\ \bibinfo
  {author} {\bibfnamefont {D.~D.}\ \bibnamefont {Awschalom}},\ }\href {\doibase
  10.1088/1367-2630/9/9/342} {\bibfield  {journal} {\bibinfo  {journal} {New J.
  Phys.}\ }\textbf {\bibinfo {volume} {9}},\ \bibinfo {pages} {342} (\bibinfo
  {year} {2007})}\BibitemShut {NoStop}%
\bibitem [{\citenamefont {Glas}\ \emph {et~al.}(2007)\citenamefont {Glas},
  \citenamefont {Harmand},\ and\ \citenamefont {Patriarche}}]{Glas2007}%
  \BibitemOpen
  \bibfield  {author} {\bibinfo {author} {\bibfnamefont {F.}~\bibnamefont
  {Glas}}, \bibinfo {author} {\bibfnamefont {J.-C.}\ \bibnamefont {Harmand}}, \
  and\ \bibinfo {author} {\bibfnamefont {G.}~\bibnamefont {Patriarche}},\
  }\href {\doibase 10.1103/PhysRevLett.99.146101} {\bibfield  {journal}
  {\bibinfo  {journal} {Phys. Rev. Lett.}\ }\textbf {\bibinfo {volume} {99}},\
  \bibinfo {pages} {146101} (\bibinfo {year} {2007})}\BibitemShut {NoStop}%
\bibitem [{\citenamefont {Patriarche}\ \emph {et~al.}(2008)\citenamefont
  {Patriarche}, \citenamefont {Glas}, \citenamefont {Tchernycheva},
  \citenamefont {Sartel}, \citenamefont {Largeau}, \citenamefont {Harmand},\
  and\ \citenamefont {Cirlin}}]{Patriarche2008}%
  \BibitemOpen
  \bibfield  {author} {\bibinfo {author} {\bibfnamefont {G.}~\bibnamefont
  {Patriarche}}, \bibinfo {author} {\bibfnamefont {F.}~\bibnamefont {Glas}},
  \bibinfo {author} {\bibfnamefont {M.}~\bibnamefont {Tchernycheva}}, \bibinfo
  {author} {\bibfnamefont {C.}~\bibnamefont {Sartel}}, \bibinfo {author}
  {\bibfnamefont {L.}~\bibnamefont {Largeau}}, \bibinfo {author} {\bibfnamefont
  {J.-C.}\ \bibnamefont {Harmand}}, \ and\ \bibinfo {author} {\bibfnamefont
  {G.~E.}\ \bibnamefont {Cirlin}},\ }\href {\doibase 10.1021/nl080319y}
  {\bibfield  {journal} {\bibinfo  {journal} {Nano Lett.}\ }\textbf {\bibinfo
  {volume} {8}},\ \bibinfo {pages} {1638} (\bibinfo {year} {2008})}\BibitemShut
  {NoStop}%
\bibitem [{\citenamefont {Zhang}\ \emph {et~al.}(2010)\citenamefont {Zhang},
  \citenamefont {Luo}, \citenamefont {Zunger}, \citenamefont {Akopian},
  \citenamefont {Zwiller},\ and\ \citenamefont {Harmand}}]{Zhang2010a}%
  \BibitemOpen
  \bibfield  {author} {\bibinfo {author} {\bibfnamefont {L.}~\bibnamefont
  {Zhang}}, \bibinfo {author} {\bibfnamefont {J.-W.}\ \bibnamefont {Luo}},
  \bibinfo {author} {\bibfnamefont {A.}~\bibnamefont {Zunger}}, \bibinfo
  {author} {\bibfnamefont {N.}~\bibnamefont {Akopian}}, \bibinfo {author}
  {\bibfnamefont {V.}~\bibnamefont {Zwiller}}, \ and\ \bibinfo {author}
  {\bibfnamefont {J.-C.}\ \bibnamefont {Harmand}},\ }\href {\doibase
  10.1021/nl102109s} {\bibfield  {journal} {\bibinfo  {journal} {Nano Lett.}\
  }\textbf {\bibinfo {volume} {10}},\ \bibinfo {pages} {4055} (\bibinfo {year}
  {2010})}\BibitemShut {NoStop}%
\bibitem [{\citenamefont {Krogstrup}\ \emph {et~al.}(2011)\citenamefont
  {Krogstrup}, \citenamefont {Curiotto}, \citenamefont {Johnson}, \citenamefont
  {Aagesen}, \citenamefont {Nyg\aa{}rd},\ and\ \citenamefont
  {Chatain}}]{Krogstrup2011}%
  \BibitemOpen
  \bibfield  {author} {\bibinfo {author} {\bibfnamefont {P.}~\bibnamefont
  {Krogstrup}}, \bibinfo {author} {\bibfnamefont {S.}~\bibnamefont {Curiotto}},
  \bibinfo {author} {\bibfnamefont {E.}~\bibnamefont {Johnson}}, \bibinfo
  {author} {\bibfnamefont {M.}~\bibnamefont {Aagesen}}, \bibinfo {author}
  {\bibfnamefont {J.}~\bibnamefont {Nyg\aa{}rd}}, \ and\ \bibinfo {author}
  {\bibfnamefont {D.}~\bibnamefont {Chatain}},\ }\href {\doibase
  10.1103/PhysRevLett.106.125505} {\bibfield  {journal} {\bibinfo  {journal}
  {Phys. Rev. Lett.}\ }\textbf {\bibinfo {volume} {106}},\ \bibinfo {pages}
  {125505} (\bibinfo {year} {2011})}\BibitemShut {NoStop}%
\bibitem [{\citenamefont {Rieger}\ \emph {et~al.}(2013)\citenamefont {Rieger},
  \citenamefont {Lepsa}, \citenamefont {\mbox{Th.} Sch\"apers},\ and\
  \citenamefont {Gr\"utzmacher}}]{Rieger2013}%
  \BibitemOpen
  \bibfield  {author} {\bibinfo {author} {\bibfnamefont {T.}~\bibnamefont
  {Rieger}}, \bibinfo {author} {\bibfnamefont {M.~I.}\ \bibnamefont {Lepsa}},
  \bibinfo {author} {\bibnamefont {\mbox{Th.} Sch\"apers}}, \ and\ \bibinfo
  {author} {\bibfnamefont {D.}~\bibnamefont {Gr\"utzmacher}},\ }\href {\doibase
  https://doi.org/10.1016/j.jcrysgro.2012.12.035} {\bibfield  {journal}
  {\bibinfo  {journal} {J. Cryst. Growth}\ }\textbf {\bibinfo {volume} {378}},\
  \bibinfo {pages} {506 } (\bibinfo {year} {2013})}\BibitemShut {NoStop}%
\bibitem [{\citenamefont {Schroth}\ \emph {et~al.}(2015)\citenamefont
  {Schroth}, \citenamefont {K\"ohl}, \citenamefont {Hornung}, \citenamefont
  {Dimakis}, \citenamefont {Somaschini}, \citenamefont {Geelhaar},
  \citenamefont {Biermanns}, \citenamefont {Bauer}, \citenamefont {Lazarev},
  \citenamefont {Pietsch},\ and\ \citenamefont {Baumbach}}]{Schroth2015}%
  \BibitemOpen
  \bibfield  {author} {\bibinfo {author} {\bibfnamefont {P.}~\bibnamefont
  {Schroth}}, \bibinfo {author} {\bibfnamefont {M.}~\bibnamefont {K\"ohl}},
  \bibinfo {author} {\bibfnamefont {J.-W.}\ \bibnamefont {Hornung}}, \bibinfo
  {author} {\bibfnamefont {E.}~\bibnamefont {Dimakis}}, \bibinfo {author}
  {\bibfnamefont {C.}~\bibnamefont {Somaschini}}, \bibinfo {author}
  {\bibfnamefont {L.}~\bibnamefont {Geelhaar}}, \bibinfo {author}
  {\bibfnamefont {A.}~\bibnamefont {Biermanns}}, \bibinfo {author}
  {\bibfnamefont {S.}~\bibnamefont {Bauer}}, \bibinfo {author} {\bibfnamefont
  {S.}~\bibnamefont {Lazarev}}, \bibinfo {author} {\bibfnamefont
  {U.}~\bibnamefont {Pietsch}}, \ and\ \bibinfo {author} {\bibfnamefont
  {T.}~\bibnamefont {Baumbach}},\ }\href {\doibase
  10.1103/PhysRevLett.114.055504} {\bibfield  {journal} {\bibinfo  {journal}
  {Phys. Rev. Lett.}\ }\textbf {\bibinfo {volume} {114}},\ \bibinfo {pages}
  {055504} (\bibinfo {year} {2015})}\BibitemShut {NoStop}%
\bibitem [{\citenamefont {Fontcuberta~i
  Morral}(2016)}]{FontcubertaiMorral2016}%
  \BibitemOpen
  \bibfield  {author} {\bibinfo {author} {\bibfnamefont {A.}~\bibnamefont
  {Fontcuberta~i Morral}},\ }\href {http://dx.doi.org/10.1038/531308a}
  {\bibfield  {journal} {\bibinfo  {journal} {Nature}\ }\textbf {\bibinfo
  {volume} {531}},\ \bibinfo {pages} {308} (\bibinfo {year}
  {2016})}\BibitemShut {NoStop}%
\bibitem [{\citenamefont {Jacobsson}\ \emph {et~al.}(2016)\citenamefont
  {Jacobsson}, \citenamefont {Panciera}, \citenamefont {Tersoff}, \citenamefont
  {Reuter}, \citenamefont {Lehmann}, \citenamefont {Hofmann}, \citenamefont
  {Dick},\ and\ \citenamefont {Ross}}]{Jacobsson2016}%
  \BibitemOpen
  \bibfield  {author} {\bibinfo {author} {\bibfnamefont {D.}~\bibnamefont
  {Jacobsson}}, \bibinfo {author} {\bibfnamefont {F.}~\bibnamefont {Panciera}},
  \bibinfo {author} {\bibfnamefont {J.}~\bibnamefont {Tersoff}}, \bibinfo
  {author} {\bibfnamefont {M.~C.}\ \bibnamefont {Reuter}}, \bibinfo {author}
  {\bibfnamefont {S.}~\bibnamefont {Lehmann}}, \bibinfo {author} {\bibfnamefont
  {S.}~\bibnamefont {Hofmann}}, \bibinfo {author} {\bibfnamefont {K.~A.}\
  \bibnamefont {Dick}}, \ and\ \bibinfo {author} {\bibfnamefont {F.~M.}\
  \bibnamefont {Ross}},\ }\href {http://dx.doi.org/10.1038/nature17148}
  {\bibfield  {journal} {\bibinfo  {journal} {Nature}\ }\textbf {\bibinfo
  {volume} {531}},\ \bibinfo {pages} {317} (\bibinfo {year}
  {2016})}\BibitemShut {NoStop}%
\bibitem [{\citenamefont {Zhang}\ \emph {et~al.}(2018)\citenamefont {Zhang},
  \citenamefont {Sun}, \citenamefont {Sanchez}, \citenamefont {Ramsteiner},
  \citenamefont {Aagesen}, \citenamefont {Wu}, \citenamefont {Kim},
  \citenamefont {Jurczak}, \citenamefont {Huo}, \citenamefont {Lauhon},\ and\
  \citenamefont {Liu}}]{Zhang2018}%
  \BibitemOpen
  \bibfield  {author} {\bibinfo {author} {\bibfnamefont {Y.}~\bibnamefont
  {Zhang}}, \bibinfo {author} {\bibfnamefont {Z.}~\bibnamefont {Sun}}, \bibinfo
  {author} {\bibfnamefont {A.~M.}\ \bibnamefont {Sanchez}}, \bibinfo {author}
  {\bibfnamefont {M.}~\bibnamefont {Ramsteiner}}, \bibinfo {author}
  {\bibfnamefont {M.}~\bibnamefont {Aagesen}}, \bibinfo {author} {\bibfnamefont
  {J.}~\bibnamefont {Wu}}, \bibinfo {author} {\bibfnamefont {D.}~\bibnamefont
  {Kim}}, \bibinfo {author} {\bibfnamefont {P.}~\bibnamefont {Jurczak}},
  \bibinfo {author} {\bibfnamefont {S.}~\bibnamefont {Huo}}, \bibinfo {author}
  {\bibfnamefont {L.~J.}\ \bibnamefont {Lauhon}}, \ and\ \bibinfo {author}
  {\bibfnamefont {H.}~\bibnamefont {Liu}},\ }\href {\doibase
  10.1021/acs.nanolett.7b03366} {\bibfield  {journal} {\bibinfo  {journal}
  {Nano Letters}\ }\textbf {\bibinfo {volume} {18}},\ \bibinfo {pages} {81}
  (\bibinfo {year} {2018})}\BibitemShut {NoStop}%
\bibitem [{\citenamefont {De}\ and\ \citenamefont {Pryor}(2010)}]{De2010}%
  \BibitemOpen
  \bibfield  {author} {\bibinfo {author} {\bibfnamefont {A.}~\bibnamefont
  {De}}\ and\ \bibinfo {author} {\bibfnamefont {C.~E.}\ \bibnamefont {Pryor}},\
  }\href {\doibase 10.1103/PhysRevB.81.155210} {\bibfield  {journal} {\bibinfo
  {journal} {Phys. Rev. B}\ }\textbf {\bibinfo {volume} {81}},\ \bibinfo
  {pages} {155210} (\bibinfo {year} {2010})}\BibitemShut {NoStop}%
\bibitem [{\citenamefont {Cheiwchanchamnangij}\ and\ \citenamefont
  {Lambrecht}(2011)}]{Chei2011}%
  \BibitemOpen
  \bibfield  {author} {\bibinfo {author} {\bibfnamefont {T.}~\bibnamefont
  {Cheiwchanchamnangij}}\ and\ \bibinfo {author} {\bibfnamefont {W.~R.~L.}\
  \bibnamefont {Lambrecht}},\ }\href {\doibase 10.1103/PhysRevB.84.035203}
  {\bibfield  {journal} {\bibinfo  {journal} {Phys. Rev. B}\ }\textbf {\bibinfo
  {volume} {84}},\ \bibinfo {pages} {035203} (\bibinfo {year}
  {2011})}\BibitemShut {NoStop}%
\bibitem [{\citenamefont {Intronati}\ \emph {et~al.}(2013)\citenamefont
  {Intronati}, \citenamefont {Tamborenea}, \citenamefont {Weinmann},\ and\
  \citenamefont {Jalabert}}]{Intronati2013}%
  \BibitemOpen
  \bibfield  {author} {\bibinfo {author} {\bibfnamefont {G.~A.}\ \bibnamefont
  {Intronati}}, \bibinfo {author} {\bibfnamefont {P.~I.}\ \bibnamefont
  {Tamborenea}}, \bibinfo {author} {\bibfnamefont {D.}~\bibnamefont
  {Weinmann}}, \ and\ \bibinfo {author} {\bibfnamefont {R.~A.}\ \bibnamefont
  {Jalabert}},\ }\href {\doibase 10.1103/PhysRevB.88.045303} {\bibfield
  {journal} {\bibinfo  {journal} {Phys. Rev. B}\ }\textbf {\bibinfo {volume}
  {88}},\ \bibinfo {pages} {045303} (\bibinfo {year} {2013})}\BibitemShut
  {NoStop}%
\bibitem [{\citenamefont {Gmitra}\ and\ \citenamefont
  {Fabian}(2016)}]{Gmitra2016}%
  \BibitemOpen
  \bibfield  {author} {\bibinfo {author} {\bibfnamefont {M.}~\bibnamefont
  {Gmitra}}\ and\ \bibinfo {author} {\bibfnamefont {J.}~\bibnamefont
  {Fabian}},\ }\href {\doibase 10.1103/PhysRevB.94.165202} {\bibfield
  {journal} {\bibinfo  {journal} {Phys. Rev. B}\ }\textbf {\bibinfo {volume}
  {94}},\ \bibinfo {pages} {165202} (\bibinfo {year} {2016})}\BibitemShut
  {NoStop}%
\bibitem [{\citenamefont {Campos}\ \emph {et~al.}(2018)\citenamefont {Campos},
  \citenamefont {Faria~Junior}, \citenamefont {Gmitra}, \citenamefont
  {Sipahi},\ and\ \citenamefont {Fabian}}]{Campos2018}%
  \BibitemOpen
  \bibfield  {author} {\bibinfo {author} {\bibfnamefont {T.}~\bibnamefont
  {Campos}}, \bibinfo {author} {\bibfnamefont {P.~E.}\ \bibnamefont
  {Faria~Junior}}, \bibinfo {author} {\bibfnamefont {M.}~\bibnamefont
  {Gmitra}}, \bibinfo {author} {\bibfnamefont {G.~M.}\ \bibnamefont {Sipahi}},
  \ and\ \bibinfo {author} {\bibfnamefont {J.}~\bibnamefont {Fabian}},\ }\href
  {\doibase 10.1103/PhysRevB.97.245402} {\bibfield  {journal} {\bibinfo
  {journal} {Phys. Rev. B}\ }\textbf {\bibinfo {volume} {97}},\ \bibinfo
  {pages} {245402} (\bibinfo {year} {2018})}\BibitemShut {NoStop}%
\bibitem [{\citenamefont {Faria~Junior}\ \emph {et~al.}(2016)\citenamefont
  {Faria~Junior}, \citenamefont {Campos}, \citenamefont {Bastos}, \citenamefont
  {Gmitra}, \citenamefont {Fabian},\ and\ \citenamefont
  {Sipahi}}]{FariaJunior2016}%
  \BibitemOpen
  \bibfield  {author} {\bibinfo {author} {\bibfnamefont {P.~E.}\ \bibnamefont
  {Faria~Junior}}, \bibinfo {author} {\bibfnamefont {T.}~\bibnamefont
  {Campos}}, \bibinfo {author} {\bibfnamefont {C.~M.~O.}\ \bibnamefont
  {Bastos}}, \bibinfo {author} {\bibfnamefont {M.}~\bibnamefont {Gmitra}},
  \bibinfo {author} {\bibfnamefont {J.}~\bibnamefont {Fabian}}, \ and\ \bibinfo
  {author} {\bibfnamefont {G.~M.}\ \bibnamefont {Sipahi}},\ }\href {\doibase
  10.1103/PhysRevB.93.235204} {\bibfield  {journal} {\bibinfo  {journal} {Phys.
  Rev. B}\ }\textbf {\bibinfo {volume} {93}},\ \bibinfo {pages} {235204}
  (\bibinfo {year} {2016})}\BibitemShut {NoStop}%
\bibitem [{\citenamefont {Furthmeier}\ \emph {et~al.}(2016)\citenamefont
  {Furthmeier}, \citenamefont {Dirnberger}, \citenamefont {Gmitra},
  \citenamefont {Bayer}, \citenamefont {Forsch}, \citenamefont {Hubmann},
  \citenamefont {Sch\"uller}, \citenamefont {Reiger}, \citenamefont {Fabian},
  \citenamefont {Korn},\ and\ \citenamefont {Bougeard}}]{Furthmeier2016}%
  \BibitemOpen
  \bibfield  {author} {\bibinfo {author} {\bibfnamefont {S.}~\bibnamefont
  {Furthmeier}}, \bibinfo {author} {\bibfnamefont {F.}~\bibnamefont
  {Dirnberger}}, \bibinfo {author} {\bibfnamefont {M.}~\bibnamefont {Gmitra}},
  \bibinfo {author} {\bibfnamefont {A.}~\bibnamefont {Bayer}}, \bibinfo
  {author} {\bibfnamefont {M.}~\bibnamefont {Forsch}}, \bibinfo {author}
  {\bibfnamefont {J.}~\bibnamefont {Hubmann}}, \bibinfo {author} {\bibfnamefont
  {C.}~\bibnamefont {Sch\"uller}}, \bibinfo {author} {\bibfnamefont
  {E.}~\bibnamefont {Reiger}}, \bibinfo {author} {\bibfnamefont
  {J.}~\bibnamefont {Fabian}}, \bibinfo {author} {\bibfnamefont
  {T.}~\bibnamefont {Korn}}, \ and\ \bibinfo {author} {\bibfnamefont
  {D.}~\bibnamefont {Bougeard}},\ }\href {\doibase
  http://dx.doi.org/10.1038/ncomms12413 10.1038/ncomms12413} {\bibfield
  {journal} {\bibinfo  {journal} {Nat. Commun.}\ }\textbf {\bibinfo {volume}
  {7}},\ \bibinfo {pages} {12413} (\bibinfo {year} {2016})}\BibitemShut
  {NoStop}%
\bibitem [{\citenamefont {Scher\"ubl}\ \emph {et~al.}(2016)\citenamefont
  {Scher\"ubl}, \citenamefont {F\"ul\"op}, \citenamefont {Madsen},
  \citenamefont {Nyg\aa{}rd},\ and\ \citenamefont {Csonka}}]{Scheruebl2016}%
  \BibitemOpen
  \bibfield  {author} {\bibinfo {author} {\bibfnamefont {Z.}~\bibnamefont
  {Scher\"ubl}}, \bibinfo {author} {\bibfnamefont {G.}~\bibnamefont
  {F\"ul\"op}}, \bibinfo {author} {\bibfnamefont {M.~H.}\ \bibnamefont
  {Madsen}}, \bibinfo {author} {\bibfnamefont {J.}~\bibnamefont {Nyg\aa{}rd}},
  \ and\ \bibinfo {author} {\bibfnamefont {S.}~\bibnamefont {Csonka}},\ }\href
  {\doibase 10.1103/PhysRevB.94.035444} {\bibfield  {journal} {\bibinfo
  {journal} {Phys. Rev. B}\ }\textbf {\bibinfo {volume} {94}},\ \bibinfo
  {pages} {035444} (\bibinfo {year} {2016})}\BibitemShut {NoStop}%
\bibitem [{\citenamefont {Jespersen}\ \emph {et~al.}(2018)\citenamefont
  {Jespersen}, \citenamefont {Krogstrup}, \citenamefont {Lunde}, \citenamefont
  {Tanta}, \citenamefont {Kanne}, \citenamefont {Johnson},\ and\ \citenamefont
  {Nyg\aa{}rd}}]{Jespersen2018}%
  \BibitemOpen
  \bibfield  {author} {\bibinfo {author} {\bibfnamefont {T.~S.}\ \bibnamefont
  {Jespersen}}, \bibinfo {author} {\bibfnamefont {P.}~\bibnamefont
  {Krogstrup}}, \bibinfo {author} {\bibfnamefont {A.~M.}\ \bibnamefont
  {Lunde}}, \bibinfo {author} {\bibfnamefont {R.}~\bibnamefont {Tanta}},
  \bibinfo {author} {\bibfnamefont {T.}~\bibnamefont {Kanne}}, \bibinfo
  {author} {\bibfnamefont {E.}~\bibnamefont {Johnson}}, \ and\ \bibinfo
  {author} {\bibfnamefont {J.}~\bibnamefont {Nyg\aa{}rd}},\ }\href {\doibase
  10.1103/PhysRevB.97.041303} {\bibfield  {journal} {\bibinfo  {journal} {Phys.
  Rev. B}\ }\textbf {\bibinfo {volume} {97}},\ \bibinfo {pages} {041303}
  (\bibinfo {year} {2018})}\BibitemShut {NoStop}%
\bibitem [{\citenamefont {Hansen}\ \emph {et~al.}(2005)\citenamefont {Hansen},
  \citenamefont {Bj\"{o}rk}, \citenamefont {Fasth}, \citenamefont {Thelander},\
  and\ \citenamefont {Samuelson}}]{Hansen2005}%
  \BibitemOpen
  \bibfield  {author} {\bibinfo {author} {\bibfnamefont {A.~E.}\ \bibnamefont
  {Hansen}}, \bibinfo {author} {\bibfnamefont {M.~T.}\ \bibnamefont
  {Bj\"{o}rk}}, \bibinfo {author} {\bibfnamefont {I.~C.}\ \bibnamefont
  {Fasth}}, \bibinfo {author} {\bibfnamefont {C.}~\bibnamefont {Thelander}}, \
  and\ \bibinfo {author} {\bibfnamefont {L.}~\bibnamefont {Samuelson}},\ }\href
  {http://link.aps.org/abstract/PRB/v71/e205328} {\bibfield  {journal}
  {\bibinfo  {journal} {Phys. Rev. B}\ }\textbf {\bibinfo {volume} {71}},\
  \bibinfo {pages} {205328} (\bibinfo {year} {2005})}\BibitemShut {NoStop}%
\bibitem [{\citenamefont {Dhara}\ \emph {et~al.}(2009)\citenamefont {Dhara},
  \citenamefont {Solanki}, \citenamefont {Singh}, \citenamefont {Narayanan},
  \citenamefont {Chaudhari}, \citenamefont {Gokhale}, \citenamefont
  {Bhattacharya},\ and\ \citenamefont {Deshmukh}}]{Dhara2009}%
  \BibitemOpen
  \bibfield  {author} {\bibinfo {author} {\bibfnamefont {S.}~\bibnamefont
  {Dhara}}, \bibinfo {author} {\bibfnamefont {H.~S.}\ \bibnamefont {Solanki}},
  \bibinfo {author} {\bibfnamefont {V.}~\bibnamefont {Singh}}, \bibinfo
  {author} {\bibfnamefont {A.}~\bibnamefont {Narayanan}}, \bibinfo {author}
  {\bibfnamefont {P.}~\bibnamefont {Chaudhari}}, \bibinfo {author}
  {\bibfnamefont {M.}~\bibnamefont {Gokhale}}, \bibinfo {author} {\bibfnamefont
  {A.}~\bibnamefont {Bhattacharya}}, \ and\ \bibinfo {author} {\bibfnamefont
  {M.~M.}\ \bibnamefont {Deshmukh}},\ }\href {\doibase
  10.1103/PhysRevB.79.121311} {\bibfield  {journal} {\bibinfo  {journal} {Phys.
  Rev. B}\ }\textbf {\bibinfo {volume} {79}},\ \bibinfo {pages} {121311}
  (\bibinfo {year} {2009})}\BibitemShut {NoStop}%
\bibitem [{\citenamefont {Hao}\ \emph {et~al.}(2010)\citenamefont {Hao},
  \citenamefont {Tu}, \citenamefont {Cao}, \citenamefont {Zhou}, \citenamefont
  {Li}, \citenamefont {Guo}, \citenamefont {Fung}, \citenamefont {Ji},
  \citenamefont {Guo},\ and\ \citenamefont {Lu}}]{XiaoJie2010}%
  \BibitemOpen
  \bibfield  {author} {\bibinfo {author} {\bibfnamefont {X.-J.}\ \bibnamefont
  {Hao}}, \bibinfo {author} {\bibfnamefont {T.}~\bibnamefont {Tu}}, \bibinfo
  {author} {\bibfnamefont {G.}~\bibnamefont {Cao}}, \bibinfo {author}
  {\bibfnamefont {C.}~\bibnamefont {Zhou}}, \bibinfo {author} {\bibfnamefont
  {H.-O.}\ \bibnamefont {Li}}, \bibinfo {author} {\bibfnamefont {G.-C.}\
  \bibnamefont {Guo}}, \bibinfo {author} {\bibfnamefont {W.~Y.}\ \bibnamefont
  {Fung}}, \bibinfo {author} {\bibfnamefont {Z.}~\bibnamefont {Ji}}, \bibinfo
  {author} {\bibfnamefont {G.-P.}\ \bibnamefont {Guo}}, \ and\ \bibinfo
  {author} {\bibfnamefont {W.}~\bibnamefont {Lu}},\ }\href {\doibase
  10.1021/nl101181e} {\bibfield  {journal} {\bibinfo  {journal} {Nano Lett.}\
  }\textbf {\bibinfo {volume} {10}},\ \bibinfo {pages} {2956} (\bibinfo {year}
  {2010})}\BibitemShut {NoStop}%
\bibitem [{\citenamefont {van Weperen}\ \emph {et~al.}(2015)\citenamefont {van
  Weperen}, \citenamefont {Tarasinski}, \citenamefont {Eeltink}, \citenamefont
  {Pribiag}, \citenamefont {Plissard}, \citenamefont {Bakkers}, \citenamefont
  {Kouwenhoven},\ and\ \citenamefont {Wimmer}}]{Weperen2015}%
  \BibitemOpen
  \bibfield  {author} {\bibinfo {author} {\bibfnamefont {I.}~\bibnamefont {van
  Weperen}}, \bibinfo {author} {\bibfnamefont {B.}~\bibnamefont {Tarasinski}},
  \bibinfo {author} {\bibfnamefont {D.}~\bibnamefont {Eeltink}}, \bibinfo
  {author} {\bibfnamefont {V.~S.}\ \bibnamefont {Pribiag}}, \bibinfo {author}
  {\bibfnamefont {S.~R.}\ \bibnamefont {Plissard}}, \bibinfo {author}
  {\bibfnamefont {E.~P. A.~M.}\ \bibnamefont {Bakkers}}, \bibinfo {author}
  {\bibfnamefont {L.~P.}\ \bibnamefont {Kouwenhoven}}, \ and\ \bibinfo {author}
  {\bibfnamefont {M.}~\bibnamefont {Wimmer}},\ }\href {\doibase
  10.1103/PhysRevB.91.201413} {\bibfield  {journal} {\bibinfo  {journal} {Phys.
  Rev. B}\ }\textbf {\bibinfo {volume} {91}},\ \bibinfo {pages} {201413}
  (\bibinfo {year} {2015})}\BibitemShut {NoStop}%
\bibitem [{\citenamefont {Roulleau}\ \emph {et~al.}(2010)\citenamefont
  {Roulleau}, \citenamefont {Choi}, \citenamefont {Riedi}, \citenamefont
  {Heinzel}, \citenamefont {Shorubalko}, \citenamefont {Ihn},\ and\
  \citenamefont {Ensslin}}]{Roulleau2010}%
  \BibitemOpen
  \bibfield  {author} {\bibinfo {author} {\bibfnamefont {P.}~\bibnamefont
  {Roulleau}}, \bibinfo {author} {\bibfnamefont {T.}~\bibnamefont {Choi}},
  \bibinfo {author} {\bibfnamefont {S.}~\bibnamefont {Riedi}}, \bibinfo
  {author} {\bibfnamefont {T.}~\bibnamefont {Heinzel}}, \bibinfo {author}
  {\bibfnamefont {I.}~\bibnamefont {Shorubalko}}, \bibinfo {author}
  {\bibfnamefont {T.}~\bibnamefont {Ihn}}, \ and\ \bibinfo {author}
  {\bibfnamefont {K.}~\bibnamefont {Ensslin}},\ }\href {\doibase
  10.1103/PhysRevB.81.155449} {\bibfield  {journal} {\bibinfo  {journal} {Phys.
  Rev. B}\ }\textbf {\bibinfo {volume} {81}},\ \bibinfo {pages} {155449}
  (\bibinfo {year} {2010})}\BibitemShut {NoStop}%
\bibitem [{\citenamefont {Liang}\ and\ \citenamefont {Gao}(2012)}]{Liang2012}%
  \BibitemOpen
  \bibfield  {author} {\bibinfo {author} {\bibfnamefont {D.}~\bibnamefont
  {Liang}}\ and\ \bibinfo {author} {\bibfnamefont {X.~P.}\ \bibnamefont
  {Gao}},\ }\href {\doibase 10.1021/nl301325h} {\bibfield  {journal} {\bibinfo
  {journal} {Nano Lett.}\ }\textbf {\bibinfo {volume} {12}},\ \bibinfo {pages}
  {3263} (\bibinfo {year} {2012})}\BibitemShut {NoStop}%
\bibitem [{\citenamefont {Takase}\ \emph {et~al.}(2017)\citenamefont {Takase},
  \citenamefont {Ashikawa}, \citenamefont {Zhang}, \citenamefont {Tateno},\
  and\ \citenamefont {Sasaki}}]{Takase2017}%
  \BibitemOpen
  \bibfield  {author} {\bibinfo {author} {\bibfnamefont {K.}~\bibnamefont
  {Takase}}, \bibinfo {author} {\bibfnamefont {Y.}~\bibnamefont {Ashikawa}},
  \bibinfo {author} {\bibfnamefont {G.}~\bibnamefont {Zhang}}, \bibinfo
  {author} {\bibfnamefont {K.}~\bibnamefont {Tateno}}, \ and\ \bibinfo {author}
  {\bibfnamefont {S.}~\bibnamefont {Sasaki}},\ }\href {\doibase
  10.1038/s41598-017-01080-0} {\bibfield  {journal} {\bibinfo  {journal} {Sci.
  Rep.}\ }\textbf {\bibinfo {volume} {7}},\ \bibinfo {pages} {930} (\bibinfo
  {year} {2017})}\BibitemShut {NoStop}%
\bibitem [{\citenamefont {Kammermeier}\ \emph
  {et~al.}(2016{\natexlab{b}})\citenamefont {Kammermeier}, \citenamefont
  {Wenk}, \citenamefont {Schliemann}, \citenamefont {Heedt},\ and\
  \citenamefont {\mbox{Th}. Sch\"apers}}]{Kammermeier2016}%
  \BibitemOpen
  \bibfield  {author} {\bibinfo {author} {\bibfnamefont {M.}~\bibnamefont
  {Kammermeier}}, \bibinfo {author} {\bibfnamefont {P.}~\bibnamefont {Wenk}},
  \bibinfo {author} {\bibfnamefont {J.}~\bibnamefont {Schliemann}}, \bibinfo
  {author} {\bibfnamefont {S.}~\bibnamefont {Heedt}}, \ and\ \bibinfo {author}
  {\bibnamefont {\mbox{Th}. Sch\"apers}},\ }\href {\doibase
  10.1103/PhysRevB.93.205306} {\bibfield  {journal} {\bibinfo  {journal} {Phys.
  Rev. B}\ }\textbf {\bibinfo {volume} {93}},\ \bibinfo {pages} {205306}
  (\bibinfo {year} {2016}{\natexlab{b}})}\BibitemShut {NoStop}%
\bibitem [{\citenamefont {Carter}\ \emph {et~al.}(2006)\citenamefont {Carter},
  \citenamefont {Chen},\ and\ \citenamefont {Cundiff}}]{Carter2006}%
  \BibitemOpen
  \bibfield  {author} {\bibinfo {author} {\bibfnamefont {S.~G.}\ \bibnamefont
  {Carter}}, \bibinfo {author} {\bibfnamefont {Z.}~\bibnamefont {Chen}}, \ and\
  \bibinfo {author} {\bibfnamefont {S.~T.}\ \bibnamefont {Cundiff}},\ }\href
  {\doibase 10.1103/PhysRevLett.97.136602} {\bibfield  {journal} {\bibinfo
  {journal} {Phys. Rev. Lett.}\ }\textbf {\bibinfo {volume} {97}},\ \bibinfo
  {pages} {136602} (\bibinfo {year} {2006})}\BibitemShut {NoStop}%
\bibitem [{\citenamefont {Koralek}\ \emph {et~al.}(2009)\citenamefont
  {Koralek}, \citenamefont {Weber}, \citenamefont {Orenstein}, \citenamefont
  {Bernevig}, \citenamefont {Zhang}, \citenamefont {Mack},\ and\ \citenamefont
  {Awschalom}}]{Koralek2009}%
  \BibitemOpen
  \bibfield  {author} {\bibinfo {author} {\bibfnamefont {J.~D.}\ \bibnamefont
  {Koralek}}, \bibinfo {author} {\bibfnamefont {C.~P.}\ \bibnamefont {Weber}},
  \bibinfo {author} {\bibfnamefont {J.}~\bibnamefont {Orenstein}}, \bibinfo
  {author} {\bibfnamefont {B.~A.}\ \bibnamefont {Bernevig}}, \bibinfo {author}
  {\bibfnamefont {S.-C.}\ \bibnamefont {Zhang}}, \bibinfo {author}
  {\bibfnamefont {S.}~\bibnamefont {Mack}}, \ and\ \bibinfo {author}
  {\bibfnamefont {D.~D.}\ \bibnamefont {Awschalom}},\ }\href
  {http://dx.doi.org/10.1038/nature07871} {\bibfield  {journal} {\bibinfo
  {journal} {Nature}\ }\textbf {\bibinfo {volume} {458}},\ \bibinfo {pages}
  {610} (\bibinfo {year} {2009})}\BibitemShut {NoStop}%
\bibitem [{\citenamefont {Wang}\ \emph {et~al.}(2013)\citenamefont {Wang},
  \citenamefont {Liu}, \citenamefont {Balocchi}, \citenamefont {Renucci},
  \citenamefont {Zhu}, \citenamefont {Amand}, \citenamefont {Fontaine},\ and\
  \citenamefont {Marie}}]{Wang2013b}%
  \BibitemOpen
  \bibfield  {author} {\bibinfo {author} {\bibfnamefont {G.}~\bibnamefont
  {Wang}}, \bibinfo {author} {\bibfnamefont {B.~L.}\ \bibnamefont {Liu}},
  \bibinfo {author} {\bibfnamefont {A.}~\bibnamefont {Balocchi}}, \bibinfo
  {author} {\bibfnamefont {P.}~\bibnamefont {Renucci}}, \bibinfo {author}
  {\bibfnamefont {C.~R.}\ \bibnamefont {Zhu}}, \bibinfo {author} {\bibfnamefont
  {T.}~\bibnamefont {Amand}}, \bibinfo {author} {\bibfnamefont
  {C.}~\bibnamefont {Fontaine}}, \ and\ \bibinfo {author} {\bibfnamefont
  {X.}~\bibnamefont {Marie}},\ }\href {http://dx.doi.org/10.1038/ncomms3372}
  {\bibfield  {journal} {\bibinfo  {journal} {Nat. Commun.}\ }\textbf {\bibinfo
  {volume} {4}},\ \bibinfo {pages} {2372} (\bibinfo {year} {2013})}\BibitemShut
  {NoStop}%
\bibitem [{\citenamefont {Bu\ss{}}\ \emph {et~al.}(2011)\citenamefont
  {Bu\ss{}}, \citenamefont {Rudolph}, \citenamefont {Starosielec},
  \citenamefont {Schaefer}, \citenamefont {Semond}, \citenamefont {Cordier},
  \citenamefont {Wieck},\ and\ \citenamefont {H\"agele}}]{Buss2011}%
  \BibitemOpen
  \bibfield  {author} {\bibinfo {author} {\bibfnamefont {J.~H.}\ \bibnamefont
  {Bu\ss{}}}, \bibinfo {author} {\bibfnamefont {J.}~\bibnamefont {Rudolph}},
  \bibinfo {author} {\bibfnamefont {S.}~\bibnamefont {Starosielec}}, \bibinfo
  {author} {\bibfnamefont {A.}~\bibnamefont {Schaefer}}, \bibinfo {author}
  {\bibfnamefont {F.}~\bibnamefont {Semond}}, \bibinfo {author} {\bibfnamefont
  {Y.}~\bibnamefont {Cordier}}, \bibinfo {author} {\bibfnamefont {A.~D.}\
  \bibnamefont {Wieck}}, \ and\ \bibinfo {author} {\bibfnamefont
  {D.}~\bibnamefont {H\"agele}},\ }\href {\doibase 10.1103/PhysRevB.84.153202}
  {\bibfield  {journal} {\bibinfo  {journal} {Phys. Rev. B}\ }\textbf {\bibinfo
  {volume} {84}},\ \bibinfo {pages} {153202} (\bibinfo {year}
  {2011})}\BibitemShut {NoStop}%
\bibitem [{\citenamefont {Jahangir}\ \emph {et~al.}(2012)\citenamefont
  {Jahangir}, \citenamefont {Do\ifmmode~\breve{g}\else \u{g}\fi{}an},
  \citenamefont {Kum}, \citenamefont {Manchon},\ and\ \citenamefont
  {Bhattacharya}}]{Jahangir2012}%
  \BibitemOpen
  \bibfield  {author} {\bibinfo {author} {\bibfnamefont {S.}~\bibnamefont
  {Jahangir}}, \bibinfo {author} {\bibfnamefont {F.}~\bibnamefont
  {Do\ifmmode~\breve{g}\else \u{g}\fi{}an}}, \bibinfo {author} {\bibfnamefont
  {H.}~\bibnamefont {Kum}}, \bibinfo {author} {\bibfnamefont {A.}~\bibnamefont
  {Manchon}}, \ and\ \bibinfo {author} {\bibfnamefont {P.}~\bibnamefont
  {Bhattacharya}},\ }\href {\doibase 10.1103/PhysRevB.86.035315} {\bibfield
  {journal} {\bibinfo  {journal} {Phys. Rev. B}\ }\textbf {\bibinfo {volume}
  {86}},\ \bibinfo {pages} {035315} (\bibinfo {year} {2012})}\BibitemShut
  {NoStop}%
\bibitem [{\citenamefont {Stefanowicz}\ \emph {et~al.}(2014)\citenamefont
  {Stefanowicz}, \citenamefont {Adhikari}, \citenamefont {Andrearczyk},
  \citenamefont {Faina}, \citenamefont {Sawicki}, \citenamefont {Majewski},
  \citenamefont {Dietl},\ and\ \citenamefont {Bonanni}}]{Stefanowicz2014}%
  \BibitemOpen
  \bibfield  {author} {\bibinfo {author} {\bibfnamefont {W.}~\bibnamefont
  {Stefanowicz}}, \bibinfo {author} {\bibfnamefont {R.}~\bibnamefont
  {Adhikari}}, \bibinfo {author} {\bibfnamefont {T.}~\bibnamefont
  {Andrearczyk}}, \bibinfo {author} {\bibfnamefont {B.}~\bibnamefont {Faina}},
  \bibinfo {author} {\bibfnamefont {M.}~\bibnamefont {Sawicki}}, \bibinfo
  {author} {\bibfnamefont {J.~A.}\ \bibnamefont {Majewski}}, \bibinfo {author}
  {\bibfnamefont {T.}~\bibnamefont {Dietl}}, \ and\ \bibinfo {author}
  {\bibfnamefont {A.}~\bibnamefont {Bonanni}},\ }\href {\doibase
  10.1103/PhysRevB.89.205201} {\bibfield  {journal} {\bibinfo  {journal} {Phys.
  Rev. B}\ }\textbf {\bibinfo {volume} {89}},\ \bibinfo {pages} {205201}
  (\bibinfo {year} {2014})}\BibitemShut {NoStop}%
\bibitem [{\citenamefont {Zutic}\ \emph {et~al.}(2004)\citenamefont {Zutic},
  \citenamefont {Fabian},\ and\ \citenamefont {Das~Sarma}}]{Zutic2004a}%
  \BibitemOpen
  \bibfield  {author} {\bibinfo {author} {\bibfnamefont {I.}~\bibnamefont
  {Zutic}}, \bibinfo {author} {\bibfnamefont {J.}~\bibnamefont {Fabian}}, \
  and\ \bibinfo {author} {\bibfnamefont {S.}~\bibnamefont {Das~Sarma}},\ }\href
  {http://arxiv.org/abs/cond-mat/0405528} {\bibfield  {journal} {\bibinfo
  {journal} {Rev. Mod. Phys.}\ }\textbf {\bibinfo {volume} {76}},\ \bibinfo
  {pages} {323} (\bibinfo {year} {2004})}\BibitemShut {NoStop}%
\bibitem [{\citenamefont {Wu}\ \emph {et~al.}(2010)\citenamefont {Wu},
  \citenamefont {Jiang},\ and\ \citenamefont {Weng}}]{Wu2010}%
  \BibitemOpen
  \bibfield  {author} {\bibinfo {author} {\bibfnamefont {M.}~\bibnamefont
  {Wu}}, \bibinfo {author} {\bibfnamefont {J.}~\bibnamefont {Jiang}}, \ and\
  \bibinfo {author} {\bibfnamefont {M.}~\bibnamefont {Weng}},\ }\href {\doibase
  http://dx.doi.org/10.1016/j.physrep.2010.04.002} {\bibfield  {journal}
  {\bibinfo  {journal} {Phys. Rep.}\ }\textbf {\bibinfo {volume} {493}},\
  \bibinfo {pages} {61 } (\bibinfo {year} {2010})}\BibitemShut {NoStop}%
\bibitem [{\citenamefont {Hikami}\ \emph {et~al.}(1980)\citenamefont {Hikami},
  \citenamefont {Larkin},\ and\ \citenamefont {Nagaoka}}]{nagaoka}%
  \BibitemOpen
  \bibfield  {author} {\bibinfo {author} {\bibfnamefont {S.}~\bibnamefont
  {Hikami}}, \bibinfo {author} {\bibfnamefont {A.~I.}\ \bibnamefont {Larkin}},
  \ and\ \bibinfo {author} {\bibfnamefont {Y.}~\bibnamefont {Nagaoka}},\ }\href
  {\doibase 10.1143/PTP.63.707} {\bibfield  {journal} {\bibinfo  {journal}
  {Prog. Theor. Phys.}\ }\textbf {\bibinfo {volume} {63}},\ \bibinfo {pages}
  {707} (\bibinfo {year} {1980})}\BibitemShut {NoStop}%
\bibitem [{\citenamefont {Wenk}\ and\ \citenamefont
  {Kettemann}(2010{\natexlab{b}})}]{wenkbook}%
  \BibitemOpen
  \bibfield  {author} {\bibinfo {author} {\bibfnamefont {P.}~\bibnamefont
  {Wenk}}\ and\ \bibinfo {author} {\bibfnamefont {S.}~\bibnamefont
  {Kettemann}},\ }in\ \href {http://arxiv.org/abs/1012.3575} {\emph {\bibinfo
  {booktitle} {Handbook on Nanophysics}}},\ \bibinfo {editor} {edited by\
  \bibinfo {editor} {\bibfnamefont {K.}~\bibnamefont {Sattler}}}\ (\bibinfo
  {publisher} {Francis \& Taylor},\ \bibinfo {year} {2010})\ p.~\bibinfo
  {pages} {49}\BibitemShut {NoStop}%
\bibitem [{\citenamefont {Golub}(2005)}]{Golub2005a}%
  \BibitemOpen
  \bibfield  {author} {\bibinfo {author} {\bibfnamefont {L.~E.}\ \bibnamefont
  {Golub}},\ }\href@noop {} {\bibfield  {journal} {\bibinfo  {journal} {Phys.
  Rev. B}\ }\textbf {\bibinfo {volume} {71}},\ \bibinfo {pages} {235310}
  (\bibinfo {year} {2005})}\BibitemShut {NoStop}%
\bibitem [{\citenamefont {Glazov}\ and\ \citenamefont
  {Golub}(2009)}]{Glazov2009}%
  \BibitemOpen
  \bibfield  {author} {\bibinfo {author} {\bibfnamefont {M.~M.}\ \bibnamefont
  {Glazov}}\ and\ \bibinfo {author} {\bibfnamefont {L.~E.}\ \bibnamefont
  {Golub}},\ }\href {http://stacks.iop.org/0268-1242/24/i=6/a=064007}
  {\bibfield  {journal} {\bibinfo  {journal} {Semicond. Sci. Tech.}\ }\textbf
  {\bibinfo {volume} {24}},\ \bibinfo {pages} {064007} (\bibinfo {year}
  {2009})}\BibitemShut {NoStop}%
\bibitem [{\citenamefont {Araki}\ \emph {et~al.}(2014)\citenamefont {Araki},
  \citenamefont {Khalsa},\ and\ \citenamefont {MacDonald}}]{Araki2014}%
  \BibitemOpen
  \bibfield  {author} {\bibinfo {author} {\bibfnamefont {Y.}~\bibnamefont
  {Araki}}, \bibinfo {author} {\bibfnamefont {G.}~\bibnamefont {Khalsa}}, \
  and\ \bibinfo {author} {\bibfnamefont {A.~H.}\ \bibnamefont {MacDonald}},\
  }\href {\doibase 10.1103/PhysRevB.90.125309} {\bibfield  {journal} {\bibinfo
  {journal} {Phys. Rev. B}\ }\textbf {\bibinfo {volume} {90}},\ \bibinfo
  {pages} {125309} (\bibinfo {year} {2014})}\BibitemShut {NoStop}%
\bibitem [{\citenamefont {Al'tshuler}\ and\ \citenamefont
  {Aronov}(1981)}]{AltshulerAronov1981}%
  \BibitemOpen
  \bibfield  {author} {\bibinfo {author} {\bibfnamefont {B.~L.}\ \bibnamefont
  {Al'tshuler}}\ and\ \bibinfo {author} {\bibfnamefont {A.~G.}\ \bibnamefont
  {Aronov}},\ }\href {http://jetpletters.ac.ru/ps/1510/article_23070.shtml}
  {\bibfield  {journal} {\bibinfo  {journal} {JETP Lett.}\ }\textbf {\bibinfo
  {volume} {33}},\ \bibinfo {pages} {499} (\bibinfo {year} {1981})},\ \bibinfo
  {note} {[Pis'ma Zh. Eskp. Teor. Fiz. {\bf 33}, No. 10, 515
  (1981)]}\BibitemShut {NoStop}%
\bibitem [{\citenamefont {Aleiner}\ and\ \citenamefont
  {Fal'ko}(2001)}]{Aleiner2001}%
  \BibitemOpen
  \bibfield  {author} {\bibinfo {author} {\bibfnamefont {I.~L.}\ \bibnamefont
  {Aleiner}}\ and\ \bibinfo {author} {\bibfnamefont {V.~I.}\ \bibnamefont
  {Fal'ko}},\ }\href {\doibase 10.1103/PhysRevLett.87.256801} {\bibfield
  {journal} {\bibinfo  {journal} {Phys. Rev. Lett.}\ }\textbf {\bibinfo
  {volume} {87}},\ \bibinfo {pages} {256801} (\bibinfo {year}
  {2001})}\BibitemShut {NoStop}%
\bibitem [{\citenamefont {Meyer}\ \emph {et~al.}(2002)\citenamefont {Meyer},
  \citenamefont {Fal'ko},\ and\ \citenamefont {Altshuler}}]{Meyer2002}%
  \BibitemOpen
  \bibfield  {author} {\bibinfo {author} {\bibfnamefont {J.~S.}\ \bibnamefont
  {Meyer}}, \bibinfo {author} {\bibfnamefont {V.~I.}\ \bibnamefont {Fal'ko}}, \
  and\ \bibinfo {author} {\bibfnamefont {B.~L.}\ \bibnamefont {Altshuler}},\
  }in\ \href@noop {} {\emph {\bibinfo {booktitle} {Nato Science Series II}}},\
  Vol.~\bibinfo {volume} {72},\ \bibinfo {editor} {edited by\ \bibinfo {editor}
  {\bibfnamefont {I.~V.}\ \bibnamefont {Lerner}}, \bibinfo {editor}
  {\bibfnamefont {B.~L.}\ \bibnamefont {Altshuler}}, \bibinfo {editor}
  {\bibfnamefont {V.~I.}\ \bibnamefont {Fal'ko}}, \ and\ \bibinfo {editor}
  {\bibfnamefont {T.}~\bibnamefont {Giamarchi}}}\ (\bibinfo  {publisher}
  {Kluwer Academic Publishers, Dordrecht},\ \bibinfo {year} {2002})\ p.\
  \bibinfo {pages} {117}\BibitemShut {NoStop}%
\bibitem [{\citenamefont {Faria~Junior}\ \emph {et~al.}(2017)\citenamefont
  {Faria~Junior}, \citenamefont {Xu}, \citenamefont {Chen}, \citenamefont
  {Sipahi},\ and\ \citenamefont {\ifmmode \check{Z}\else
  \v{Z}\fi{}uti\ifmmode~\acute{c}\else \'{c}\fi{}}}]{FariaJunior2017}%
  \BibitemOpen
  \bibfield  {author} {\bibinfo {author} {\bibfnamefont {P.~E.}\ \bibnamefont
  {Faria~Junior}}, \bibinfo {author} {\bibfnamefont {G.}~\bibnamefont {Xu}},
  \bibinfo {author} {\bibfnamefont {Y.-F.}\ \bibnamefont {Chen}}, \bibinfo
  {author} {\bibfnamefont {G.~M.}\ \bibnamefont {Sipahi}}, \ and\ \bibinfo
  {author} {\bibfnamefont {I.}~\bibnamefont {\ifmmode \check{Z}\else
  \v{Z}\fi{}uti\ifmmode~\acute{c}\else \'{c}\fi{}}},\ }\href {\doibase
  10.1103/PhysRevB.95.115301} {\bibfield  {journal} {\bibinfo  {journal} {Phys.
  Rev. B}\ }\textbf {\bibinfo {volume} {95}},\ \bibinfo {pages} {115301}
  (\bibinfo {year} {2017})}\BibitemShut {NoStop}%
\bibitem [{\citenamefont {Pershin}\ and\ \citenamefont
  {Slipko}(2010)}]{Pershin2010}%
  \BibitemOpen
  \bibfield  {author} {\bibinfo {author} {\bibfnamefont {Y.~V.}\ \bibnamefont
  {Pershin}}\ and\ \bibinfo {author} {\bibfnamefont {V.~A.}\ \bibnamefont
  {Slipko}},\ }\href {\doibase 10.1103/PhysRevB.82.125325} {\bibfield
  {journal} {\bibinfo  {journal} {Phys. Rev. B}\ }\textbf {\bibinfo {volume}
  {82}},\ \bibinfo {pages} {125325} (\bibinfo {year} {2010})}\BibitemShut
  {NoStop}%
\bibitem [{\citenamefont {Knap}\ \emph {et~al.}(1996)\citenamefont {Knap},
  \citenamefont {Skierbiszewski}, \citenamefont {Zduniak}, \citenamefont
  {Litwin-Staszewska}, \citenamefont {Bertho}, \citenamefont {Kobbi},
  \citenamefont {Robert}, \citenamefont {Pikus}, \citenamefont {Pikus},
  \citenamefont {Iordanskii}, \citenamefont {Mosser}, \citenamefont
  {Zekentes},\ and\ \citenamefont {\mbox{Yu}. B.~Lyanda-Geller}}]{Knap1996}%
  \BibitemOpen
  \bibfield  {author} {\bibinfo {author} {\bibfnamefont {W.}~\bibnamefont
  {Knap}}, \bibinfo {author} {\bibfnamefont {C.}~\bibnamefont
  {Skierbiszewski}}, \bibinfo {author} {\bibfnamefont {A.}~\bibnamefont
  {Zduniak}}, \bibinfo {author} {\bibfnamefont {E.}~\bibnamefont
  {Litwin-Staszewska}}, \bibinfo {author} {\bibfnamefont {D.}~\bibnamefont
  {Bertho}}, \bibinfo {author} {\bibfnamefont {F.}~\bibnamefont {Kobbi}},
  \bibinfo {author} {\bibfnamefont {J.~L.}\ \bibnamefont {Robert}}, \bibinfo
  {author} {\bibfnamefont {G.~E.}\ \bibnamefont {Pikus}}, \bibinfo {author}
  {\bibfnamefont {F.~G.}\ \bibnamefont {Pikus}}, \bibinfo {author}
  {\bibfnamefont {S.~V.}\ \bibnamefont {Iordanskii}}, \bibinfo {author}
  {\bibfnamefont {V.}~\bibnamefont {Mosser}}, \bibinfo {author} {\bibfnamefont
  {K.}~\bibnamefont {Zekentes}}, \ and\ \bibinfo {author} {\bibnamefont
  {\mbox{Yu}. B.~Lyanda-Geller}},\ }\href
  {http://link.aps.org/doi/10.1103/PhysRevB.53.3912} {\bibfield  {journal}
  {\bibinfo  {journal} {Phys. Rev. B}\ }\textbf {\bibinfo {volume} {53}},\
  \bibinfo {pages} {3912} (\bibinfo {year} {1996})}\BibitemShut {NoStop}%
\bibitem [{\citenamefont {Wenk}(2011)}]{wenkdiss}%
  \BibitemOpen
  \bibfield  {author} {\bibinfo {author} {\bibfnamefont {P.}~\bibnamefont
  {Wenk}},\ }\emph {\bibinfo {title} {{Itinerant Spin Dynamics in Structures of
  Reduced Dimensionality}}},\ \href@noop {} {Ph.D. thesis},\ \bibinfo  {school}
  {Jacobs University Bremen} (\bibinfo {year} {2011})\BibitemShut {NoStop}%
\bibitem [{\citenamefont {Hachiya}\ \emph {et~al.}(2014)\citenamefont
  {Hachiya}, \citenamefont {Usaj},\ and\ \citenamefont {Egues}}]{Hachiya2014}%
  \BibitemOpen
  \bibfield  {author} {\bibinfo {author} {\bibfnamefont {M.~O.}\ \bibnamefont
  {Hachiya}}, \bibinfo {author} {\bibfnamefont {G.}~\bibnamefont {Usaj}}, \
  and\ \bibinfo {author} {\bibfnamefont {J.~C.}\ \bibnamefont {Egues}},\ }\href
  {\doibase 10.1103/PhysRevB.89.125310} {\bibfield  {journal} {\bibinfo
  {journal} {Phys. Rev. B}\ }\textbf {\bibinfo {volume} {89}},\ \bibinfo
  {pages} {125310} (\bibinfo {year} {2014})}\BibitemShut {NoStop}%
\bibitem [{\citenamefont {Froltsov}(2001)}]{Froltsov2001}%
  \BibitemOpen
  \bibfield  {author} {\bibinfo {author} {\bibfnamefont {V.~A.}\ \bibnamefont
  {Froltsov}},\ }\href {\doibase 10.1103/PhysRevB.64.045311} {\bibfield
  {journal} {\bibinfo  {journal} {Phys. Rev. B}\ }\textbf {\bibinfo {volume}
  {64}},\ \bibinfo {pages} {045311} (\bibinfo {year} {2001})}\BibitemShut
  {NoStop}%
\bibitem [{\citenamefont {Kurdak}\ \emph {et~al.}(1992)\citenamefont {Kurdak},
  \citenamefont {Chang}, \citenamefont {Chin},\ and\ \citenamefont
  {Chang}}]{Kurdak1992}%
  \BibitemOpen
  \bibfield  {author} {\bibinfo {author} {\bibfnamefont {{\c{C}}.}~\bibnamefont
  {Kurdak}}, \bibinfo {author} {\bibfnamefont {A.~M.}\ \bibnamefont {Chang}},
  \bibinfo {author} {\bibfnamefont {A.}~\bibnamefont {Chin}}, \ and\ \bibinfo
  {author} {\bibfnamefont {T.~Y.}\ \bibnamefont {Chang}},\ }\href {\doibase
  10.1103/PhysRevB.46.6846} {\bibfield  {journal} {\bibinfo  {journal} {Phys.
  Rev. B}\ }\textbf {\bibinfo {volume} {46}},\ \bibinfo {pages} {6846}
  (\bibinfo {year} {1992})}\BibitemShut {NoStop}%
\bibitem [{\citenamefont {Bringer}\ and\ \citenamefont {\mbox{Th}.
  Sch\"apers}(2011)}]{Bringer2011}%
  \BibitemOpen
  \bibfield  {author} {\bibinfo {author} {\bibfnamefont {A.}~\bibnamefont
  {Bringer}}\ and\ \bibinfo {author} {\bibnamefont {\mbox{Th}. Sch\"apers}},\
  }\href {\doibase 10.1103/PhysRevB.83.115305} {\bibfield  {journal} {\bibinfo
  {journal} {Phys. Rev. B}\ }\textbf {\bibinfo {volume} {83}},\ \bibinfo
  {pages} {115305} (\bibinfo {year} {2011})}\BibitemShut {NoStop}%
\bibitem [{\citenamefont {Degtyarev}\ \emph {et~al.}(2017)\citenamefont
  {Degtyarev}, \citenamefont {Khazanova},\ and\ \citenamefont
  {Demarina}}]{Degtyarev2017}%
  \BibitemOpen
  \bibfield  {author} {\bibinfo {author} {\bibfnamefont {V.~E.}\ \bibnamefont
  {Degtyarev}}, \bibinfo {author} {\bibfnamefont {S.~V.}\ \bibnamefont
  {Khazanova}}, \ and\ \bibinfo {author} {\bibfnamefont {N.~V.}\ \bibnamefont
  {Demarina}},\ }\href {https://doi.org/10.1038/s41598-017-03415-3} {\bibfield
  {journal} {\bibinfo  {journal} {Sci. Rep.}\ }\textbf {\bibinfo {volume}
  {7}},\ \bibinfo {pages} {3411} (\bibinfo {year} {2017})}\BibitemShut
  {NoStop}%
\bibitem [{\citenamefont {Iordanskii}\ \emph {et~al.}(1994)\citenamefont
  {Iordanskii}, \citenamefont {\mbox{Yu}. B.~Lyanda-Geller},\ and\
  \citenamefont {Pikus}}]{Iordanskii1994}%
  \BibitemOpen
  \bibfield  {author} {\bibinfo {author} {\bibfnamefont {S.~V.}\ \bibnamefont
  {Iordanskii}}, \bibinfo {author} {\bibnamefont {\mbox{Yu}.
  B.~Lyanda-Geller}}, \ and\ \bibinfo {author} {\bibfnamefont {G.~E.}\
  \bibnamefont {Pikus}},\ }\href
  {http://www.jetpletters.ac.ru/ps/1323/article\_20010.pdf} {\bibfield
  {journal} {\bibinfo  {journal} {JETP Lett.}\ }\textbf {\bibinfo {volume}
  {60}},\ \bibinfo {pages} {206} (\bibinfo {year} {1994})},\ \bibinfo {note}
  {[Pis'ma Zh. Eksp. Teor. Fiz. {\bf 60}, 199 (1994)]}\BibitemShut {NoStop}%
\bibitem [{\citenamefont {Altshuler}\ and\ \citenamefont
  {Aronov}(1981)}]{Altshuler1981}%
  \BibitemOpen
  \bibfield  {author} {\bibinfo {author} {\bibfnamefont {B.~L.}\ \bibnamefont
  {Altshuler}}\ and\ \bibinfo {author} {\bibfnamefont {A.~G.}\ \bibnamefont
  {Aronov}},\ }\href {\doibase 10.1016/0038-1098(81)91153-4} {\bibfield
  {journal} {\bibinfo  {journal} {Solid State Communications}\ }\textbf
  {\bibinfo {volume} {38}},\ \bibinfo {pages} {11} (\bibinfo {year}
  {1981})}\BibitemShut {NoStop}%
\bibitem [{\citenamefont {Kikkawa}\ and\ \citenamefont
  {Awschalom}(1998)}]{Kikkawa1998}%
  \BibitemOpen
  \bibfield  {author} {\bibinfo {author} {\bibfnamefont {J.~M.}\ \bibnamefont
  {Kikkawa}}\ and\ \bibinfo {author} {\bibfnamefont {D.~D.}\ \bibnamefont
  {Awschalom}},\ }\href {\doibase 10.1103/PhysRevLett.80.4313} {\bibfield
  {journal} {\bibinfo  {journal} {Phys. Rev. Lett.}\ }\textbf {\bibinfo
  {volume} {80}},\ \bibinfo {pages} {4313} (\bibinfo {year}
  {1998})}\BibitemShut {NoStop}%
\bibitem [{\citenamefont {Dzhioev}\ \emph {et~al.}(2002)\citenamefont
  {Dzhioev}, \citenamefont {Kavokin}, \citenamefont {Korenev}, \citenamefont
  {Lazarev}, \citenamefont {Meltser}, \citenamefont {Stepanova}, \citenamefont
  {Zakharchenya}, \citenamefont {Gammon},\ and\ \citenamefont
  {Katzer}}]{Dzhioev2002}%
  \BibitemOpen
  \bibfield  {author} {\bibinfo {author} {\bibfnamefont {R.~I.}\ \bibnamefont
  {Dzhioev}}, \bibinfo {author} {\bibfnamefont {K.~V.}\ \bibnamefont
  {Kavokin}}, \bibinfo {author} {\bibfnamefont {V.~L.}\ \bibnamefont
  {Korenev}}, \bibinfo {author} {\bibfnamefont {M.~V.}\ \bibnamefont
  {Lazarev}}, \bibinfo {author} {\bibfnamefont {B.~Y.}\ \bibnamefont
  {Meltser}}, \bibinfo {author} {\bibfnamefont {M.~N.}\ \bibnamefont
  {Stepanova}}, \bibinfo {author} {\bibfnamefont {B.~P.}\ \bibnamefont
  {Zakharchenya}}, \bibinfo {author} {\bibfnamefont {D.}~\bibnamefont
  {Gammon}}, \ and\ \bibinfo {author} {\bibfnamefont {D.~S.}\ \bibnamefont
  {Katzer}},\ }\href {\doibase 10.1103/PhysRevB.66.245204} {\bibfield
  {journal} {\bibinfo  {journal} {Phys. Rev. B}\ }\textbf {\bibinfo {volume}
  {66}},\ \bibinfo {pages} {245204} (\bibinfo {year} {2002})}\BibitemShut
  {NoStop}%
\end{thebibliography}%
\end{document}